\documentclass[aps,prl,superscriptaddress,twocolumn]{revtex4-2}

\usepackage{amssymb}
\usepackage{graphicx}
\usepackage{dcolumn}
\usepackage{bm}
\usepackage{amsmath}
\usepackage{ulem}
\usepackage{textcomp}
\usepackage{float}
\usepackage[usenames]{color}

\usepackage[unicode=true,bookmarks=true,bookmarksnumbered=false,bookmarksopen=false,breaklinks=false,pdfborder={0 0 1},backref=false,colorlinks=true]{hyperref}

\hypersetup{linkcolor=magenta,urlcolor=blue,citecolor=blue,pdfstartview={FitH},hyperfootnotes=false,unicode=true}

\def\be{\begin{equation}}
\def\ee{\end{equation}}
\def\bea{\begin{eqnarray}}
\def\eea{\end{eqnarray}}

\begin{document}
\title{Emergence of Topological States in Relaxation Dynamics of Interacting Bosons}

\author{Wang Huang}
\thanks{These authors contributed equally to this work}
\affiliation{Department of Physics, National University of Defense Technology, Changsha 410073, P. R. China}

\author{Xu-Chen Yang}
\thanks{These authors contributed equally to this work}
\affiliation{Department of Physics, National University of Defense Technology, Changsha 410073, P. R. China}
\affiliation{Hunan Key Laboratory of Extreme Matter and Applications, National University of Defense Technology, Changsha 410073, P. R.  China}

\author{Rui Cao}
\thanks{These authors contributed equally to this work}
\affiliation{Department of Physics, National University of Defense Technology, Changsha 410073, P. R. China}

\author{Ying-Hai Wu}
\email{yinghaiwu88@hust.edu.cn}
\affiliation{School of Physics and Wuhan National High Magnetic Field Center, Huazhong University of Science and Technology, Wuhan 430074, P. R.  China}

\author{Jianmin Yuan}
\affiliation{Institute of Atomic and Molecular Physics, Jilin University, Changchun 130012, P. R. China}
\affiliation{Department of Physics, National University of Defense Technology, Changsha 410073, P. R. China}

\author{Yongqiang Li}
\email{li\_yq@nudt.edu.cn}
\affiliation{Department of Physics, National University of Defense Technology, Changsha 410073, P. R. China}
\affiliation{Hunan Key Laboratory of Extreme Matter and Applications, National University of Defense Technology, Changsha 410073, P. R.  China}

\begin{abstract}
Topological concepts have been employed to understand a wide range of phenomena in physics. While a plethora of ground states are known to have topological underpinning, it is still quite unclear if and how topology manifests itself in the relaxation dynamics of strongly correlated many-body systems. In this paper, we study time evolution of interacting bosons in a finite one-dimensional superlattice and uncover surprising emergent topological phenomena in the long-time stationary states. Beginning with simple product states, the system evolves into stationary states with high energy whose string correlation and entanglement are investigated. Based on extensive numerical simulations and effective-model analysis, it is demonstrated that the stationary states are nonthermal for a wide range of parameters, and they exhibit certain features that are characteristic of the symmetry-protected topological ground state of the Hamiltonian. In contrast, no topological feature is found in the stationary state as long as the system thermalizes. This difference is further corroborated by the distinct behaviour of quantum entanglement and edge states of the system. Our theoretical prediction can be examined by current experimental techniques and paves the way for a more comprehensive understanding of topological phases in nonequilibrium settings.
\end{abstract}

\date{\today}

\maketitle

{\it Introduction.} In non-relativistic quantum mechanics, the time evolution of an isolated interacting system is deterministic as dictated by its Hamiltonian~\cite{Polkovnikov2010ColloquiumND,Deutsch18,Ueda2020QuantumET,bertini2021finite}. It is expected that relaxation dynamics of a nonequilibrium initial state results in a steady state whose local observables are described by the generalized Gibbs ensemble \cite{Marcos07,Marcos08,Marcos09,James18} or Gibbs ensemble~\cite{PhysRevE.50.888,Srednicki1996ThermalFI,Srednicki1998TheAT,PhysRevA.43.2046}. Experimental investigations of relaxation dynamics are challenging because a sufficiently large number of quantum objects should be isolated from the environment and precisely manipulated for a considerable amount of time. In this regard, ultracold atomic gases provide a versatile platform for exploring nonequilibrium quantum dynamics, where the system can be driven far from equilibrium by sudden changes in the Hamiltonian~\cite{Mitra2018}.
To date, most studies along this direction employed many-body systems that are trivial from the topological perspective~\cite{Eisert_2015,Kaufman2016QuantumTT,Nichols2018SpinTI,Jepsen2020SpinTI,Wei2021QuantumGM,Zhou2021ThermalizationDO,Christakis2022ProbingSC,Li_2023}.

It is by now well established that the properties of some many-body states should be described using topological concepts. The most common setting is to explore topology in the ground states of many-body systems that are well separated from the excited states~\cite{RevModPhys.82.3045,RevModPhys.83.1057,RevModPhys.91.015005,Rachel2018InteractingTI,Zhang2018TopologicalQM}. Remarkable experimental progresses on ground-state topology of interacting systems have been reported in several quantum simulators~\cite{deLsleuc2018ObservationOA,Sompet2021RealizingTS,Semeghini_2021,Kiczynski2022EngineeringTS}. However, far-from-equilibrium dynamics and long-time stationary state of interacting topological systems are largely unexplored~\cite{Cai19,Zhou23,Kwan23,Viebahn23,Katz24}. For an isolated system, its total energy remains constant during time evolution~\cite{schuckert2023observation-old}. Consequently, the time-evolved stationary state is a high-lying excited state that has substantial overlap with an appreciable number of eigenstates [see Fig.~\ref{figure-1}(a)], and the physics is expected to be quite different from the ground state. It raises the question as to what extent topological fingerprints of the Hamiltonian can be revealed during time evolution, and whether an isolated system initially prepared in a nonequilibrium state can reach a new equilibrium with interesting topological phenomena. These issues are also intimately connected with ongoing experiments using ultracold atoms.

\begin{figure}[h!]
\includegraphics[trim = 0mm 0mm 0mm 0mm, clip=true, width=0.475\textwidth]{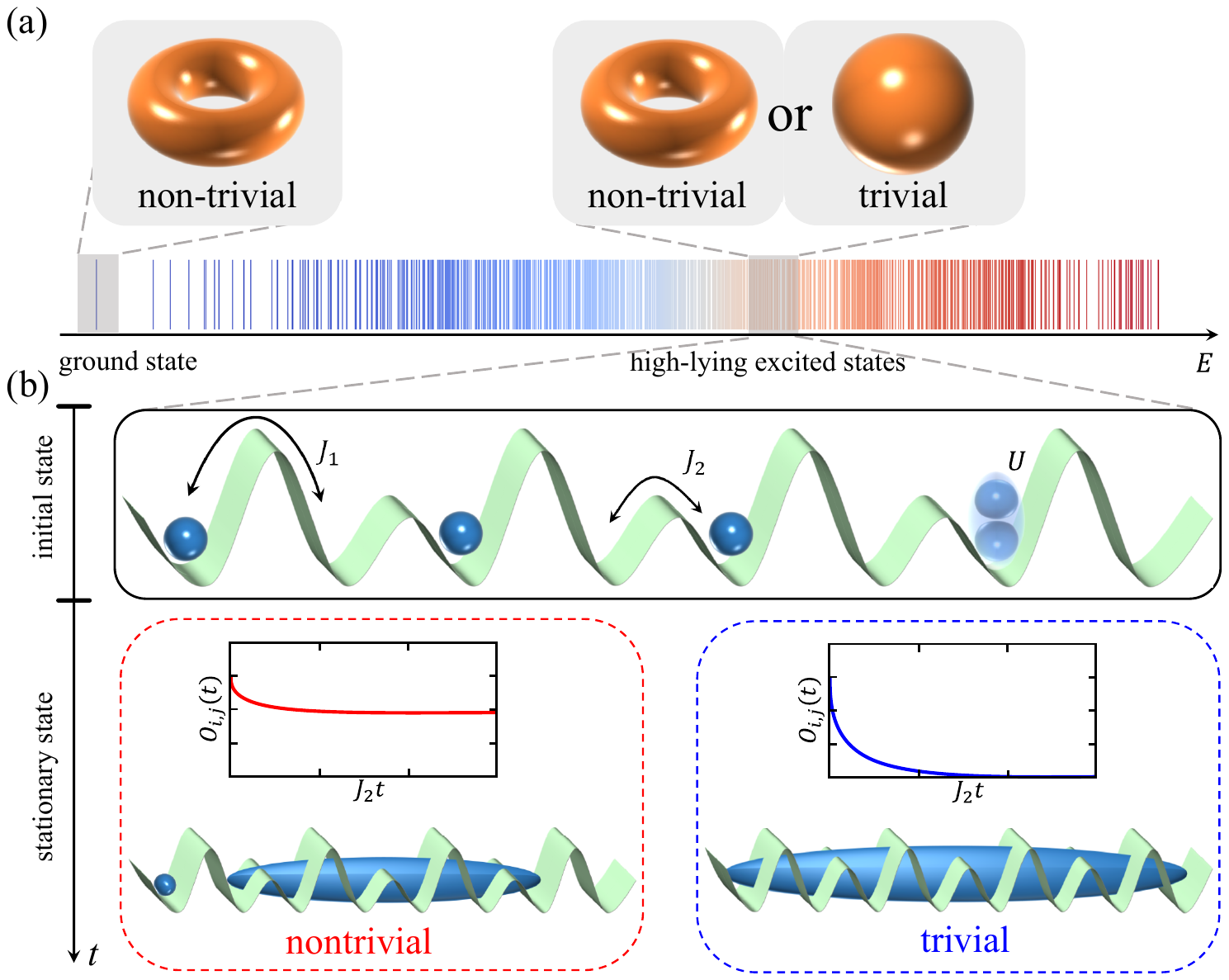}
\caption{Emergent topology in the relaxation dynamics of an interacting system. (a) The most commonly studied scenario of topological phases is based on the properties of the many-body ground state that are separated from the excited states by a gap. In contrast, it is not transparent if and how the high-lying excited states that form a continuum can be defined as topological. (b) Schematic of the one-dimensional superlattice Bose-Hubbard model and emergence of topological states that are far from the ground state in its relaxation dynamics. The system has alternating hopping $J_{1,2}$ and onsite repulsion $U$ and is initialized in certain simple product states (upper panel). Depending on the values of $J_{1,2}$ and $U$, the long-time stationary state may be topologically nontrivial with long-range string correlation or trivial without such correlation (lower panel).}
\label{figure-1}
\end{figure}

In this paper, we provide compelling evidence for the emergence of topological states in the far-from-equilibrium relaxation dynamics of interacting bosons in a finite one-dimensional superlattice. As illustrated in Fig.~\ref{figure-1}(b), the lattice potential is modulated such that the tunneling between two neighboring sites have different ratios depending on the height of the potential barrier. For our purpose, the system can be described by a superlattice Bose-Hubbard model with alternating hopping terms and onsite interactions. This is a bosonic analogue of the Su-Schrieffer-Heeger model~\cite{PhysRevLett.42.1698}, whose many-body ground state has been studied in several previous works~\cite{grusdt2013topological,Zhu2013Topological}. Using the staggered-immersion cooling method~\cite{Yang2019CoolingAE,Wang2022InterrelatedTA}, the system can be initially prepared in certain product states that are far from equilibrium and topologically trivial. For a wide range of parameters, the system relaxes and reaches nonthermal stationary states that have nontrivial topological properties reminiscent of the many-body ground state, as revealed by nonlocal string correlation, entanglement, and edge state. Previous works mainly focused on topological response of the noninteracting systems before relaxing back to equilibrium with the time-evolved state normally being trivial~\cite{Caio_2015, D_Alessio_2015, PhysRevLett.117.126803,PhysRevLett.117.235302, PhysRevLett.118.185701, PhysRevLett.121.250601,PhysRevB.97.060304,Zhang_2018,flaschner2018observation,tarnowski2019measuring}, or short-time quench dynamics of the interacting topological systems~\cite{McGinley_2018,McGinley_2019,PhysRevB.103.224308,PhysRevResearch.2.033259}. In contrast, we reveal an unknown interaction-induced many-body phenomenon in long-time relaxation dynamics, i.e. a topologically nontrivial stationary state emerges from a trivial product state. An experimental protocol is proposed to verify our theoretical results using currently available techniques.

{\it Model and method.} We study nonequilibrium dynamics of interacting bosons in a superlattice shown in Fig.~\ref{figure-1}(b). Each unit cell contains two lattice sites with equal potential depths such that the system possesses bond-inversion symmetry. In the tight-binding limit, it is described by the superlattice Bose-Hubbard Hamiltonian~\cite{grusdt2013topological}
\begin{eqnarray}
\label{eq:Ham}
\hat{H} &=& -J_1\sum_{i,\text{odd}} \hat{b}^\dagger_{i}\hat{b}_{i+1} - J_2\sum_{i,\text{even}} \hat{b}^\dagger_{i}\hat{b}_{i+1} + \rm{H.c.} \nonumber \\
&\phantom{=}& +\frac{U}{2}\sum_{i=1}^L\hat{n}_i(\hat{n}_i-1).
\end{eqnarray}
Here, $L$ is the number of sites, $\hat{b}^{\dagger}_{i}$ ($\hat{b}_{i}$) is the creation (annihilation) operator on site $i$, $\hat{n}_{i} = \hat{b}^{\dagger}_{i} \hat{b}_i$ is the particle number operator, $J_{1}$ ($J_{2}$) is the intracell (intercell) tunneling strength, and $U$ is the onsite repulsive interaction. The energy of our system is measured in units of $J_{2}$. Depending on the system parameters and filling factors, superfluid, topologically trivial and nontrivial Mott insulators may be realized~\cite{grusdt2013topological,Zhu2013Topological}. The nontrivial Mott phase is a symmetry-protected topological (SPT) state that is connected to the celebrated Haldane phase~\cite{deLsleuc2018ObservationOA,Haldane1983NonlinearFT,ChenX2012}. It exhibits a hidden order that can be characterized using the nonlocal string order parameter
\begin{eqnarray}
\label{eq:str}
\hat{O}_{i,j} &=& -4\delta\hat{n}_{i} \prod^{j-1}_{k=i+1} \text{e}^{\text{i}\pi\delta\hat{n}_{k}} \delta\hat{n}_{j},
\end{eqnarray}
with $\delta\hat{n}_{i} = \bar{n}-\hat{n}_{i}$ ($\bar{n}$ is the average particle density)~\cite{Torre06,Erez08,Mazza14}. The expectation value with respect to the ground state remains finite and approaches a nonzero constant as $|i-j|\rightarrow \infty$ in the SPT phase~\cite{Nijs89,Kennedy92}. The phase boundary between topologically trivial and nontrivial phases is $J_{1}/J_{2}=1$ for the noninteracting case, but is shifted to $J_{1}/J_{2}<1$ by many-body interactions~\cite{SM}. In subsequent calculations, we mainly focus on two representative cases, which are $J_{1}/J_{2}=0.2, U=8$ in the topological regime and $J_{1}/J_{2}=0.8, U=6$ in the trivial regime.

To study the many-body relaxation dynamics of the interacting system, we employ both exact diagonalization (ED) and the time dependent variational principle (TDVP) in the framework of matrix product states (MPS). The cases with $L \leq 12$ are studied using the ED programs in the Quspin package~\cite{Phillip2017QuSpin}. The two-site and one-site hybrid TDVP is applied to the cases with $L>12$~\cite{Haegeman2011Time}. The TDVP method is favourable because it preserves unitarity for a considerable amount of time using moderate bond dimensions~\cite{Goto2019Performance,Sebastian2019Time}. In the two-site steps, we choose truncation error $\varepsilon= 10^{-10}$ and time step $J\delta t =0.05$. Each lattice site is allowed to host at most four bosons. The canonical ensemble properties are computed using the matrix product purification method~\cite{Barthel2016Matrix,Feiguin2005Finite}, where a finite-temperature state is obtained by imaginary time evolution from an initial state with infinite temperature. Both TDVP and finite temperature simulations are performed using the ITensor library~\cite{Matthew2022The}.

{\it Time-evolved stationary state.}  In the limits of $J_{1,2}/U \rightarrow 0$ and $J_1/J_2\rightarrow 0$ or $J_2/J_1\rightarrow 0$, the many-body Hamiltonian in Eq.~(\ref{eq:Ham}) is integrable and time evolution is trivial. Instead, we focus on the nonequilibrium dynamics in the regime where $J_{1,2}/U$ and $J_1/J_2$ are finite, where one may naively expect that the system thermalizes in the long-time limit. The system is initialized in trivial product states $\left|\psi_{0}\right\rangle=\left|a_{1} a_{2}\cdots\right\rangle$, where the particle number on each site $a_{i}=0,1$ ($\sum_{i} a_{i}=N$). We choose open boundary conditions for the system and focus on the half-filled cases with the number of bosons being $N=L/2$.

\begin{figure*}
\includegraphics[trim = 0mm 0mm 0mm 0mm, clip=true, width=0.95\textwidth]{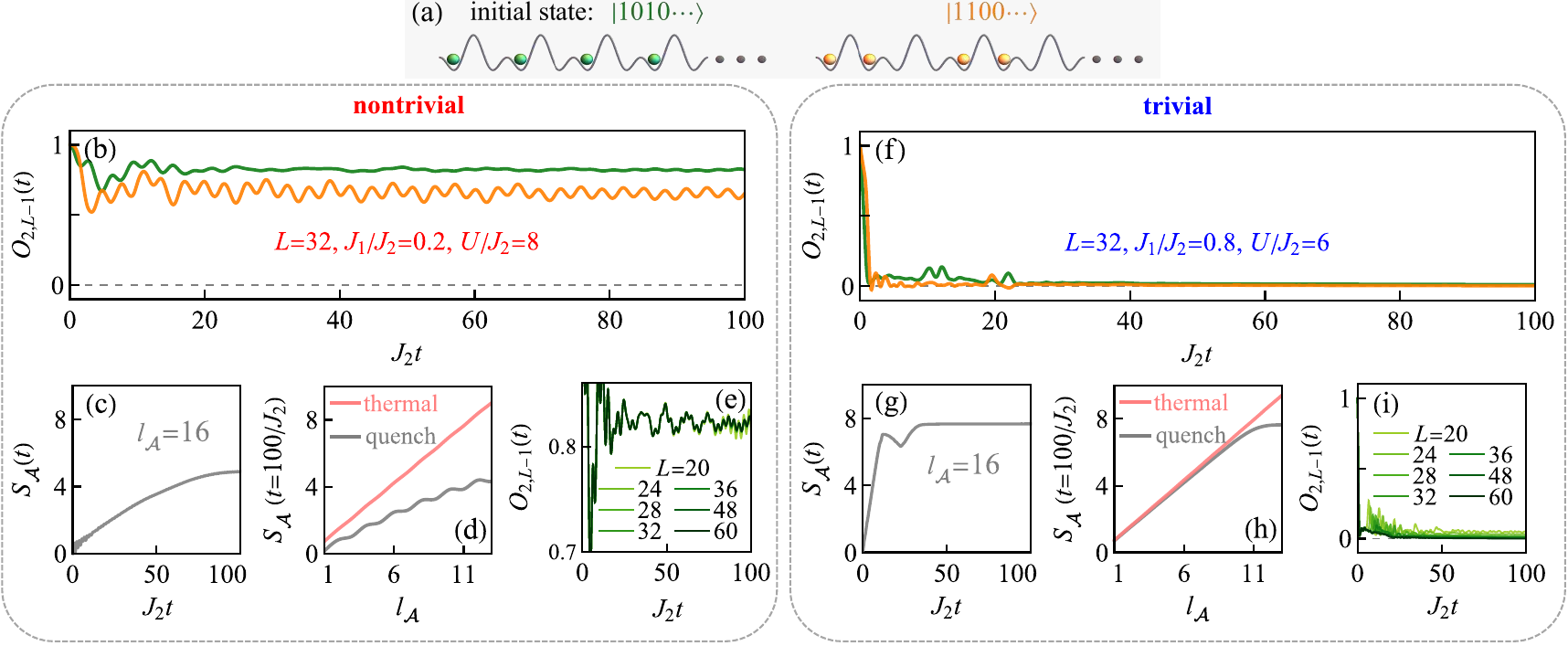}
\caption{Relaxation dynamics in topologically nontrivial and trivial cases. (a) Schematics of the topologically trivial product states $\left|1010\cdots\right\rangle$ (green) and $\left|1100\cdots\right\rangle$ (orange) that are used as initial states in our simulations. (b-e) Numerical results for the nontrivial case with $J_1/J_2=0.2$ and $U/J_2=8$. (b) The string correlation $O_{2,L-1}(t)$ of the time-evolved stationary state attains an appreciable value, which is distinct from the thermal averaged value $O_{2,L-1}^{\rm thermal} \approx 0$ at the mean energy of the initial state (represented by the dashed line). (c) Time evolution of the second-order Renyi entanglement entropy $S_{\mathcal A}(t)$ at the center of the $L=32$ system. (d) The Renyi entropy at $t=100/J_2$ and the thermal entropy given by the canonical ensemble for the $L=32$ system. The discrepancy suggests that the system has not thermalized. (e) $O_{2,L-1}(t)$ has quite weak dependence on $L$ so its nonzero value should not be finite-size effects. (f-i) Numerical results for the trivial case with $J_1/J_2=0.8$ and $U/J_2=6$. Each panel presents the same quantity as its counterpart in panels (b-e). The time-evolved stationary state has no long-range string correlation, which is consistent with the value $O_{2,L-1}^{\rm thermal} \approx 0$ for a thermalized system. The time and subsystem size dependence of the Renyi entropy also suggest that thermalization occurs. Finite-size effects are also excluded by studying multiple $L$. Numerical simulations are performed for system size up to $L=60$ and bond dimension up to $3000$~\cite{SM}.}
\label{figure-2}
\end{figure*}

In light of the state-of-the-art experimental method for state preparation, we first consider a quench from the initial state $\left|\psi_0\right\rangle=\left|1010\cdots\right\rangle$ or $\left|1100\cdots\right\rangle$ [see Fig.~\ref{figure-2}(a)] using a Hamiltonian whose ground state is topological ($J_1/J_2=0.2$, $U/J_2=8$). It is found that the many-body system relaxes and reaches a long-time stationary state that is far from the ground state. Surprisingly, the time-evolved high-lying state has long-range string order. As shown in Fig.~\ref{figure-2}(b), the correlation between the second and $(L-1)$-th sites $O_{2,L-1}(t)=\left\langle\psi(t)\right|\hat{O}_{2,L-1}\left|\psi(t)\right\rangle$ oscillates at the beginning but quickly attains a stable value with negligible fluctuations. The variation of $O_{i,j}(t)$ with $|i-j|$ is also consistent with the existence of long-range order~\cite{SM}. This behaviour is observed for all system sizes that we have studied [see Fig.~\ref{figure-2}(e)], so it should be a genuine phenomenon but not a finite-size effect. Moreover, the string order is remarkably robust when the initial state is varied~\cite{SM}. For the TDVP calculations, multiple bond dimensions have been tested such that the maximal one captures the entanglement of the time-evolved state~\cite{SM}. For the small $L=12$ system, long-range string order is observed for time up to $t=1000/J_2$ using ED~\cite{SM}.

The existence of long-range string correlation is associated with the absence of thermalization. At the energy corresponding to the initial state $\left|\psi_0\right\rangle$, thermal average of the string correlation is $O_{2,L-1}^{\rm thermal} \approx 0$. If the system reaches thermal equilibrium, $O_{2,L-1}(t)$ should approach zero in the long-time limit, but we clearly obtain nonzero values depending on the initial states. Thermalization can also be probed using the second-order Renyi entanglement entropy $S_{\mathcal A}(t) = -\text{log} \left[\text{Tr}\left(\rho^{2}_{\mathcal A}(t)\right)\right]$, where $\mathcal A$ is a subsystem with $l_{\mathcal A}$ sites and $\rho_{\mathcal A}(t)$ is the associated reduced density matrix. The half-system entropy increases with time and begins to saturate to a constant at the late time $t=100/J_2$ [see Fig.~\ref{figure-2}(c)]. The saturated entropy is noticeably smaller than the thermal entropy given by canonical ensemble as shown in Fig.~\ref{figure-2}(d), which suggests that the time-evolved stationary state is nonthermal~\cite{Calabrese_2005}.

Next we study the topologically trivial case with $J_1/J_2=0.8$ and $U/J_2=6$. In contrast to previous results, the system does thermalize and memory about the initial states is lost when it reaches equilibrium. In Fig.~\ref{figure-2}(f) and (i), the string correlation $O_{2,L-1}(t)$ quickly decays to nearly zero values for both initial states. The dynamics of $S_{\mathcal A}(t)$ for $l_{\mathcal A}=L/2$ is presented in Fig.~\ref{figure-2}(g). It increases with time linearly for a certain interval and eventually reaches a maximum value that is constrained by the MPS bond dimension. The subsystem size dependence of $S_{\mathcal A}(t=100/J_2)$ is displayed in Fig.~\ref{figure-2}(h). It grows linearly with $l_{\mathcal A}$ and agrees with the thermal entropy given by the canonical ensemble~\cite{Kaufman2016QuantumTT}. These features clearly distinguish this case from the previous one.

It is intriguing that the nonequilibrium phase transition and thermalization is absent for hardcore interaction $U=\infty$ but can only be observed in the soft-core case~\cite{SM}. To further understand this phenomenon, perturbation theory is employed to construct an effective model at half-filling for the $J_{1,2}/U\ll 1$ regime in which terms up to second order of $J_{1,2}/U$ are kept. Time evolution using the effective model and the original model in Eq.~(\ref{eq:Ham}) agrees with each other remarkably well in the soft-core cases~\cite{SM}. But for the hardcore case, it is instead found that both local and nonlocal observables exhibit persistent oscillations up to the time $t=100/J_2$~\cite{SM}.

\begin{figure}[t]
\includegraphics[trim = 0mm 0mm 0mm 0mm, clip=true, width=0.475\textwidth]{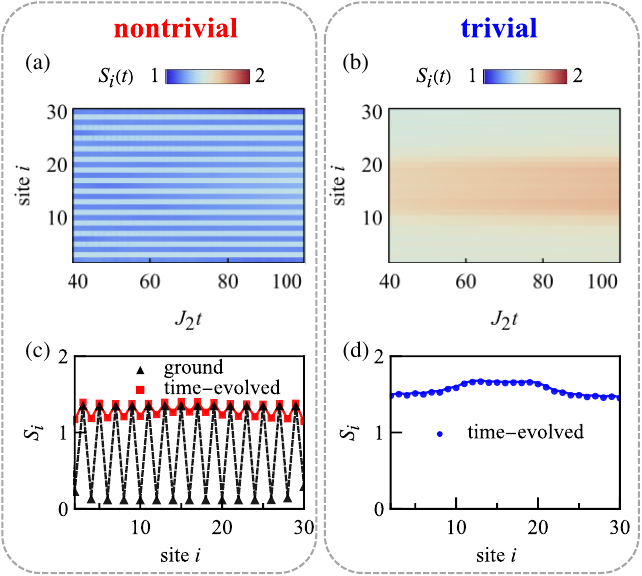}
\caption{Dynamics of the two-site entropy. (a) For the nontrivial case with $J_1/J_2=0.2$ and $U/J_2=8$, the two-site entropy oscillates between large and small values on odd and even lattice sites in the bulk of the system. (b) For the trivial case with $J_1/J_2=0.8$ and $U/J_2=6$, the two-site entropy is almost homogeneous as one would expect for a thermal state. The line cut at $t=100/J_{2}$ in panels (a) and (b) are shown respectively in panels (c) and (d). The same quantity for the nontrivial many-body ground state is shown in panel (c) for comparison. The oscillations in the ground state and the nontrivial time-evolved stationary state are similar. All panels are for the $L=36$ system.}
\label{figure-3}
\end{figure}

In addition to the string order parameter, we have also studied other quantities of the stationary state to substantiate its topological nature. It is found that the parity and density-wave orders vanish in the bulk so the long-range string correlation is not caused by some trivial states~\cite{SM}. For two adjacent lattice sites $i$ and $i+1$, their reduced density matrix $\rho_{i,i+1}(t)$ can be used to define the two-site entanglement entropy $S_{i}(t)=-\text{Tr}\left[\rho_{i,i+1}(t)\text{ln}\rho_{i,i+1}(t)\right]$~\cite{PhysRevLett.96.116401}. In the nontrivial case, $S_{i}(t)$ oscillates between large and small values with a period of two sites for time up to $t=100/J_2$ [see Fig.~\ref{figure-3}(a)(c)]. When the system is trivial, $S_{i}(t)$ reaches its maximum value quickly and is basically homogeneous in real space [see Fig.~\ref{figure-3}(b)(d)]. These features are analogous to those of the many-body ground state~\cite{Torre06,Erez08,Mazza14} so the time-evolved state shares similar topological features as the ground state. The dynamics of entanglement spectrum has also been investigated and multiple level crossings are observed~\cite{SM}, which is similar to previous results on free fermions~\cite{PhysRevLett.121.250601}.

{\it Experimental protocol.} Our proposal can be readily realized using ultracold bosons trapped in a one-dimensional spin-dependent optical superlattice~\cite{Anderlini_2007}. The inhomogeneity induced by the Gaussian beams can be compensated by a tailored light profile to generate a flat box potential with hard walls~\cite{2017A,2021Quantum}. The superlattice should be adjusted to create balanced double wells so the system is described by Eq.~(\ref{eq:Ham}) with alternating tunneling strengths $J_1$ and $J_2$ controlled by the superlattice depth. It is essential to prepare the system in simple product state at half-filling in which every two sites connected by the $J_2$ hopping in the bulk has only one boson~\cite{SM}. Actually, high-fidelity preparation of our initial state, whose odd (even) sites are singly occupied (empty) [see Fig.~\ref{figure-2}(a)], has already been achieved in previous experiments using the staggered-immersion cooling method~\cite{Yang2019CoolingAE,Wang2022InterrelatedTA}.

\begin{figure}[t]
\includegraphics[trim = 0mm 0mm 0mm 0mm, clip=true, width=0.485\textwidth]{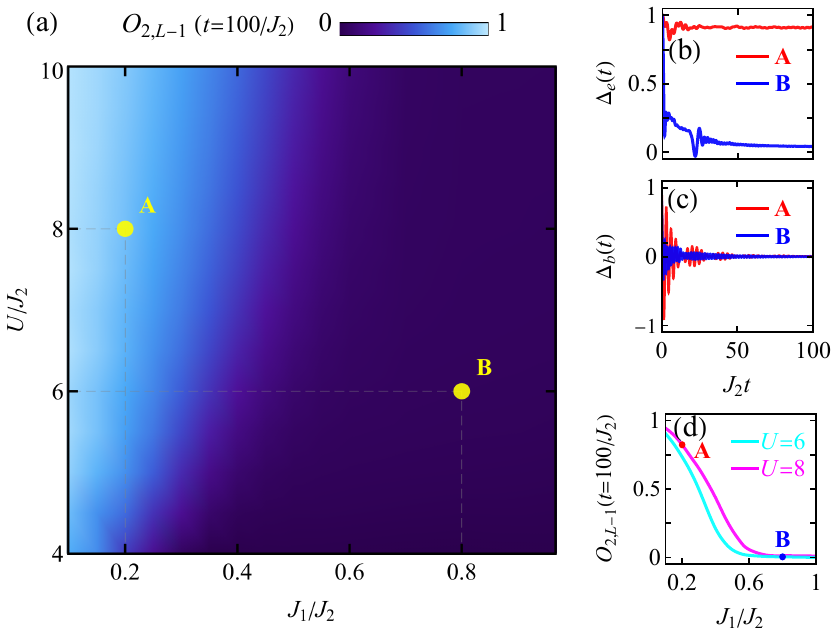}
\caption{ Nonequilibrium phase diagram of the superlattice Bose-Hubbard model. (a) Color map of the string correlation $O_{2,L-1}(t=100/J_2)$ in the $L=36$ system for a wide range of $J_1/J_2$ and $U/J_2$. The point A (B) indicates the topologically nontrivial (trivial) case that has been studied in detail. (b)(c) Time evolution of the imbalance $\Delta_e(t)$ at the edge and $\Delta_b(t)$ in the bulk. The two quantities exhibit very different behavior for the A point but are similar for the B point. (d) The long-time limit of $O_{2,L-1}(t)$ as a function of the hopping strengths for two different interactions.}
\label{figure-4}
\end{figure}

To further assess the experimental feasibility, we have explored the nonequilibirum phase diagram for a wide range of parameters. By inspecting the string order parameter at $t=100/J_2$ (typical timescale realized experimentally), it is found that topologically nontrivial stationary states appear in a large portion of the phase diagram, as shown in Fig.~\ref{figure-4}(a)(d). This suggests that nontrivial relaxation dynamics should be quite accessible. In view of the single-site resolution provided by quantum gas microscope~\cite{Gross_2021}, we expect that nonlocal string correlation~\cite{Sompet2021RealizingTS} and entanglement entropy~\cite{Kaufman2016QuantumTT} can be measured in future experiments.

In addition, time evolution of the edge states could also reveal useful information, as experiments on superconducting circuits and trapped ions have already demonstrated~\cite{Cai19,Katz24}. It should be fruitful to measure the imbalance $\Delta_{e}(t)\equiv n_\text{1}(t)-n_\text{L}(t)\neq 0$ associated with the two edges and the averaged imbalance $\Delta_{b}(t)\equiv \frac{2}{L-2}\sum^{L-1}_{i=2}[n_\text{i,odd}(t)-n_\text{i,even}(t)]=0$ in the bulk, where $n_i(t)=\langle \psi (t)| {\hat n}_i |\psi(t) \rangle$. If the system reaches topologically nontrivial stationary state, $\Delta_{e}(t)$ has an appreciable value throughout the time evolution but $\Delta_{b}(t)$ decays to zero after a short period [see red lines in Fig.~\ref{figure-4}(b)(c)]. This can be taken as evidence for topologically protected edge states in the far-from-equilibrium dynamics. In contrast, both quantities rapidly decay to negligible values in the trivial case [see blue lines in Fig.~\ref{figure-4}(b)(c)].

{\it Discussion.} We have uncovered surprising emergence of topological states in the far-from-equilibrium long-time evolution of strongly interacting systems. This study goes beyond previous works that mainly studied topological systems consist of noninteracting particles~\cite{Caio_2015,D_Alessio_2015,flaschner2018observation,tarnowski2019measuring} or short-time evolution of interacting systems~\cite{McGinley_2018,McGinley_2019,PhysRevB.103.224308}, and is related to ongoing experiments on ultracold atoms. Topologically distinct behaviour in the relaxation dynamics are observed as a trivial initial state can evolve to a stationary state that exhibits topological properties in certain parameter regimes. While product states were used as initial states here, other high-lying initial states may also lead to intricate topological physics. It is also possible to induce nonequilibrium topological critical phenomena and study their universality classes and scaling laws. Time evolution and nonequilibrium topology of interacting systems in higher dimensions should also be very interesting. Finally, whether and how nonlocal observables thermalize after a sufficiently long time in large systems calls for more investigations~\cite{Marcos09,PhysRevLett.100.175702}. These questions are very difficult to address from either the experimental or theoretical side and beyond the scope of this work.

We acknowledge useful discussions with Zhen-Sheng Yuan, Xiong-Jun Liu, Bing Yang, and Jun-Jun Xu. This work is supported by the National Natural Science Foundation of China under ation of China under Grants Nos. 12374252, 12074431, and 12174130, and the Excellent Youth Foundation of Hunan Scientific Committee under Grant No. 2021JJ10044. Numerical calculations were performed on the TianHe-1A cluster of the National Supercomputer Center at Tianjin.

%

\clearpage
\onecolumngrid

\setcounter{figure}{0}
\setcounter{table}{0}
\setcounter{equation}{0}
\renewcommand{\thefigure}{S\arabic{figure}}
\renewcommand{\thetable}{S\arabic{table}}
\renewcommand{\theequation}{S\arabic{equation}}

\centerline{\Large{\bf Supplemental Material}}

\maketitle

\renewcommand{\theenumi}{\Roman{enumi}}  %
\renewcommand{\theenumii}{\Alph{enumii}} %
\renewcommand{\theenumiii}{\arabic{enumiii}} %

\section*{Contents}
\addcontentsline{toc}{chapter}{Table of Contents}
\begin{enumerate}
  \item \hyperref[chapter:groundstate]{Many-body ground states of the superlattice Bose-Hubbard model\hfill 1}
  \item \hyperref[chapter:dynamics]{Relaxation dynamics of the superlattice Bose-Hubbard model\hfill 3}
    \begin{enumerate}
      \item \hyperref[section:tdvp_ed]{Comparison of TDVP with ED\hfill 4}
      \item \hyperref[section:dynamics_sto]{Dynamics of the string correlation\hfill 4}
      \item \hyperref[section:dynamics_sto_Id_dw]{Dynamics of parity and density-wave correlations\hfill 5}
      \item \hyperref[section:dynamics_initial_state]{Initial-state dependence of relaxation dynamics \hfill 6}
      \item \hyperref[section:dynamics_ee]{Dynamics of the entanglement entropy\hfill 8}
      \item \hyperref[section:dynamics_es]{Dynamics of the entanglement spectrum\hfill 8}
    \end{enumerate}
  \item \hyperref[chapter:analysis]{Analytical results and effective models\hfill 9}
      \begin{enumerate}
      \item \hyperref[section:non_trivial_limit]{The nontrivial limit $J_1/J_2\to 0$\hfill 9}
      \item \hyperref[section:trivial_limit]{The trivial limit $J_2/J_1\to 0$\hfill 10}
      \item \hyperref[section:effective]{Effective model in the strongly correlated regime\hfill 10}
        \begin{enumerate}
            \item \hyperref[subsection:first_order]{First-order effective model\hfill 11}
            \item \hyperref[subsection:second_order]{Second-order effective model\hfill 12}
        \end{enumerate}
      \end{enumerate}
\end{enumerate}

\vspace{1em}

In this Supplemental Material, we provide more details about the ground-state properties, and far-from-equilibrium many-body dynamics, and construct effective models to provide additional insight into the physics. Without loss of generality and unless otherwise specified, we focus on the two cases studied in the main text, i.e., $J_1/J_2=0.2$ and $U/J_2=8$ in the nontrivial regime, and $J_1/J_2=0.8$ and $U/J_2=6$ in the trivial regime.

\section{Many-body ground states of the superlattice Bose-Hubbard model}\label{chapter:groundstate}

\begin{figure}[htb!]
\includegraphics[width=0.8\textwidth]{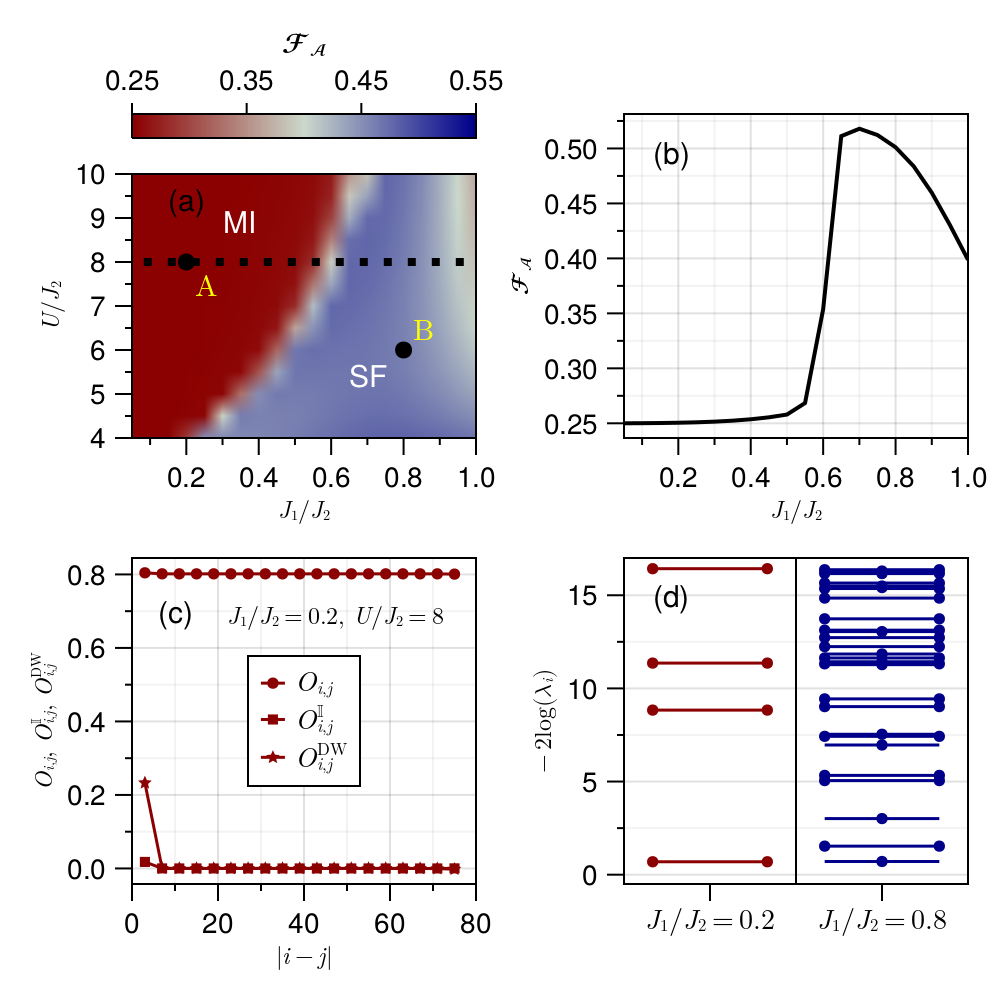}
\caption{{\bf Ground-state properties of the one-dimensional superlattice Bose-Hubbard model.} (a) Phase diagram of the half-filled system as a function of $J_1/J_2$ and $U/J_2$. The lattice size is $L=80$ and open boundary condition is used. There is one MI phase and one SF phase whose boundary is determined by computing the bipartite fluctuation $\mathcal{F}_{\mathcal A}$. Two representative points are labeled for subsequent studies: the nontrivial case A corresponds to $J_1/J_2=0.2,U/J_{2}=8$ and the trivial case B corresponds to $J_1/J_2=0.8,U/J_{2}=6$. (b) The line cut of $\mathcal{F}_{\mathcal A}$ as a function of $J_1/J_2$ for $U/J_2=8$ [the dotted line in (a)]. (c) The string correlation $O_{i,j}$, parity correlation $O^{\mathbb{I}}_{i,j}$, and density-wave correlation $O^{\rm DW}_{i,j}$ for the nontrivial MI phase versus the lattice site distance $|i-j|$. $O_{i,j}$ is long-ranged but $O^{\mathbb{I}}_{i,j}$ and $O^{\rm DW}_{i,j}$ are not. (d) Entanglement spectrum for the nontrivial (red lines and dots) and the trivial cases (blue lines and dots). There is a two-fold degeneracy for each entanglement level in the nontrivial case.}
     \label{fig:bi_fluc}
\end{figure}

This section reviews the ground-state properties of the one-dimensional superlattice Bose-Hubbard model. We choose open boundary condition for the system and focus on the $J_1/J_2<1$ regime at half filling. It has been established that this regime supports two possible phases: a Mott insulator (MI) for strong interaction and a superfluid (SF) for weak interaction~\cite{grusdt2013topological,Zhu2013Topological}. For the $L=80$ case, the many-body phase diagram obtained using density matrix renormalization group (DMRG)~\cite{PhysRevLett.69.2863} simulations is presented in Fig.~\ref{fig:bi_fluc}(a). The boundary between the MI and SF phases is determined by the bipartite particle-number fluctuations
\begin{eqnarray}
\mathcal{F}_{\mathcal A} = \left( \sum_{i\in {\mathcal A}}\hat{n}_i \right)^2 - \sum_{i\in {\mathcal A}}\hat{n}^2_i, \label{eq:bi_fluc}
\end{eqnarray}
where $\mathcal{A}$ the left subsystem of the chain. This quantity is suppressed in the MI phase but has appreciable values in the SF phase. For a fixed choice of $U/J_{2}$, a quantum phase transition can be induced by tuning $J_1/J_2$, as shown in Fig.~\ref{fig:bi_fluc}(b)~\cite{PhysRevB.82.012405,PhysRevLett.108.116401}.

The MI phase is topologically nontrivial that can be characterized by nonlocal string correlation and entanglement spectrum. To begin with, one may inspect the hardcore limit $U\rightarrow\infty$. Using the Jordan-Wigner transformation, the nontrivial phase can be smoothly connected to the Haldane phase of a bond-alternating spin-1/2 antiferromagnetic Heisenberg model~\cite{PhysRevB.75.144420,PhysRevB.87.054402,Bahovadinov_2019}. It is well known that the Haldane phase has a hidden order that can be probed using a nonlocal string correlation~\cite{Nijs89}. In the superlattice Bose-Hubbard model, we can construct the operator
\begin{eqnarray}
    \hat{O}_{i,j}=-4\delta\hat{n}_i\prod_{k=i+1}^{j-1} e^{\text{i}\pi\delta\hat{n}_k} \delta\hat{n}_j,\label{eq:sto}
\end{eqnarray}
to reveal the string order, where $\delta\hat{n}_i=\overline{n}-\hat{n}_i$ is the density fluctuations with respect to the average filling and the prefactor $4$ is chosen for proper normalization~\cite{Torre06}. As shown in Fig.~\ref{fig:bi_fluc}(c), the expectation value of ${\hat O}_{i,j}$ in the ground state is clearly nonzero for the whole range of $|i-j|$, which signifies the realization of the Haldane phase. Besides the string correlation, we have also studied the parity order parameter
 \begin{eqnarray}
\hat{O}^{\mathbb{I}} _{i,j} = \mathbb{I}_i \prod_{k=i+1}^{j-1} e^{\text{i}\pi\delta\hat{n}_k} \mathbb{I}_{j},
\label{eq:stoId}
\end{eqnarray}
($\mathbb{I}_i$ is the identity operator on lattice site $i$) and the density-wave order parameter
\begin{eqnarray}
\hat{O}^{\rm DW}_{i,j} = (-1)^{|i-j|}\delta \hat{n}_i \delta \hat{n}_{j}
\label{eq:O_dw},
\end{eqnarray}
that were introduced in Refs.~\cite{Erez08,PhysRevB.86.125441,Torre06}. Both of them vanish in the MI phase [see Fig.~\ref{fig:bi_fluc}(c)], so the long-range string correlation of the ground state cannot be attributed to properties of a trivial state. Entanglement spectrum is a very useful tool for studying topological phases~\cite{PhysRevLett.101.010504}. The chain is divided into two subsystems $\mathcal{A}$ and $\mathcal{B}$ with equal length and a Schmidt decomposition of the ground state $\left|\Psi_{\rm GS}\right\rangle$ yields
\begin{eqnarray}
    \left|\Psi_{\rm GS}\right\rangle = \sum_\alpha \lambda_\alpha \left|{\mathcal A}_{\alpha}\right\rangle \left|{\mathcal B}_{\alpha}\right\rangle.
\end{eqnarray}
Here $\lambda_\alpha \left(\alpha=1,\dots,\chi\right)$ are the Schmidt values and $\left|{\mathcal A}_{\alpha}\right\rangle$ and $\left|{\mathcal B}_{\alpha}\right\rangle$ are the associated Schmidt vectors. Entanglement spectrum is computed from the Schmidt values via $-2\log{\lambda_{\alpha}}$. For the MI phase, each level in the entanglement spectrum is two-fold degenerate as one can see from the left panels of Fig.~\ref{fig:bi_fluc}(d)~\cite{PhysRevB.81.064439,PhysRevLett.113.020401}. This degeneracy disappears when the system enters the SF phase as shown in the right panels of Fig.~\ref{fig:bi_fluc}(d).

\begin{figure}[htb!]
    \centering
    \includegraphics[width=1\textwidth]{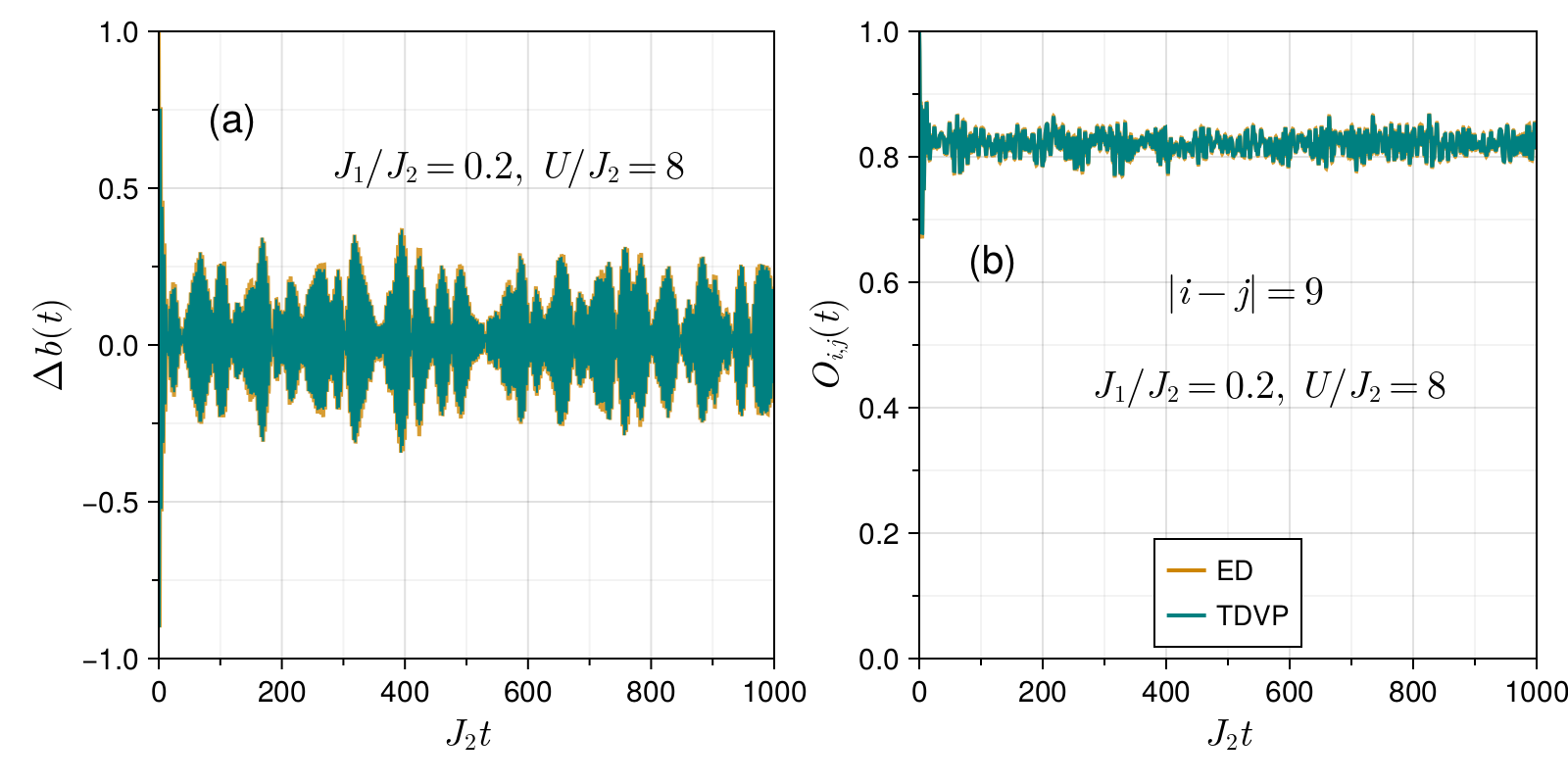}
\caption{\textbf{Comparison of TDVP with ED for the $L=12$ system in the nontrivial case}. (a) Time evolution of the bulk imbalance $\Delta_b (t)$. (b) Time evolution of the string correlation $O_{ij}(t)$. Its value is clearly nonzero up to $t=1000/J_2$. Both quantities exhibit oscillations that are mainly caused by finite-size effects and would be gradually suppressed as the system size increases.}
    \label{fig:ed_tdvp}
\end{figure}

\section{Relaxation dynamics of the superlattice Bose-Hubbard model}\label{chapter:dynamics}
\label{secA2}

This section provides more numerical results about the relaxation dynamics of the superlattice Bose-Hubbard model obtained using the time-dependent variational principle (TDVP) and exact diagonalization (ED). In Sec.~\ref{secA2A}, ED results up to $t=1000/J_{2}$ are presented and compared with TDVP results. In Sec.~\ref{secA2B}, the dependence of the string correlation on matrix-product-state (MPS) bond dimension and lattice-site distance are investigated. In Sec.~\ref{secA2C}, the dynamics of the parity and density-wave orders are presented to exclude trivial origin of our observations in the stationary states. In Sec.~\ref{secA2D}, the second-order Renyi entropy is computed to check if the stationary state has thermalized. In Sec.~\ref{secA2E}, the dynamics of entanglement spectrum is employed as an alternative method to probe the topological nature of the stationary states.

\subsection{Comparison of TDVP with ED}\label{section:tdvp_ed}
\label{secA2A}

For smaller systems, long-time dynamics can also be investigated by time-dependent ED~\cite{Phillip2017QuSpin}. To verify the reliability of TDVP, the dynamics of the interacting bosonic system is compared against the results of ED. We take a $L=12$ chain as an example, and study the long-time dynamics in the topologically nontrivial regime. We choose the initial state to be $\left|\psi \left(0\right)\right\rangle=\left|1010\dots\right\rangle$ and a time step of $\Delta t=0.05/J_2$ within ED and TDVP. As shown in Fig.~\ref{fig:ed_tdvp}, the comparisons between ED and TVDP are in good agreement even up to $t=1000/J_2$. In addition, we find the string correlation of the $L=12$ chain persists for long time, indicating the stability of nonthermal features in the many-body relaxation dynamics. We remark here that the persistent oscillations of observations are mainly a result of the finite-size effect and suppressed for larger system size.

\begin{figure}[H]
\centering
\includegraphics[width=1\textwidth]{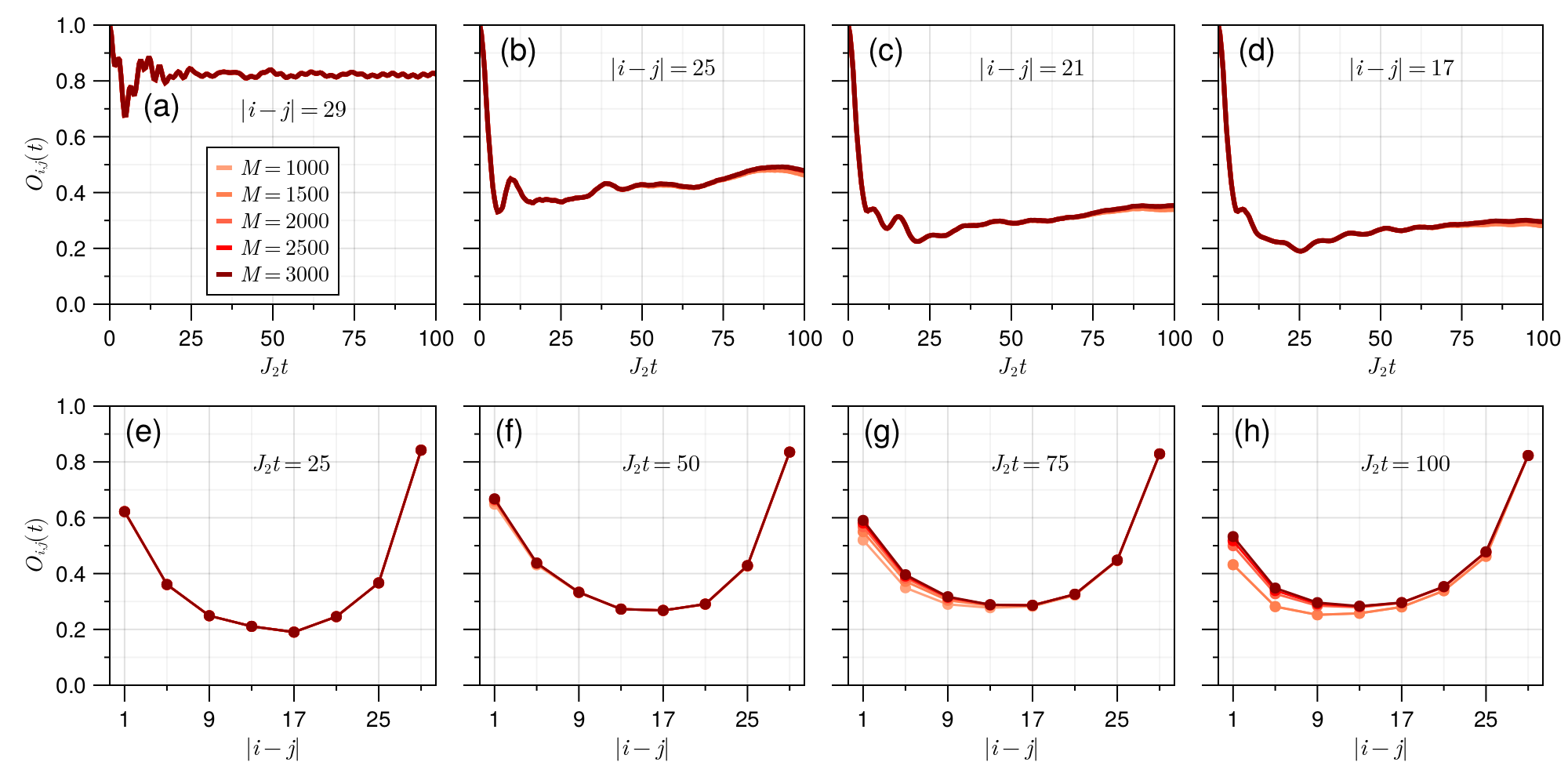}
\caption{\textbf{String correlation for the $L=32$ system in the nontrivial case}. (a-d) Time evolution of $O_{i,j}(t)$ for four different lengths $|i-j|=29$, $25$, $21$, and $17$. (e-h) $O_{i,j}(t)$ as a function of the length $|i-j|$ at four different time slices. The results are obtained using bond dimensions between $1000$ to $3000$.}
\label{fig:sto_length_M}
\end{figure}

\subsection{Dynamics of the string correlation}\label{section:dynamics_sto}
\label{secA2B}

Numerical simulations based on MPS are fundamentally variational, so it is important to verify that good convergence has been achieved. Time evolution of the string correlation at four different lengths $|i-j|$ are presented in Fig.~\ref{fig:sto_length_M}(a-d). It decays rapidly at the beginning but is clearly distinct from zero in the long-time limit. The values of $O_{i,j}(t)$ at different bond dimensions almost coincide, so we conclude that $M=3000$ is sufficient for our purpose. The string correlation versus string length $|i-j|$ at four different time slices are presented in Fig.~\ref{fig:sto_length_M}(e-h). An interesting feature is that $O_{i,j}(t)$ first decreases (to a nonzero minimum) and then increases as the length $|i-j|$ becomes large. It has been observed in a previous experiment that the string correlation exhibits length dependence in the Haldane phase realized using Fermi-Hubbard ladder~\cite{Sompet2021RealizingTS}.

\begin{figure}[H]
\centering
\includegraphics[width=1.0\textwidth]{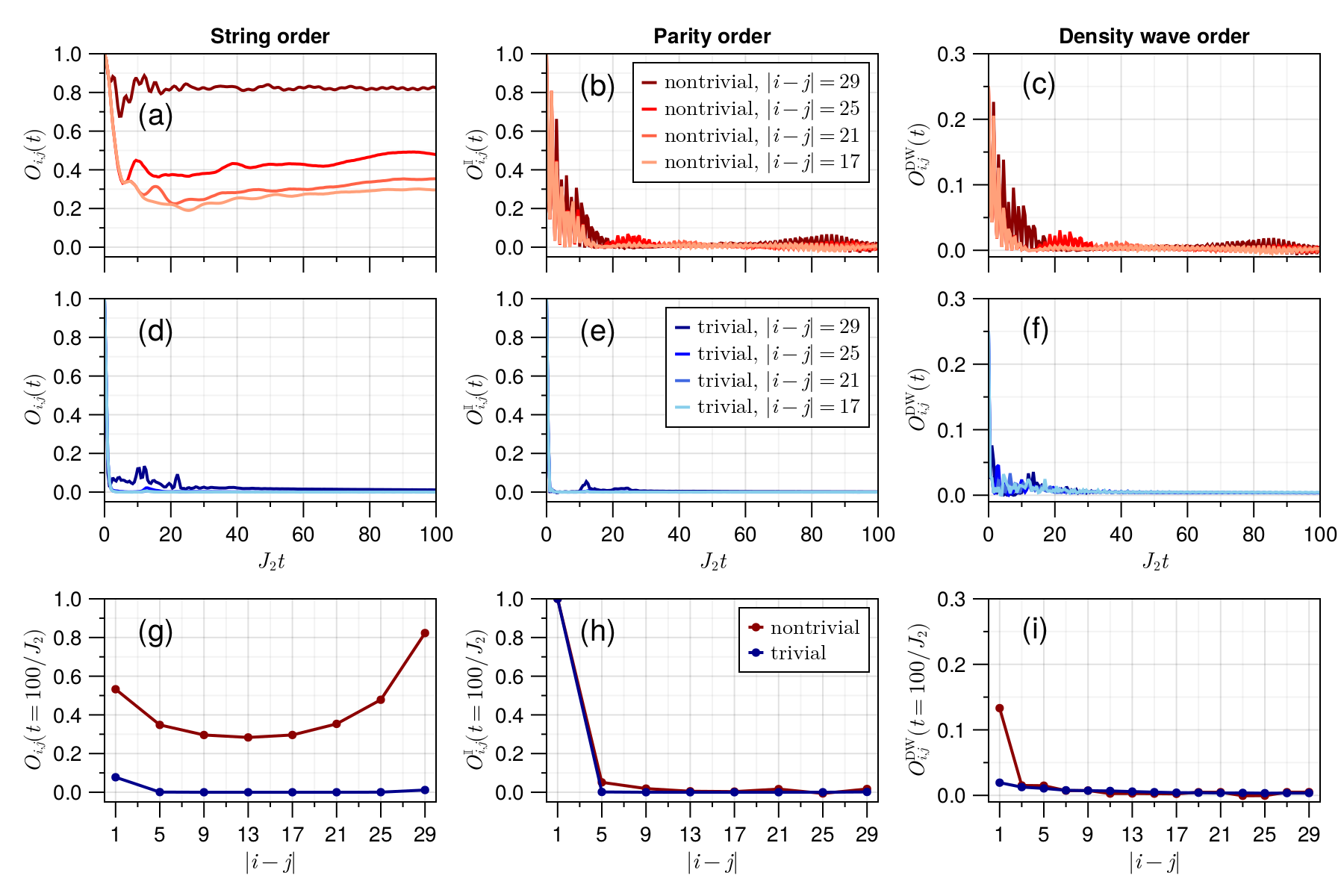}
\caption{\textbf{Dynamics of the string, parity, and density-wave correlations in the $L=32$ system.} (a-c) Time evolution of the three quantities at four different lengths $|i-j|$ in the nontrivial case. (d-f) Time evolution of the three quantities at four different lengths $|i-j|$ in the trivial case. In all cases, two ends of the chain are not used when computing the density-wave correlation to eliminate the impact of bosons localized at the edges. (g-i) The three quantities at $t=100/J_2$ versus the length $|i-j|$. In the nontrivial case, a trivial product state evolves to an emergent topological state with long-range string correlation but no parity or density-wave correlations. In the trivial case, there is no long-range correlation in the time-evolved stationary state. The results are obtained using bond dimension up to $3000$.}
\label{fig:sto_parity_dw}
\end{figure}

\subsection{Dynamics of parity and density-wave correlations}\label{section:dynamics_sto_Id_dw}
\label{secA2C}

Besides the Haldane phase, some trivial states can also have nonzero string correlation. It is important to explore and rule out these trivial origins so our observation can be attributed to the presence of symmetry-protected topological (SPT) order. To this end, we have computed the parity correlation function $\hat{O}^{\mathbb{I}} _{i,j}$ and density-wave correlation function $\hat{O}^{\rm DW}_{i,j}$. The initial state $\left|1010\dots\right\rangle$ has long-range string, parity, and density-wave correlations as shown in Fig.~\ref{fig:sto_parity_dw}(a-c), respectively. For these three panels, the Hamiltonian belongs to the nontrivial regime. Time-evolved stationary state at large $t$ still has long-range string correlation, but the other two quantities vanish. This distinction provides strong support for the topological nature of the stationary state. In contrast, when the Hamiltonian is trivial and the system thermalizes during time evolution, the initial state would have a {\it effective} high temperature so the stationary state is expected to be featureless with no symmetry-breaking order. This picture is confirmed by the numerical results in Fig.~\ref{fig:sto_parity_dw}(d-f). The three correlations for different lengths $|i-j|$ in the nontrivial and trivial regimes are presented in Fig.~\ref{fig:sto_parity_dw}(g-i). The string correlation in the nontrivial case has a nonzero value for the whole range of $|i-j|$ but all other quantities vanish at large $|i-j|$.

\begin{figure}[H]
    \centering
    \includegraphics[width=1\textwidth]{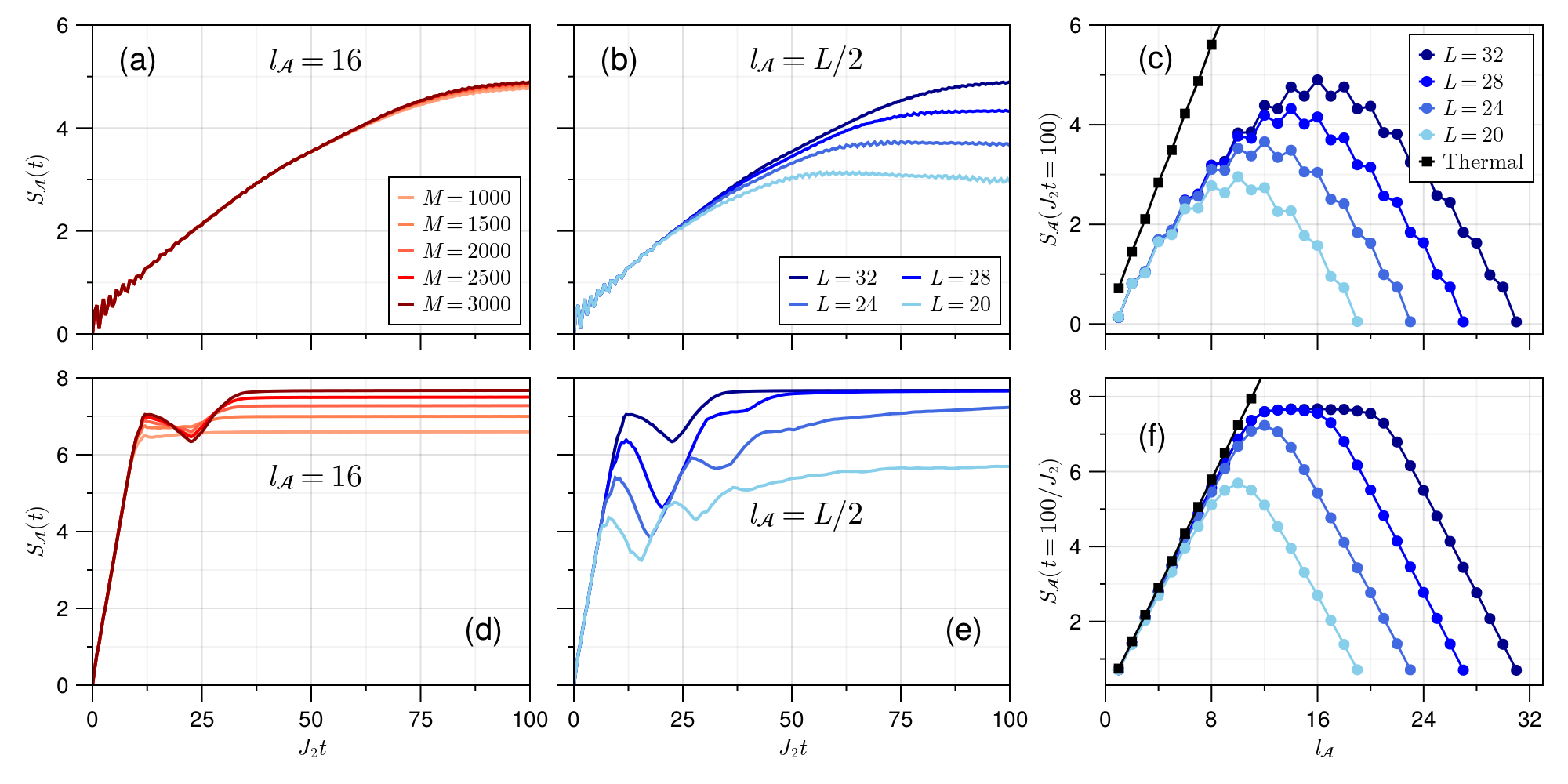}
    \caption{\textbf{Dynamics of the entanglement entropy.} The first row is for the nontrivial case and the second row is for the trivial case. (a) The variations with MPS bond dimension of the entropy at the center of the $L=32$ system. (b) The half system entropy for several different $L$'s. (c) The subsystem size dependence of the late-time entropy. Thermal entropy given by the canonical thermal ensembles is shown as black lines with rectangle markers for comparison. (d-f) Each panel shows the same quantity as the panel above it. The results in (b), (c), (e), and (f) are obtained using bond dimension $3000$.}
    \label{fig:renyi_M_t}
\end{figure}

\subsection{Initial-state dependence of relaxation dynamics}
\label{section:dynamics_initial_state}
\label{secA2D}

\begin{figure}[H]
    \centering
    \includegraphics[width=1\textwidth]{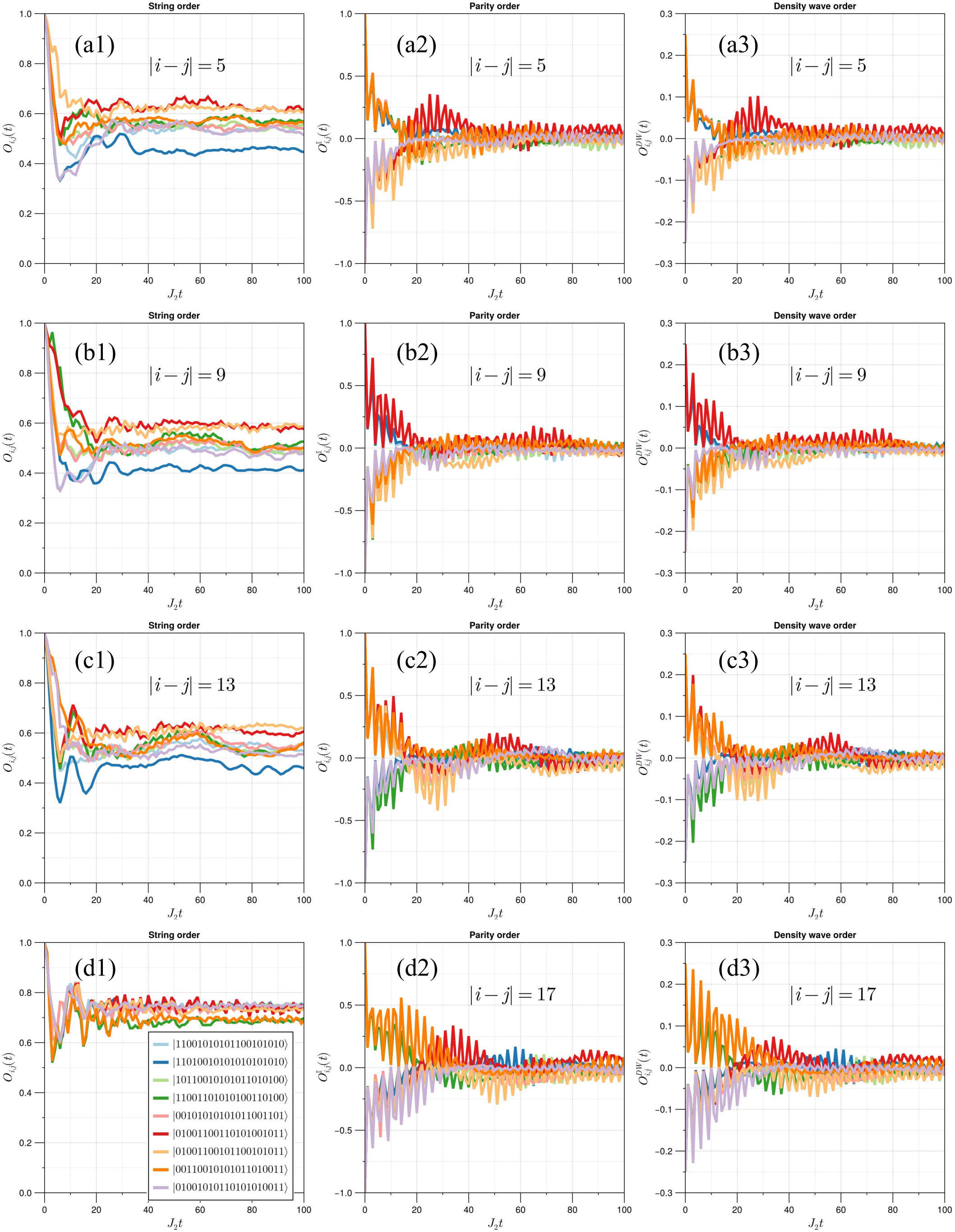}
    \caption{\textbf{Dynamics of the string, parity, and density-wave correlations.} From top to bottom, time evolution of the three quantities for four different lengths $|i-j|$ in the nontrivial regime. Parameters: $L=20$, $U/J_2=8$, and $J_1/J_2=0.2$. The results are obtained with bond dimension up to $1000$.}
    \label{fig:initial_depend}
\end{figure}

As is widely recognized, initial states typically influence many-body dynamics in time evolution. It is pertinent to examine the impact of the initial state on string, parity and density-wave correlations. With the performance of numerical calculations in different initial states, the long-range correlations remain robust to the initial states once singly occupied on the two sites adjoined by $J_2$-bonds. As shown in Fig.~\ref{fig:initial_depend}, the string correlations of the long-time evolved stationary states exhibit nonzero values for different lengths, while parity and density-wave orders decay to zero quickly. Note here that these initial states satisfy half filling in the bulk naturally, reminiscent of ground-state topology. The filling at the edges is not essential to this bulk phenomenon, which is attributed to the distinct relaxation times of the bulk and boundary states. The robustness of initial states is beneficial to the experimental research for observing the exotic dynamics.

\begin{figure}[H]
    \centering
    \includegraphics[width=0.5\textwidth]{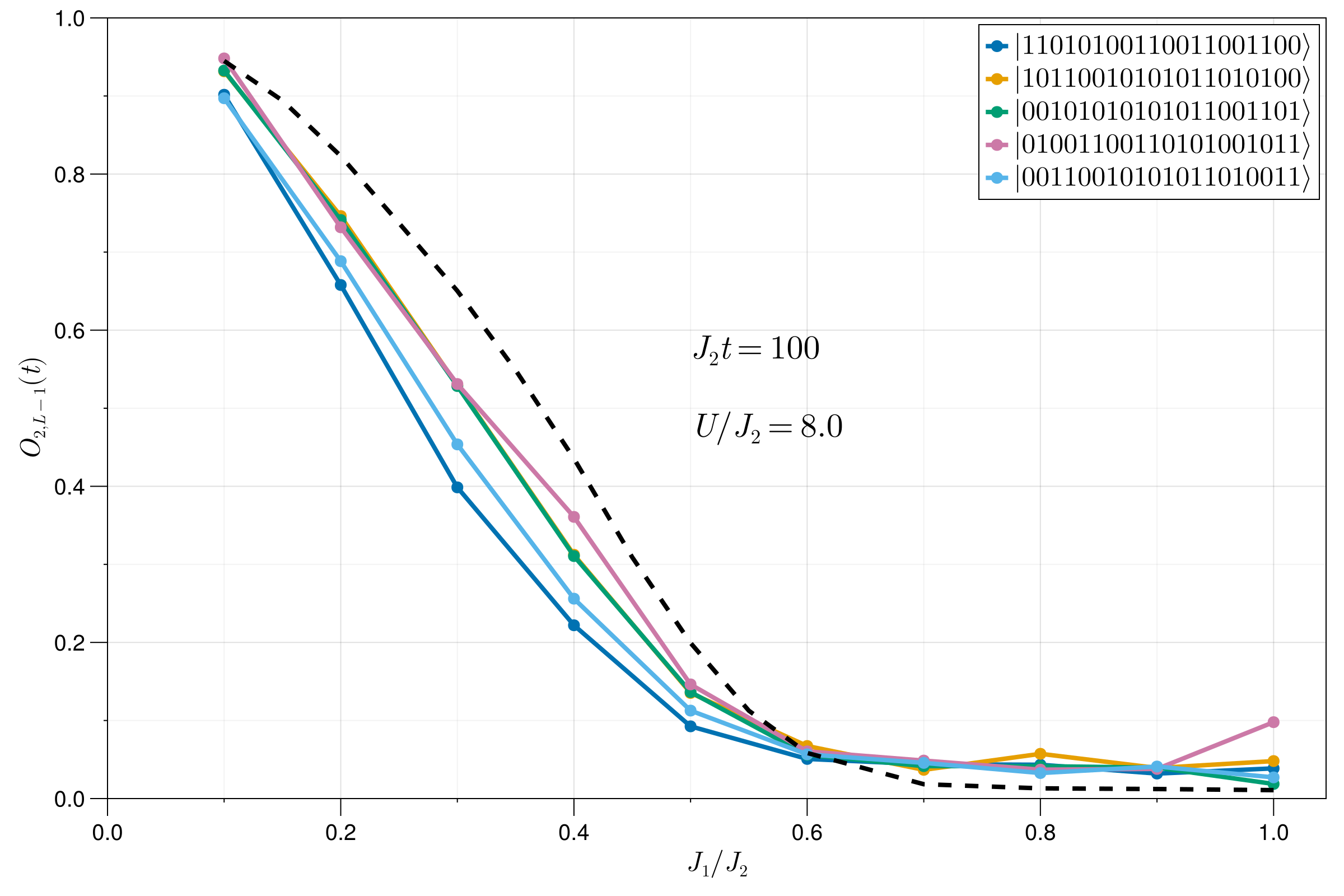}
    \caption{\textbf{String order as a function of the hopping strengths for different initial states.} String order of the long-time stationary state at $t=100/J_2$. The interaction strength is fixed at $U/J_2=8$ and the hopping strength $J_1/J_2$ is varied. The solid lines correspond to the $L=20$ system with multiple initial states given in the inset and the dashed line corresponds to the $L=32$ system with initial state $\left|1010\cdots\right\rangle$. The MPS bond dimension is 1000.}
    \label{fig:initial_depend_phase}
\end{figure}

{ We have performed additional calculations to map out the phase diagram for different initial states. It is found that the phase diagram is remarkably robust in the timescale $t=100/J_2$, which is a typical value accessible in ultracold atom experiments. Numerical results about the string order are shown in Fig.~\ref{fig:initial_depend_phase}. We have checked many different values of $J_1/J_2$ with $U/J_2=8$. All initial states are chosen to satisfy the constraint mentioned above: there is only one boson in every two sites connected by a $J_2$ hopping in the bulk. The phase boundary is almost independent of the initial state.}

\subsection{Dynamics of the entanglement entropy}\label{section:dynamics_ee}
\label{secA2E}

It has been pointed out that local observables may not be sufficient to uncover the existence or absence of thermalization. Instead, bipartite entanglement entropy could serve as a more powerful tool for analyzing thermalization. We have computed the second-order Renyi entropy $S_{\mathcal A}(t) = -\text{log} \left[\text{Tr}\left(\rho^{2}_{\mathcal{A}}(t)\right)\right]$ for multiple systems, where $\mathcal{A}$ is a subsystem with $l_{\mathcal A}$ sites and $\rho_{\mathcal A}(t)$ is the associated reduced density matrix. The data presented below are for the $L=32$ case.

The dynamical buildup of entanglement entropy in the topologically nontrivial and trivial regimes are very different. In the former case, the entropy increases with time for a transient period and then begins to saturate at late time [see Fig.~\ref{fig:renyi_M_t}(a)]. By inspecting multiple bond dimensions, we conclude that the saturation is not an artificial truncation caused by finite bond dimension. The saturated value is actually a constant and $M=1500$ is sufficient for the relaxation dynamics when $t \leq 100/J_2$. In the latter case, the entropy also saturates after rapid initial growth, and the maximal entropy increases with the bond dimension [see Fig.~\ref{fig:renyi_M_t}(d)]. This is not surprising because a high-lying state that thermalizes in time evolution should lead to volume law entropy that can only be captured by very large bond dimension. The system-size dependence of the entropy is displayed in Fig.~\ref{fig:renyi_M_t}(b) and (e). In the nontrivial case, $t=100/J_2$ is sufficient to observe saturation in all cases so we set this as the upper limit in our simulations. We have also studied the evolution of the entropy with the subsystem length $l_{\mathcal A}$ at $t=100/J_2$. As shown in Fig.~\ref{fig:renyi_M_t}(f), the entropy grows linearly with $l_{\mathcal A}$ in the trivial case and is basically identical to the thermal entropy given by the canonical ensemble before it saturates in the middle of the system. However, one can see from Fig.~\ref{fig:renyi_M_t}(c) that the entropy is noticeably smaller than the thermal entropy in the nontrivial case. The data in Fig.~\ref{fig:renyi_M_t}(b) and (c) corroborate the nonthermal nature of the time-evolved stationary state in the nontrivial case.

\subsection{Dynamics of the entanglement spectrum}\label{section:dynamics_es}
\label{secA2F}

\begin{figure}[H]
    \centering
    \includegraphics[width=1\textwidth]{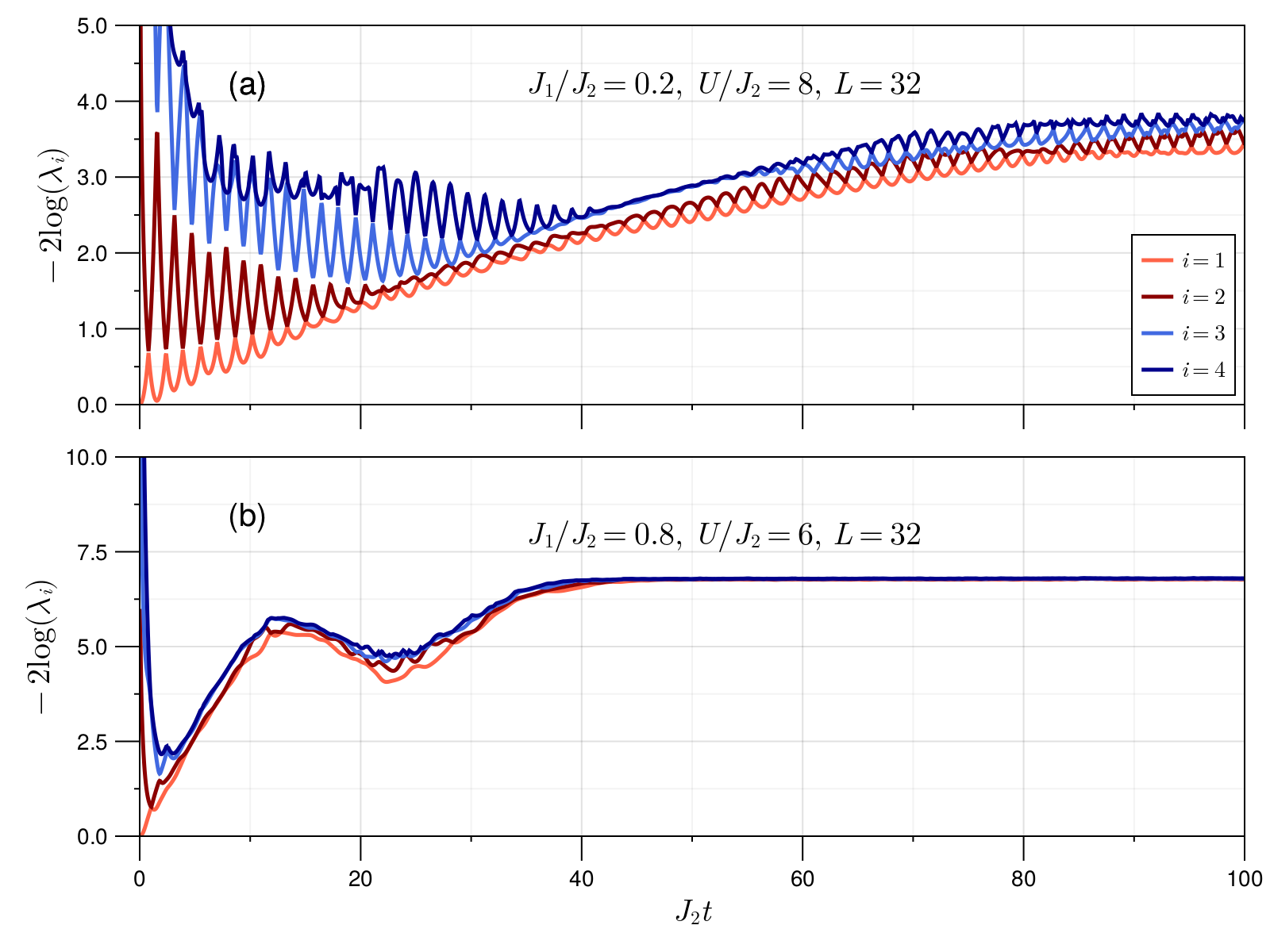}
    \caption{\textbf{Dynamics of the entanglement spectrum in the $L=32$ system}. (a) Time evolution of the largest four entanglement levels for the nontrivial case. (b) The same quantities as in panel (a) for the trivial case. The results are obtained using bond dimension $2500$.}
    \label{fig:ent_sp}
\end{figure}

One may naively anticipate that the two-fold degeneracy in the entanglement spectrum of the ground state can also be found in the time-evolved stationary state. This is not the case because the Schmidt values generally oscillate with time rather than stay constant. Nevertheless, it has been proposed that level crossings of the Schmidt values is a useful fingerprint of topological states in time evolution~\cite{PhysRevLett.121.250601}. The largest four Schmidt values of the $L=32$ system are plotted in Fig.~\ref{fig:ent_sp}. For the nontrivial case in panel (a), level crossings between $\lambda_1$ and $\lambda_2$ and between $\lambda_3$ and $\lambda_4$ occur at certain time slices during time evolution. On the contrary, the four levels become degenerate at $t \sim 40/J_{2}$ and remain constant thereafter for the trivial case in panel (b). It would be very interesting to explore the entanglement spectrum using advanced experimental techniques in ultracold atoms~\cite{PhysRevX.6.041033, Islam_2015, PhysRevLett.109.020505}.

\section{Analytical results and effective models}\label{chapter:analysis}

This section aims to unveil physics of the superlattice Bose-Hubbard model using analytical results in certain limits. An inspection of the wave functions in the nontrivial and trivial limits provides important insights about the string correlation. We also construct an effective model for our system using perturbation theory. It is demonstrated that the relaxation dynamics and nonequilibrium physics can be understood by keeping up to second-order terms.

\subsection{The nontrivial limit $J_1/J_2 \rightarrow 0$}\label{section:non_trivial_limit}

In the limit of $J_1/J_2 \rightarrow 0$, Eq.~(1) in main text becomes
\begin{eqnarray}
    \hat{H}_{J_2} = -J_2\sum_{i,\text{even}} \hat{b}^{\dagger}_{i}\hat{b}_{i+1} + {\rm{H.c.}} + \frac{U}{2} \sum^{L}_{i=1} \hat{n}_i(\hat{n}_i-1).
\label{J1_limit}
\end{eqnarray}
Time evolution of the initial state $\left|1010\cdots\right\rangle$ under this Hamiltonian is significantly simplified. The two sites at the ends are decoupled from the bulk while the other sites break into pairs consist of two adjacent sites. It is useful to introduce the Bell pairs $\left|\pm\right\rangle_i=(\left|01\right\rangle\pm\left|10\right\rangle)_{i,i+1}/\sqrt{2}$ on sites $i$ and $i+1$. The ground state of the coupled two sites is $\left|+\right\rangle_i$, so the ground state of the full system is
\begin{eqnarray}
    \left|\Psi_{J_2} \right\rangle=\left|1\right\rangle\otimes\left|+\right\rangle_2\otimes\left|+\right\rangle_4\otimes\cdots\otimes\left|0\right\rangle.
\end{eqnarray}
and its string correlation $O_{ij}=1$. The time-evolved state is
\begin{eqnarray}
    |\psi_{J_2}\left(t\right)\rangle= \left|1\right\rangle \otimes \left[\frac{e^{iJ_2 t}|+\rangle_2+e^{-iJ_2 t}|-\rangle_2}{\sqrt{2}} \right]\otimes \cdots\otimes\left|0\right\rangle,
\end{eqnarray}
for which the string correlation is still $O_{i,j}\left(t\right)=1$. While this analysis is interesting, it is not exactly the same as the physics discussed in the main text. This can be seen from the average imbalance in the bulk
\begin{eqnarray}
\Delta_{b}(t) \equiv \frac{2}{L-2} \sum^{L-1}_{i=2} \left[ n_\text{i,odd}(t)-n_\text{i,even}(t) \right] \propto \cos{2J_2t} \, ,
\end{eqnarray}
whose oscillation is in sharp contrast to the results in Fig. 4(c) of the main text.

\subsection{The trivial limit $J_2/J_1 \rightarrow 0$}\label{section:trivial_limit}

In the limit of $J_2/J_1\rightarrow 0$, Eq.~(1) in the main text becomes
\begin{eqnarray}
    \hat{H}_{J_1} = -J_1\sum_{i,\text{odd}} \hat{b}^{\dagger}_{i} \hat{b}_{i+1} + {\rm{H.c.}} + \frac{U}{2} \sum^{L}_{i=1} \hat{n}_i(\hat{n}_i-1).
\label{J2_limit}
\end{eqnarray}
An even numbered site is coupled with its left neighbor but decoupled from its right neighbor, so the system breaks into $L/2$ copies of two-site pairs. The ground state can be written as $\left|\Psi_{J_1}\right\rangle=\left|+\right\rangle_1\otimes\left|+\right\rangle_3\otimes\cdots$. It has no string correlation as one would expect. Time evolution of the initial state $\left|1010\cdots\right\rangle$ yields
\begin{eqnarray}
    |\psi_{J_1}\left(t\right)\rangle=  \left[\frac{e^{iJ_1 t}|+\rangle_1-e^{-iJ_1 t}|-\rangle_1}{\sqrt{2}} \right]\otimes \left[\frac{e^{iJ_1 t}|+\rangle_3-e^{-iJ_1 t}|-\rangle_3}{\sqrt{2}} \right]\otimes \cdots.
\end{eqnarray}
One can see that the string correlation
\begin{eqnarray}
O_{i,j}\left(t\right) = \frac{1}{2}+\frac{1}{2}\cos \left(4J_1t\right)
\end{eqnarray}
and the bulk imbalance $\Delta_b(t)\propto \cos{2J_1t}$. Both quantities oscillate with time, which is different from the ground state and the results in Fig.~4(c) of the main text.

\subsection{Effective model in the strongly correlated regime}\label{section:effective}

The two limits discussed above fail to provide an accurate account of the relaxation dynamics that we are interested in. This is hardly surprising given that the system is actually integrable in both cases. To overcome this challenge, effective models for our system have been derived using the projection operator method~\cite{Essler_Frahm_Göhmann_Klümper_Korepin_2005,auerbach2012interacting}. Deep in the Mott phase with $U \gg J_{1}, J_{2}$, large onsite repulsion suppresses particle number fluctuations, so it is natural to organize perturbative calculations according to lattice site occupations. We define $\hat{P}$ ($\hat{Q}$) as the projection operator onto the subspace $\mathcal{H}_{P}$ in which each site hosts at most one boson ($\mathcal{H}_{Q}$ at least one site is doubly occupied). These operators satisfy the relations $\hat{P}^2=\hat{P}$, $\hat{Q}^2=\hat{Q}$, and $\hat{P}+\hat{Q}=1$.

The hopping terms in the original Hamiltonian are denoted as $H_{J}$ and the onsite repulsion as $H_U$. Its eigenvalue problem can be written as
\begin{eqnarray}
   \hat{H}|\psi\rangle =\left(\hat{H}_J+\hat{H}_U \right) |\psi \rangle =\left(\hat{H}_J+\hat{H}_U \right) \left( \hat{P}+\hat{Q} \right)|\psi \rangle=E|\psi \rangle.
\end{eqnarray}
If we multiply both sides by $\hat{P}$ or $\hat{Q}$, the equation becomes
\begin{eqnarray}
   \left( \hat{P}\hat{H}_{J}\hat{P}+\hat{P}\hat{H}_{J}\hat{Q}+\hat{P}\hat{H}_{U}\hat{P}+\hat{P}\hat{H}_{U}\hat{Q} \right)|\psi \rangle =E \hat{P}|\psi \rangle , \\
   \left( \hat{Q}\hat{H}_{J}\hat{P}+\hat{Q}\hat{H}_{J}\hat{Q}+\hat{Q}\hat{H}_{U}\hat{P}+\hat{Q}\hat{H}_{U}\hat{Q} \right)|\psi \rangle =E \hat{Q}|\psi \rangle. \label{Eff_Q}
\end{eqnarray}
The first one can be manipulated with the aid of the second one to yield
\begin{eqnarray}
   \left( \hat{P}\hat{H}_{J}\hat{P}+\hat{P}\hat{H}_{J}\hat{Q} \frac{1}{E-\hat{Q}\hat{H}_{U}\hat{Q}-\hat{Q}\hat{H}_{J}\hat{Q}} \hat{Q}\hat{H}_{J}\hat{P} \right) \hat{P} |\psi \rangle=E \hat{P} |\psi \rangle,
\end{eqnarray}
so we conclude that the effective Hamiltonian for the subspace $\mathcal{H}_P$ is
\begin{eqnarray}
   \hat{H}_{\rm eff}= \hat{P}\hat{H}_{J}\hat{P}+\hat{P}\hat{H}_{J}\hat{Q} \frac{1}{E-\hat{Q}\hat{H}_{U}\hat{Q}-\hat{Q}\hat{H}_{J}\hat{Q}} \hat{Q}\hat{H}_{J}\hat{P}.
   \label{Eq:effective_midddle}
\end{eqnarray}
Using the identity $\left( A-B \right) ^{-1}=A^{-1} \sum^{\infty}_{n=0} \left( BA^{-1} \right)^{n}$, it can be further expanded as
\begin{eqnarray}
   \hat{H}_{\rm eff}= \hat{P}\hat{H}_{J}\hat{P}+\hat{P}\hat{H}_{J}\hat{Q} \frac{1}{E_{P}-\hat{Q}\hat{H}_{U}\hat{Q}} \sum^{\infty}_{n=0} \left( \frac{\hat{Q}\hat{H}_{J}\hat{Q}-E+E_{P}}{E_{P}-\hat{Q}\hat{H}_{U}\hat{Q}} \right)^n \hat{Q}\hat{H}_{J}\hat{P},
    \label{eq:effe_ham}
\end{eqnarray}
where $A=E_{P}-\hat{Q}\hat{H}_{U}\hat{Q}$, $B=\hat{Q}\hat{H}_{J}\hat{Q}-E+E_{P}$, and $E_{P}$ is the energy of $H_U$ in the subspace $\mathcal{H}_{P}$.

\begin{figure}[H]
\centering
\includegraphics[width=0.9\textwidth]{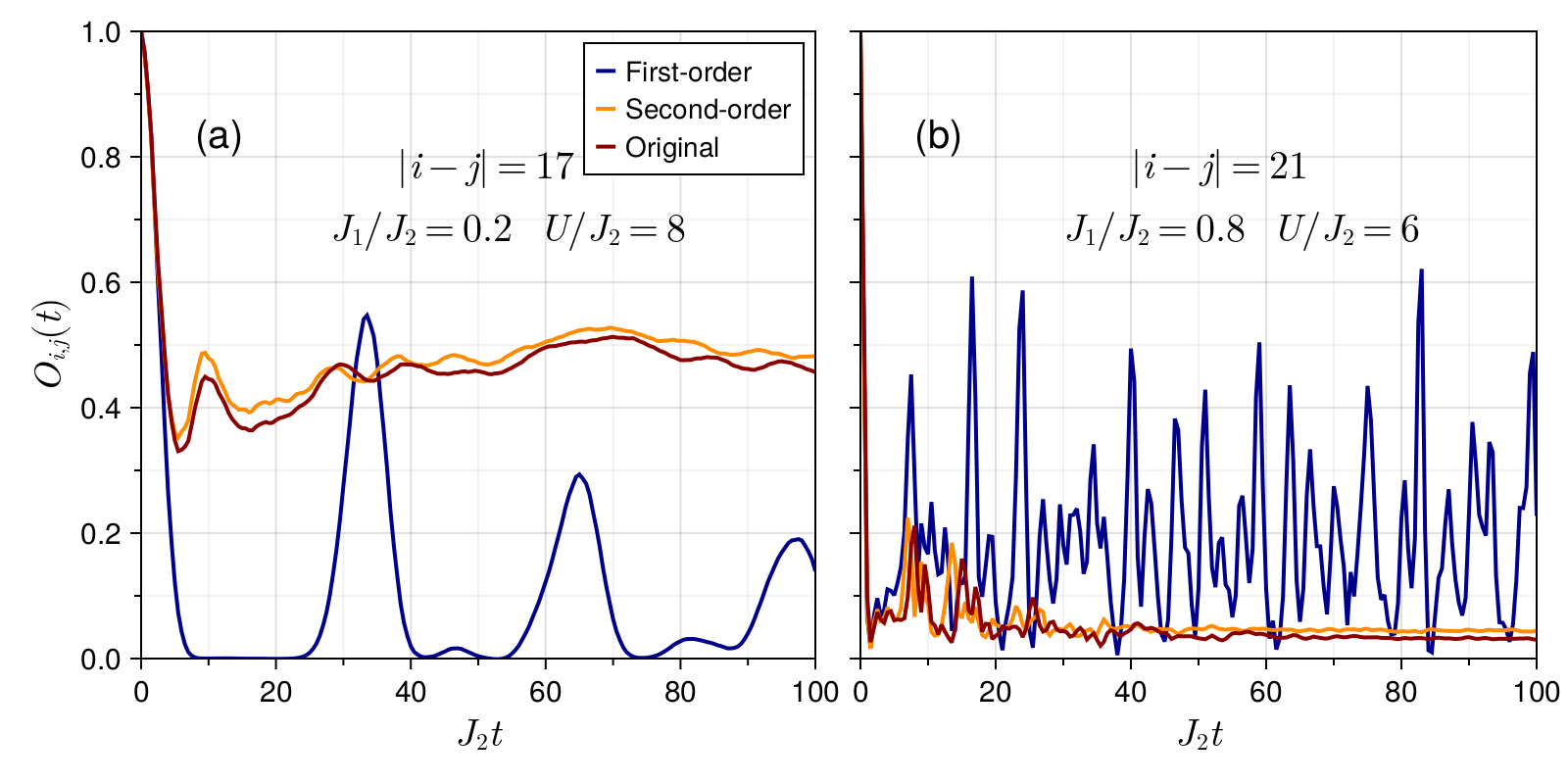}
\caption{\textbf{Comparison of the string correlations in the effective and original models for the $L=24$ system.} (a) Time evolution of $O_{i,j}(t)$ for the nontrivial case obtained using the first-, second-order effective Hamiltonians, and the original Hamiltonian. (b) The same quantities as in panel (a) for the trivial case. The predictions based on the second-order effective model and the original model are in good agreement.}
\label{fig:comp_eff_ed}
\end{figure}

\subsubsection{First-order effective model}\label{subsection:first_order}

If we only keep first-order terms, virtual hopping processes would be ignored. This means that only the hopping terms are retained and the effective Hamiltonian is simply
\begin{eqnarray}
\hat{H}_{1} = -J_1\sum_{i,\text{odd}}\hat{b}^\dagger_{i}\hat{b}^{}_{i+1}-\!J_2\sum_{i,\text{even}}\hat{b}^\dagger_{i}\hat{b}^{}_{i+1}+ {\rm{H.c.}} .
\label{first_order}
\end{eqnarray}
The system is still integrable~\cite{Marcos07,PhysRevLett.106.140405} so it is expected that the long-time average of a local observable is time-independent and can be described by generalized Gibbs ensembles. Time evolution from the initial state $|1010...\rangle$ under Eq.~(\ref{first_order}) leads to persistent oscillations of the bulk imbalance and the string correlation. As shown in Fig.~\ref{fig:comp_eff_ed}(a)(b), the string correlation $O_{i,j}\left(t\right)$ oscillates with time rather than acquire stable values in the whole regime $0<J_1/J_2<1$ and for timescale up to $t=100/J_2$, which suggests that the nonequilibrium topological state cannot be explained using first-order perturbation theory.

\subsubsection{Second-order effective model}\label{subsection:second_order}

The effective model with up to second-order virtual hopping processes is
\begin{eqnarray}
   \hat{H}_{\rm eff}= \hat{P}\hat{H}_{J}\hat{P}+\hat{P}\hat{H}_{J}\hat{Q} \frac{1}{E_{P}-\hat{Q}\hat{H}_{U}\hat{Q}} \hat{Q}\hat{H}_{J}\hat{P}.
\end{eqnarray}
A straightforward calculation using Eq.~(1) of the main text gives
\begin{eqnarray}
\hat{H}=-J_1\sum_{i,\text{odd}}\hat{b}^\dagger_{i}\hat{b}^{}_{i+1}-\!J_2\sum_{i,\text{even}}\hat{b}^\dagger_{i}\hat{b}^{}_{i+1}+\frac{4J^2_1}{U}\sum_{i,\text{odd}}\hat{n}_i\hat{n}_{i+1} +\frac{4J^2_2}{U}\sum_{i,\text{even}}\hat{n}_i\hat{n}_{i+1}+\!\frac{2J_1J_2}{U}\sum_i \hat{b}^\dagger_{i-1}\hat{n}_i \hat{b}_{i+1} + {\rm H.c.}.
\label{effe_ham}
\end{eqnarray}
For large but finite onsite repulsion, the coefficients $J^2_{1,2}/U$ and $J_1J_2/U$ of the second-order processes are nonzero. As shown in Fig.~\ref{fig:comp_eff_ed}(a)(b), remarkable agreement is found between the results obtained using Eq.~(\ref{effe_ham}) and the original Hamiltonian presented in the main text. This fact underscores the essential role of the soft-core interaction in the relaxation dynamics. The emergent behaviour observed here should be attributed to the interplay of the onsite interactions and topological characters of the Hamiltonian.


\begin{thebibliography}{93}%
    \makeatletter
    \providecommand \@ifxundefined [1]{%
     \@ifx{#1\undefined}
    }%
    \providecommand \@ifnum [1]{%
     \ifnum #1\expandafter \@firstoftwo
     \else \expandafter \@secondoftwo
     \fi
    }%
    \providecommand \@ifx [1]{%
     \ifx #1\expandafter \@firstoftwo
     \else \expandafter \@secondoftwo
     \fi
    }%
    \providecommand \natexlab [1]{#1}%
    \providecommand \enquote  [1]{``#1''}%
    \providecommand \bibnamefont  [1]{#1}%
    \providecommand \bibfnamefont [1]{#1}%
    \providecommand \citenamefont [1]{#1}%
    \providecommand \href@noop [0]{\@secondoftwo}%
    \providecommand \href [0]{\begingroup \@sanitize@url \@href}%
    \providecommand \@href[1]{\@@startlink{#1}\@@href}%
    \providecommand \@@href[1]{\endgroup#1\@@endlink}%
    \providecommand \@sanitize@url [0]{\catcode `\\12\catcode `\$12\catcode `\&12\catcode `\#12\catcode `\^12\catcode `\_12\catcode `\%12\relax}%
    \providecommand \@@startlink[1]{}%
    \providecommand \@@endlink[0]{}%
    \providecommand \url  [0]{\begingroup\@sanitize@url \@url }%
    \providecommand \@url [1]{\endgroup\@href {#1}{\urlprefix }}%
    \providecommand \urlprefix  [0]{URL }%
    \providecommand \Eprint [0]{\href }%
    \providecommand \doibase [0]{https://doi.org/}%
    \providecommand \selectlanguage [0]{\@gobble}%
    \providecommand \bibinfo  [0]{\@secondoftwo}%
    \providecommand \bibfield  [0]{\@secondoftwo}%
    \providecommand \translation [1]{[#1]}%
    \providecommand \BibitemOpen [0]{}%
    \providecommand \bibitemStop [0]{}%
    \providecommand \bibitemNoStop [0]{.\EOS\space}%
    \providecommand \EOS [0]{\spacefactor3000\relax}%
    \providecommand \BibitemShut  [1]{\csname bibitem#1\endcsname}%
    \let\auto@bib@innerbib\@empty
    \bibitem [{\citenamefont {Polkovnikov}\ \emph {et~al.}(2011)\citenamefont {Polkovnikov}, \citenamefont {Sengupta}, \citenamefont {Silva},\ and\ \citenamefont {Vengalattore}}]{Polkovnikov2010ColloquiumND}%
      \BibitemOpen
      \bibfield  {author} {\bibinfo {author} {\bibfnamefont {A.}~\bibnamefont {Polkovnikov}}, \bibinfo {author} {\bibfnamefont {K.}~\bibnamefont {Sengupta}}, \bibinfo {author} {\bibfnamefont {A.}~\bibnamefont {Silva}},\ and\ \bibinfo {author} {\bibfnamefont {M.}~\bibnamefont {Vengalattore}},\ }\bibfield  {title} {\bibinfo {title} {{Colloquium: Nonequilibrium dynamics of closed interacting quantum systems}},\ }\href {https://doi.org/10.1103/RevModPhys.83.863} {\bibfield  {journal} {\bibinfo  {journal} {Rev. Mod. Phys.}\ }\textbf {\bibinfo {volume} {83}},\ \bibinfo {pages} {863} (\bibinfo {year} {2011})}\BibitemShut {NoStop}%
    \bibitem [{\citenamefont {Deutsch}(2018)}]{Deutsch18}%
      \BibitemOpen
      \bibfield  {author} {\bibinfo {author} {\bibfnamefont {J.~M.}\ \bibnamefont {Deutsch}},\ }\bibfield  {title} {\bibinfo {title} {{Eigenstate thermalization hypothesis}},\ }\href {https://doi.org/10.1088/1361-6633/aac9f1} {\bibfield  {journal} {\bibinfo  {journal} {Rep. Prog. Phys.}\ }\textbf {\bibinfo {volume} {81}},\ \bibinfo {pages} {082001} (\bibinfo {year} {2018})}\BibitemShut {NoStop}%
    \bibitem [{\citenamefont {Ueda}(2020)}]{Ueda2020QuantumET}%
      \BibitemOpen
      \bibfield  {author} {\bibinfo {author} {\bibfnamefont {M.}~\bibnamefont {Ueda}},\ }\bibfield  {title} {\bibinfo {title} {{Quantum equilibration, thermalization and prethermalization in ultracold atoms}},\ }\href {https://doi.org/10.1038/s42254-020-0237-x} {\bibfield  {journal} {\bibinfo  {journal} {Nat. Rev. Phys.}\ }\textbf {\bibinfo {volume} {2}},\ \bibinfo {pages} {669 } (\bibinfo {year} {2020})}\BibitemShut {NoStop}%
    \bibitem [{\citenamefont {Bertini}\ \emph {et~al.}(2021)\citenamefont {Bertini}, \citenamefont {Heidrich-Meisner}, \citenamefont {Karrasch}, \citenamefont {Prosen}, \citenamefont {Steinigeweg},\ and\ \citenamefont {{\v{Z}}nidari{\v{c}}}}]{bertini2021finite}%
      \BibitemOpen
      \bibfield  {author} {\bibinfo {author} {\bibfnamefont {B.}~\bibnamefont {Bertini}}, \bibinfo {author} {\bibfnamefont {F.}~\bibnamefont {Heidrich-Meisner}}, \bibinfo {author} {\bibfnamefont {C.}~\bibnamefont {Karrasch}}, \bibinfo {author} {\bibfnamefont {T.}~\bibnamefont {Prosen}}, \bibinfo {author} {\bibfnamefont {R.}~\bibnamefont {Steinigeweg}},\ and\ \bibinfo {author} {\bibfnamefont {M.}~\bibnamefont {{\v{Z}}nidari{\v{c}}}},\ }\bibfield  {title} {\bibinfo {title} {{Finite-temperature transport in one-dimensional quantum lattice models}},\ }\href {https://doi.org/10.1103/RevModPhys.93.025003} {\bibfield  {journal} {\bibinfo  {journal} {Rev. Mod. Phys.}\ }\textbf {\bibinfo {volume} {93}},\ \bibinfo {pages} {025003} (\bibinfo {year} {2021})}\BibitemShut {NoStop}%
    \bibitem [{\citenamefont {Rigol}\ \emph {et~al.}(2007)\citenamefont {Rigol}, \citenamefont {Dunjko}, \citenamefont {Yurovsky},\ and\ \citenamefont {Olshanii}}]{Marcos07}%
      \BibitemOpen
      \bibfield  {author} {\bibinfo {author} {\bibfnamefont {M.}~\bibnamefont {Rigol}}, \bibinfo {author} {\bibfnamefont {V.}~\bibnamefont {Dunjko}}, \bibinfo {author} {\bibfnamefont {V.}~\bibnamefont {Yurovsky}},\ and\ \bibinfo {author} {\bibfnamefont {M.}~\bibnamefont {Olshanii}},\ }\bibfield  {title} {\bibinfo {title} {{Relaxation in a completely integrable many-body quantum system: An ab initio study of the dynamics of the highly excited states of 1D lattice hard-core Bosons}},\ }\href {https://doi.org/10.1103/PhysRevLett.98.050405} {\bibfield  {journal} {\bibinfo  {journal} {Phys. Rev. Lett.}\ }\textbf {\bibinfo {volume} {98}},\ \bibinfo {pages} {050405} (\bibinfo {year} {2007})}\BibitemShut {NoStop}%
    \bibitem [{\citenamefont {Rigol}\ \emph {et~al.}(2008)\citenamefont {Rigol}, \citenamefont {Dunjko},\ and\ \citenamefont {Olshanii}}]{Marcos08}%
      \BibitemOpen
      \bibfield  {author} {\bibinfo {author} {\bibfnamefont {M.}~\bibnamefont {Rigol}}, \bibinfo {author} {\bibfnamefont {V.}~\bibnamefont {Dunjko}},\ and\ \bibinfo {author} {\bibfnamefont {M.}~\bibnamefont {Olshanii}},\ }\bibfield  {title} {\bibinfo {title} {{Thermalization and its mechanism for generic isolated quantum systems}},\ }\href {https://doi.org/10.1038/nature06838} {\bibfield  {journal} {\bibinfo  {journal} {Nature}\ }\textbf {\bibinfo {volume} {452}},\ \bibinfo {pages} {854} (\bibinfo {year} {2008})}\BibitemShut {NoStop}%
    \bibitem [{\citenamefont {Rigol}(2009)}]{Marcos09}%
      \BibitemOpen
      \bibfield  {author} {\bibinfo {author} {\bibfnamefont {M.}~\bibnamefont {Rigol}},\ }\bibfield  {title} {\bibinfo {title} {{Breakdown of thermalization in finite one-dimensional systems}},\ }\href {https://doi.org/10.1103/PhysRevLett.103.100403} {\bibfield  {journal} {\bibinfo  {journal} {Phys. Rev. Lett.}\ }\textbf {\bibinfo {volume} {103}},\ \bibinfo {pages} {100403} (\bibinfo {year} {2009})}\BibitemShut {NoStop}%
    \bibitem [{\citenamefont {Garrison}\ and\ \citenamefont {Grover}(2018)}]{James18}%
      \BibitemOpen
      \bibfield  {author} {\bibinfo {author} {\bibfnamefont {J.~R.}\ \bibnamefont {Garrison}}\ and\ \bibinfo {author} {\bibfnamefont {T.}~\bibnamefont {Grover}},\ }\bibfield  {title} {\bibinfo {title} {{Does a single eigenstate encode the full Hamiltonian?}},\ }\href {https://doi.org/10.1103/PhysRevX.8.021026} {\bibfield  {journal} {\bibinfo  {journal} {Phys. Rev. X}\ }\textbf {\bibinfo {volume} {8}},\ \bibinfo {pages} {021026} (\bibinfo {year} {2018})}\BibitemShut {NoStop}%
    \bibitem [{\citenamefont {Srednicki}(1994)}]{PhysRevE.50.888}%
      \BibitemOpen
      \bibfield  {author} {\bibinfo {author} {\bibfnamefont {M.}~\bibnamefont {Srednicki}},\ }\bibfield  {title} {\bibinfo {title} {{Chaos and quantum thermalization}},\ }\href {https://doi.org/10.1103/PhysRevE.50.888} {\bibfield  {journal} {\bibinfo  {journal} {Phys. Rev. E}\ }\textbf {\bibinfo {volume} {50}},\ \bibinfo {pages} {888} (\bibinfo {year} {1994})}\BibitemShut {NoStop}%
    \bibitem [{\citenamefont {Srednicki}(1996)}]{Srednicki1996ThermalFI}%
      \BibitemOpen
      \bibfield  {author} {\bibinfo {author} {\bibfnamefont {M.}~\bibnamefont {Srednicki}},\ }\bibfield  {title} {\bibinfo {title} {{Thermal fluctuations in quantized chaotic systems}},\ }\href {https://doi.org/10.1088/0305-4470/29/4/003} {\bibfield  {journal} {\bibinfo  {journal} {J. Phys. A: Math. Gen.}\ }\textbf {\bibinfo {volume} {29}},\ \bibinfo {pages} {L75} (\bibinfo {year} {1996})}\BibitemShut {NoStop}%
    \bibitem [{\citenamefont {Srednicki}(1999)}]{Srednicki1998TheAT}%
      \BibitemOpen
      \bibfield  {author} {\bibinfo {author} {\bibfnamefont {M.}~\bibnamefont {Srednicki}},\ }\bibfield  {title} {\bibinfo {title} {{The approach to thermal equilibrium in quantized chaotic systems}},\ }\href {https://doi.org/10.1088/0305-4470/32/7/007} {\bibfield  {journal} {\bibinfo  {journal} {J. Phys. A: Math. Gen.}\ }\textbf {\bibinfo {volume} {32}},\ \bibinfo {pages} {1163} (\bibinfo {year} {1999})}\BibitemShut {NoStop}%
    \bibitem [{\citenamefont {Deutsch}(1991)}]{PhysRevA.43.2046}%
      \BibitemOpen
      \bibfield  {author} {\bibinfo {author} {\bibfnamefont {J.~M.}\ \bibnamefont {Deutsch}},\ }\bibfield  {title} {\bibinfo {title} {{Quantum statistical mechanics in a closed system}},\ }\href {https://doi.org/10.1103/PhysRevA.43.2046} {\bibfield  {journal} {\bibinfo  {journal} {Phys. Rev. A}\ }\textbf {\bibinfo {volume} {43}},\ \bibinfo {pages} {2046} (\bibinfo {year} {1991})}\BibitemShut {NoStop}%
    \bibitem [{\citenamefont {Mitra}(2018)}]{Mitra2018}%
      \BibitemOpen
      \bibfield  {author} {\bibinfo {author} {\bibfnamefont {A.}~\bibnamefont {Mitra}},\ }\bibfield  {title} {\bibinfo {title} {{Quantum quench dynamics}},\ }\href {https://doi.org/10.1146/annurev-conmatphys-031016-025451} {\bibfield  {journal} {\bibinfo  {journal} {Annu. Rev. Condens. Matter Phys.}\ }\textbf {\bibinfo {volume} {9}},\ \bibinfo {pages} {245} (\bibinfo {year} {2018})}\BibitemShut {NoStop}%
    \bibitem [{\citenamefont {Eisert}\ \emph {et~al.}(2015)\citenamefont {Eisert}, \citenamefont {Friesdorf},\ and\ \citenamefont {Gogolin}}]{Eisert_2015}%
      \BibitemOpen
      \bibfield  {author} {\bibinfo {author} {\bibfnamefont {J.}~\bibnamefont {Eisert}}, \bibinfo {author} {\bibfnamefont {M.}~\bibnamefont {Friesdorf}},\ and\ \bibinfo {author} {\bibfnamefont {C.}~\bibnamefont {Gogolin}},\ }\bibfield  {title} {\bibinfo {title} {Quantum many-body systems out of equilibrium},\ }\href {https://doi.org/10.1038/nphys3215} {\bibfield  {journal} {\bibinfo  {journal} {Nature Physics}\ }\textbf {\bibinfo {volume} {11}},\ \bibinfo {pages} {124} (\bibinfo {year} {2015})}\BibitemShut {NoStop}%
    \bibitem [{\citenamefont {Kaufman}\ \emph {et~al.}(2016)\citenamefont {Kaufman}, \citenamefont {Tai}, \citenamefont {Lukin}, \citenamefont {Rispoli}, \citenamefont {Schittko}, \citenamefont {Preiss},\ and\ \citenamefont {Greiner}}]{Kaufman2016QuantumTT}%
      \BibitemOpen
      \bibfield  {author} {\bibinfo {author} {\bibfnamefont {A.~M.}\ \bibnamefont {Kaufman}}, \bibinfo {author} {\bibfnamefont {M.~E.}\ \bibnamefont {Tai}}, \bibinfo {author} {\bibfnamefont {A.}~\bibnamefont {Lukin}}, \bibinfo {author} {\bibfnamefont {M.}~\bibnamefont {Rispoli}}, \bibinfo {author} {\bibfnamefont {R.}~\bibnamefont {Schittko}}, \bibinfo {author} {\bibfnamefont {P.~M.}\ \bibnamefont {Preiss}},\ and\ \bibinfo {author} {\bibfnamefont {M.}~\bibnamefont {Greiner}},\ }\bibfield  {title} {\bibinfo {title} {{Quantum thermalization through entanglement in an isolated many-body system}},\ }\href {https://doi.org/10.1126/science.aaf6725} {\bibfield  {journal} {\bibinfo  {journal} {Science}\ }\textbf {\bibinfo {volume} {353}},\ \bibinfo {pages} {794} (\bibinfo {year} {2016})}\BibitemShut {NoStop}%
    \bibitem [{\citenamefont {Nichols}\ \emph {et~al.}(2018)\citenamefont {Nichols}, \citenamefont {Cheuk}, \citenamefont {Okan}, \citenamefont {Hartke}, \citenamefont {Mendez}, \citenamefont {Senthil}, \citenamefont {Khatami}, \citenamefont {Zhang},\ and\ \citenamefont {Zwierlein}}]{Nichols2018SpinTI}%
      \BibitemOpen
      \bibfield  {author} {\bibinfo {author} {\bibfnamefont {M.~A.}\ \bibnamefont {Nichols}}, \bibinfo {author} {\bibfnamefont {L.~W.}\ \bibnamefont {Cheuk}}, \bibinfo {author} {\bibfnamefont {M.}~\bibnamefont {Okan}}, \bibinfo {author} {\bibfnamefont {T.}~\bibnamefont {Hartke}}, \bibinfo {author} {\bibfnamefont {E.}~\bibnamefont {Mendez}}, \bibinfo {author} {\bibfnamefont {T.}~\bibnamefont {Senthil}}, \bibinfo {author} {\bibfnamefont {E.}~\bibnamefont {Khatami}}, \bibinfo {author} {\bibfnamefont {H.}~\bibnamefont {Zhang}},\ and\ \bibinfo {author} {\bibfnamefont {M.~W.}\ \bibnamefont {Zwierlein}},\ }\bibfield  {title} {\bibinfo {title} {{Spin transport in a Mott insulator of ultracold fermions}},\ }\href {https://doi.org/10.1126/science.aat4387} {\bibfield  {journal} {\bibinfo  {journal} {Science}\ }\textbf {\bibinfo {volume} {363}},\ \bibinfo {pages} {383 } (\bibinfo {year} {2018})}\BibitemShut {NoStop}%
    \bibitem [{\citenamefont {Jepsen}\ \emph {et~al.}(2020)\citenamefont {Jepsen}, \citenamefont {Amato-Grill}, \citenamefont {Dimitrova}, \citenamefont {Ho}, \citenamefont {Demler},\ and\ \citenamefont {Ketterle}}]{Jepsen2020SpinTI}%
      \BibitemOpen
      \bibfield  {author} {\bibinfo {author} {\bibfnamefont {P.~N.}\ \bibnamefont {Jepsen}}, \bibinfo {author} {\bibfnamefont {J.}~\bibnamefont {Amato-Grill}}, \bibinfo {author} {\bibfnamefont {I.}~\bibnamefont {Dimitrova}}, \bibinfo {author} {\bibfnamefont {W.~W.}\ \bibnamefont {Ho}}, \bibinfo {author} {\bibfnamefont {E.~A.}\ \bibnamefont {Demler}},\ and\ \bibinfo {author} {\bibfnamefont {W.}~\bibnamefont {Ketterle}},\ }\bibfield  {title} {\bibinfo {title} {{Spin transport in a tunable Heisenberg model realized with ultracold atoms}},\ }\href {https://doi.org/10.1038/s41586-020-3033-y} {\bibfield  {journal} {\bibinfo  {journal} {Nature}\ }\textbf {\bibinfo {volume} {588}},\ \bibinfo {pages} {403 } (\bibinfo {year} {2020})}\BibitemShut {NoStop}%
    \bibitem [{\citenamefont {Wei}\ \emph {et~al.}(2022)\citenamefont {Wei}, \citenamefont {Rubio-Abadal}, \citenamefont {Ye}, \citenamefont {Machado}, \citenamefont {Kemp}, \citenamefont {Srakaew}, \citenamefont {Hollerith}, \citenamefont {Rui}, \citenamefont {Gopalakrishnan}, \citenamefont {Yao}, \citenamefont {Bloch},\ and\ \citenamefont {Zeiher}}]{Wei2021QuantumGM}%
      \BibitemOpen
      \bibfield  {author} {\bibinfo {author} {\bibfnamefont {D.}~\bibnamefont {Wei}}, \bibinfo {author} {\bibfnamefont {A.}~\bibnamefont {Rubio-Abadal}}, \bibinfo {author} {\bibfnamefont {B.}~\bibnamefont {Ye}}, \bibinfo {author} {\bibfnamefont {F.}~\bibnamefont {Machado}}, \bibinfo {author} {\bibfnamefont {J.}~\bibnamefont {Kemp}}, \bibinfo {author} {\bibfnamefont {K.}~\bibnamefont {Srakaew}}, \bibinfo {author} {\bibfnamefont {S.}~\bibnamefont {Hollerith}}, \bibinfo {author} {\bibfnamefont {J.}~\bibnamefont {Rui}}, \bibinfo {author} {\bibfnamefont {S.}~\bibnamefont {Gopalakrishnan}}, \bibinfo {author} {\bibfnamefont {N.~Y.}\ \bibnamefont {Yao}}, \bibinfo {author} {\bibfnamefont {I.}~\bibnamefont {Bloch}},\ and\ \bibinfo {author} {\bibfnamefont {J.}~\bibnamefont {Zeiher}},\ }\bibfield  {title} {\bibinfo {title} {{Quantum gas microscopy of Kardar-Parisi-Zhang superdiffusion}},\ }\href {https://doi.org/10.1126/science.abk2397} {\bibfield  {journal} {\bibinfo  {journal} {Science}\ }\textbf {\bibinfo {volume} {376}},\ \bibinfo {pages} {716 } (\bibinfo {year} {2022})}\BibitemShut {NoStop}%
    \bibitem [{\citenamefont {Zhou}\ \emph {et~al.}(2022)\citenamefont {Zhou}, \citenamefont {Su}, \citenamefont {Halimeh}, \citenamefont {Ott}, \citenamefont {Sun}, \citenamefont {Hauke}, \citenamefont {Yang}, \citenamefont {Yuan}, \citenamefont {urgen Berges},\ and\ \citenamefont {Pan}}]{Zhou2021ThermalizationDO}%
      \BibitemOpen
      \bibfield  {author} {\bibinfo {author} {\bibfnamefont {Z.-Y.}\ \bibnamefont {Zhou}}, \bibinfo {author} {\bibfnamefont {G.-X.}\ \bibnamefont {Su}}, \bibinfo {author} {\bibfnamefont {J.~C.}\ \bibnamefont {Halimeh}}, \bibinfo {author} {\bibfnamefont {R.}~\bibnamefont {Ott}}, \bibinfo {author} {\bibfnamefont {H.}~\bibnamefont {Sun}}, \bibinfo {author} {\bibfnamefont {P.}~\bibnamefont {Hauke}}, \bibinfo {author} {\bibfnamefont {B.}~\bibnamefont {Yang}}, \bibinfo {author} {\bibfnamefont {Z.-S.}\ \bibnamefont {Yuan}}, \bibinfo {author} {\bibfnamefont {J.}~\bibnamefont {urgen Berges}},\ and\ \bibinfo {author} {\bibfnamefont {J.-W.}\ \bibnamefont {Pan}},\ }\bibfield  {title} {\bibinfo {title} {{Thermalization dynamics of a gauge theory on a quantum simulator}},\ }\href {https://doi.org/10.1126/science.abl6277} {\bibfield  {journal} {\bibinfo  {journal} {Science}\ }\textbf {\bibinfo {volume} {377}},\ \bibinfo {pages} {311 } (\bibinfo {year} {2022})}\BibitemShut {NoStop}%
    \bibitem [{\citenamefont {Christakis}\ \emph {et~al.}(2023)\citenamefont {Christakis}, \citenamefont {Rosenberg}, \citenamefont {Raj}, \citenamefont {Chi}, \citenamefont {Morningstar}, \citenamefont {Huse}, \citenamefont {Yan},\ and\ \citenamefont {Bakr}}]{Christakis2022ProbingSC}%
      \BibitemOpen
      \bibfield  {author} {\bibinfo {author} {\bibfnamefont {L.}~\bibnamefont {Christakis}}, \bibinfo {author} {\bibfnamefont {J.~S.}\ \bibnamefont {Rosenberg}}, \bibinfo {author} {\bibfnamefont {R.}~\bibnamefont {Raj}}, \bibinfo {author} {\bibfnamefont {S.}~\bibnamefont {Chi}}, \bibinfo {author} {\bibfnamefont {A.}~\bibnamefont {Morningstar}}, \bibinfo {author} {\bibfnamefont {D.~A.}\ \bibnamefont {Huse}}, \bibinfo {author} {\bibfnamefont {Z.~Z.}\ \bibnamefont {Yan}},\ and\ \bibinfo {author} {\bibfnamefont {W.~S.}\ \bibnamefont {Bakr}},\ }\bibfield  {title} {\bibinfo {title} {{Probing site-resolved correlations in a spin system of ultracold molecules}},\ }\href {https://doi.org/10.1038/s41586-022-05558-4} {\bibfield  {journal} {\bibinfo  {journal} {Nature}\ }\textbf {\bibinfo {volume} {614}},\ \bibinfo {pages} {64} (\bibinfo {year} {2023})}\BibitemShut {NoStop}%
    \bibitem [{\citenamefont {Li}\ \emph {et~al.}(2023)\citenamefont {Li}, \citenamefont {Matsuda}, \citenamefont {Miller}, \citenamefont {Carroll}, \citenamefont {Tobias}, \citenamefont {Higgins},\ and\ \citenamefont {Ye}}]{Li_2023}%
      \BibitemOpen
      \bibfield  {author} {\bibinfo {author} {\bibfnamefont {J.-R.}\ \bibnamefont {Li}}, \bibinfo {author} {\bibfnamefont {K.}~\bibnamefont {Matsuda}}, \bibinfo {author} {\bibfnamefont {C.}~\bibnamefont {Miller}}, \bibinfo {author} {\bibfnamefont {A.~N.}\ \bibnamefont {Carroll}}, \bibinfo {author} {\bibfnamefont {W.~G.}\ \bibnamefont {Tobias}}, \bibinfo {author} {\bibfnamefont {J.~S.}\ \bibnamefont {Higgins}},\ and\ \bibinfo {author} {\bibfnamefont {J.}~\bibnamefont {Ye}},\ }\bibfield  {title} {\bibinfo {title} {{Tunable itinerant spin dynamics with polar molecules}},\ }\href {https://doi.org/10.1038/s41586-022-05479-2} {\bibfield  {journal} {\bibinfo  {journal} {Nature}\ }\textbf {\bibinfo {volume} {614}},\ \bibinfo {pages} {70 } (\bibinfo {year} {2023})}\BibitemShut {NoStop}%
    \bibitem [{\citenamefont {Hasan}\ and\ \citenamefont {Kane}(2010)}]{RevModPhys.82.3045}%
      \BibitemOpen
      \bibfield  {author} {\bibinfo {author} {\bibfnamefont {M.~Z.}\ \bibnamefont {Hasan}}\ and\ \bibinfo {author} {\bibfnamefont {C.~L.}\ \bibnamefont {Kane}},\ }\bibfield  {title} {\bibinfo {title} {{Colloquium: Topological insulators}},\ }\href {https://doi.org/10.1103/RevModPhys.82.3045} {\bibfield  {journal} {\bibinfo  {journal} {Rev. Mod. Phys.}\ }\textbf {\bibinfo {volume} {82}},\ \bibinfo {pages} {3045} (\bibinfo {year} {2010})}\BibitemShut {NoStop}%
    \bibitem [{\citenamefont {Qi}\ and\ \citenamefont {Zhang}(2011)}]{RevModPhys.83.1057}%
      \BibitemOpen
      \bibfield  {author} {\bibinfo {author} {\bibfnamefont {X.-L.}\ \bibnamefont {Qi}}\ and\ \bibinfo {author} {\bibfnamefont {S.-C.}\ \bibnamefont {Zhang}},\ }\bibfield  {title} {\bibinfo {title} {{Topological insulators and superconductors}},\ }\href {https://doi.org/10.1103/RevModPhys.83.1057} {\bibfield  {journal} {\bibinfo  {journal} {Rev. Mod. Phys.}\ }\textbf {\bibinfo {volume} {83}},\ \bibinfo {pages} {1057} (\bibinfo {year} {2011})}\BibitemShut {NoStop}%
    \bibitem [{\citenamefont {Cooper}\ \emph {et~al.}(2019)\citenamefont {Cooper}, \citenamefont {Dalibard},\ and\ \citenamefont {Spielman}}]{RevModPhys.91.015005}%
      \BibitemOpen
      \bibfield  {author} {\bibinfo {author} {\bibfnamefont {N.~R.}\ \bibnamefont {Cooper}}, \bibinfo {author} {\bibfnamefont {J.}~\bibnamefont {Dalibard}},\ and\ \bibinfo {author} {\bibfnamefont {I.~B.}\ \bibnamefont {Spielman}},\ }\bibfield  {title} {\bibinfo {title} {{Topological bands for ultracold atoms}},\ }\href {https://doi.org/10.1103/RevModPhys.91.015005} {\bibfield  {journal} {\bibinfo  {journal} {Rev. Mod. Phys.}\ }\textbf {\bibinfo {volume} {91}},\ \bibinfo {pages} {015005} (\bibinfo {year} {2019})}\BibitemShut {NoStop}%
    \bibitem [{\citenamefont {Rachel}(2018)}]{Rachel2018InteractingTI}%
      \BibitemOpen
      \bibfield  {author} {\bibinfo {author} {\bibfnamefont {S.}~\bibnamefont {Rachel}},\ }\bibfield  {title} {\bibinfo {title} {{Interacting topological insulators: a review}},\ }\href {https://doi.org/10.1088/1361-6633/aad6a6} {\bibfield  {journal} {\bibinfo  {journal} {Rep. Prog. Phys.}\ }\textbf {\bibinfo {volume} {81}},\ \bibinfo {pages} {116501} (\bibinfo {year} {2018})}\BibitemShut {NoStop}%
    \bibitem [{\citenamefont {Zhang}\ \emph {et~al.}(2018{\natexlab{a}})\citenamefont {Zhang}, \citenamefont {Zhu}, \citenamefont {Zhao}, \citenamefont {Yan},\ and\ \citenamefont {Zhu}}]{Zhang2018TopologicalQM}%
      \BibitemOpen
      \bibfield  {author} {\bibinfo {author} {\bibfnamefont {D.-W.}\ \bibnamefont {Zhang}}, \bibinfo {author} {\bibfnamefont {Y.-Q.}\ \bibnamefont {Zhu}}, \bibinfo {author} {\bibfnamefont {Y.~X.}\ \bibnamefont {Zhao}}, \bibinfo {author} {\bibfnamefont {H.}~\bibnamefont {Yan}},\ and\ \bibinfo {author} {\bibfnamefont {S.-L.}\ \bibnamefont {Zhu}},\ }\bibfield  {title} {\bibinfo {title} {{Topological quantum matter with cold atoms}},\ }\href {https://doi.org/10.1080/00018732.2019.1594094} {\bibfield  {journal} {\bibinfo  {journal} {Adv. Phys.}\ }\textbf {\bibinfo {volume} {67}},\ \bibinfo {pages} {253 } (\bibinfo {year} {2018}{\natexlab{a}})}\BibitemShut {NoStop}%
    \bibitem [{\citenamefont {de~L{\'e}s{\'e}leuc}\ \emph {et~al.}(2019)\citenamefont {de~L{\'e}s{\'e}leuc}, \citenamefont {Lienhard}, \citenamefont {Scholl}, \citenamefont {Barredo}, \citenamefont {Weber}, \citenamefont {Lang}, \citenamefont {B{\"u}chler}, \citenamefont {Lahaye},\ and\ \citenamefont {Browaeys}}]{deLsleuc2018ObservationOA}%
      \BibitemOpen
      \bibfield  {author} {\bibinfo {author} {\bibfnamefont {S.}~\bibnamefont {de~L{\'e}s{\'e}leuc}}, \bibinfo {author} {\bibfnamefont {V.}~\bibnamefont {Lienhard}}, \bibinfo {author} {\bibfnamefont {P.}~\bibnamefont {Scholl}}, \bibinfo {author} {\bibfnamefont {D.}~\bibnamefont {Barredo}}, \bibinfo {author} {\bibfnamefont {S.~T.}\ \bibnamefont {Weber}}, \bibinfo {author} {\bibfnamefont {N.}~\bibnamefont {Lang}}, \bibinfo {author} {\bibfnamefont {H.~P.}\ \bibnamefont {B{\"u}chler}}, \bibinfo {author} {\bibfnamefont {T.}~\bibnamefont {Lahaye}},\ and\ \bibinfo {author} {\bibfnamefont {A.}~\bibnamefont {Browaeys}},\ }\bibfield  {title} {\bibinfo {title} {{Observation of a symmetry-protected topological phase of interacting bosons with Rydberg atoms}},\ }\href {https://doi.org/10.1126/science.aav9105} {\bibfield  {journal} {\bibinfo  {journal} {Science}\ }\textbf {\bibinfo {volume} {365}},\ \bibinfo {pages} {775 } (\bibinfo {year} {2019})}\BibitemShut {NoStop}%
    \bibitem [{\citenamefont {Sompet}\ \emph {et~al.}(2022)\citenamefont {Sompet}, \citenamefont {Hirthe}, \citenamefont {Bourgund}, \citenamefont {Chalopin}, \citenamefont {Bibo}, \citenamefont {Koepsell}, \citenamefont {Bojovi{\'c}}, \citenamefont {Verresen}, \citenamefont {Pollmann}, \citenamefont {Salomon}, \citenamefont {Gross}, \citenamefont {Hilker},\ and\ \citenamefont {Bloch}}]{Sompet2021RealizingTS}%
      \BibitemOpen
      \bibfield  {author} {\bibinfo {author} {\bibfnamefont {P.}~\bibnamefont {Sompet}}, \bibinfo {author} {\bibfnamefont {S.}~\bibnamefont {Hirthe}}, \bibinfo {author} {\bibfnamefont {D.}~\bibnamefont {Bourgund}}, \bibinfo {author} {\bibfnamefont {T.}~\bibnamefont {Chalopin}}, \bibinfo {author} {\bibfnamefont {J.}~\bibnamefont {Bibo}}, \bibinfo {author} {\bibfnamefont {J.}~\bibnamefont {Koepsell}}, \bibinfo {author} {\bibfnamefont {P.}~\bibnamefont {Bojovi{\'c}}}, \bibinfo {author} {\bibfnamefont {R.}~\bibnamefont {Verresen}}, \bibinfo {author} {\bibfnamefont {F.}~\bibnamefont {Pollmann}}, \bibinfo {author} {\bibfnamefont {G.}~\bibnamefont {Salomon}}, \bibinfo {author} {\bibfnamefont {C.}~\bibnamefont {Gross}}, \bibinfo {author} {\bibfnamefont {T.~A.}\ \bibnamefont {Hilker}},\ and\ \bibinfo {author} {\bibfnamefont {I.}~\bibnamefont {Bloch}},\ }\bibfield  {title} {\bibinfo {title} {{Realizing the symmetry-protected Haldane phase in Fermi-Hubbard ladders}},\ }\href {https://doi.org/10.1038/s41586-022-04688-z} {\bibfield  {journal} {\bibinfo  {journal} {Nature}\ }\textbf {\bibinfo {volume} {606}},\ \bibinfo {pages} {484 } (\bibinfo {year} {2022})}\BibitemShut {NoStop}%
    \bibitem [{\citenamefont {Semeghini}\ \emph {et~al.}(2021)\citenamefont {Semeghini}, \citenamefont {Levine}, \citenamefont {Keesling}, \citenamefont {Ebadi}, \citenamefont {Wang}, \citenamefont {Bluvstein}, \citenamefont {Verresen}, \citenamefont {Pichler}, \citenamefont {Kalinowski}, \citenamefont {Samajdar}, \citenamefont {Omran}, \citenamefont {Sachdev}, \citenamefont {Vishwanath}, \citenamefont {Greiner}, \citenamefont {Vuletić},\ and\ \citenamefont {Lukin}}]{Semeghini_2021}%
      \BibitemOpen
      \bibfield  {author} {\bibinfo {author} {\bibfnamefont {G.}~\bibnamefont {Semeghini}}, \bibinfo {author} {\bibfnamefont {H.}~\bibnamefont {Levine}}, \bibinfo {author} {\bibfnamefont {A.}~\bibnamefont {Keesling}}, \bibinfo {author} {\bibfnamefont {S.}~\bibnamefont {Ebadi}}, \bibinfo {author} {\bibfnamefont {T.~T.}\ \bibnamefont {Wang}}, \bibinfo {author} {\bibfnamefont {D.}~\bibnamefont {Bluvstein}}, \bibinfo {author} {\bibfnamefont {R.}~\bibnamefont {Verresen}}, \bibinfo {author} {\bibfnamefont {H.}~\bibnamefont {Pichler}}, \bibinfo {author} {\bibfnamefont {M.}~\bibnamefont {Kalinowski}}, \bibinfo {author} {\bibfnamefont {R.}~\bibnamefont {Samajdar}}, \bibinfo {author} {\bibfnamefont {A.}~\bibnamefont {Omran}}, \bibinfo {author} {\bibfnamefont {S.}~\bibnamefont {Sachdev}}, \bibinfo {author} {\bibfnamefont {A.}~\bibnamefont {Vishwanath}}, \bibinfo {author} {\bibfnamefont {M.}~\bibnamefont {Greiner}}, \bibinfo {author} {\bibfnamefont {V.}~\bibnamefont {Vuletić}},\ and\ \bibinfo {author} {\bibfnamefont {M.~D.}\ \bibnamefont {Lukin}},\ }\bibfield  {title} {\bibinfo {title} {Probing topological spin liquids on a programmable quantum simulator},\ }\href {https://doi.org/10.1126/science.abi8794} {\bibfield  {journal} {\bibinfo  {journal} {Science}\ }\textbf {\bibinfo {volume} {374}},\ \bibinfo {pages} {1242} (\bibinfo {year} {2021})}\BibitemShut {NoStop}%
    \bibitem [{\citenamefont {Kiczynski}\ \emph {et~al.}(2022)\citenamefont {Kiczynski}, \citenamefont {Gorman}, \citenamefont {Geng}, \citenamefont {Donnelly}, \citenamefont {Chung}, \citenamefont {He}, \citenamefont {Keizer},\ and\ \citenamefont {Simmons}}]{Kiczynski2022EngineeringTS}%
      \BibitemOpen
      \bibfield  {author} {\bibinfo {author} {\bibfnamefont {M.}~\bibnamefont {Kiczynski}}, \bibinfo {author} {\bibfnamefont {S.~K.}\ \bibnamefont {Gorman}}, \bibinfo {author} {\bibfnamefont {H.}~\bibnamefont {Geng}}, \bibinfo {author} {\bibfnamefont {M.~B.}\ \bibnamefont {Donnelly}}, \bibinfo {author} {\bibfnamefont {Y.}~\bibnamefont {Chung}}, \bibinfo {author} {\bibfnamefont {Y.}~\bibnamefont {He}}, \bibinfo {author} {\bibfnamefont {J.~G.}\ \bibnamefont {Keizer}},\ and\ \bibinfo {author} {\bibfnamefont {M.~Y.}\ \bibnamefont {Simmons}},\ }\bibfield  {title} {\bibinfo {title} {{Engineering topological states in atom-based semiconductor quantum dots}},\ }\href {https://doi.org/10.1038/s41586-022-04706-0} {\bibfield  {journal} {\bibinfo  {journal} {Nature}\ }\textbf {\bibinfo {volume} {606}},\ \bibinfo {pages} {694 } (\bibinfo {year} {2022})}\BibitemShut {NoStop}%
    \bibitem [{\citenamefont {Cai}\ \emph {et~al.}(2019)\citenamefont {Cai}, \citenamefont {Han}, \citenamefont {Mei}, \citenamefont {Xu}, \citenamefont {Ma}, \citenamefont {Li}, \citenamefont {Wang}, \citenamefont {Song}, \citenamefont {Xue}, \citenamefont {Yin}, \citenamefont {Jia},\ and\ \citenamefont {Sun}}]{Cai19}%
      \BibitemOpen
      \bibfield  {author} {\bibinfo {author} {\bibfnamefont {W.}~\bibnamefont {Cai}}, \bibinfo {author} {\bibfnamefont {J.}~\bibnamefont {Han}}, \bibinfo {author} {\bibfnamefont {F.}~\bibnamefont {Mei}}, \bibinfo {author} {\bibfnamefont {Y.}~\bibnamefont {Xu}}, \bibinfo {author} {\bibfnamefont {Y.}~\bibnamefont {Ma}}, \bibinfo {author} {\bibfnamefont {X.}~\bibnamefont {Li}}, \bibinfo {author} {\bibfnamefont {H.}~\bibnamefont {Wang}}, \bibinfo {author} {\bibfnamefont {Y.~P.}\ \bibnamefont {Song}}, \bibinfo {author} {\bibfnamefont {Z.-Y.}\ \bibnamefont {Xue}}, \bibinfo {author} {\bibfnamefont {Z.-q.}\ \bibnamefont {Yin}}, \bibinfo {author} {\bibfnamefont {S.}~\bibnamefont {Jia}},\ and\ \bibinfo {author} {\bibfnamefont {L.}~\bibnamefont {Sun}},\ }\bibfield  {title} {\bibinfo {title} {Observation of topological magnon insulator states in a superconducting circuit},\ }\href {https://doi.org/10.1103/PhysRevLett.123.080501} {\bibfield  {journal} {\bibinfo  {journal} {Phys. Rev. Lett.}\ }\textbf {\bibinfo {volume} {123}},\ \bibinfo {pages} {080501} (\bibinfo {year} {2019})}\BibitemShut {NoStop}%
    \bibitem [{\citenamefont {Zhou}\ \emph {et~al.}(2023)\citenamefont {Zhou}, \citenamefont {Cappellini}, \citenamefont {Tusi}, \citenamefont {Franchi}, \citenamefont {Parravicini}, \citenamefont {Repellin}, \citenamefont {Greschner}, \citenamefont {Inguscio}, \citenamefont {Giamarchi}, \citenamefont {Filippone}, \citenamefont {Catani},\ and\ \citenamefont {Fallani}}]{Zhou23}%
      \BibitemOpen
      \bibfield  {author} {\bibinfo {author} {\bibfnamefont {T.-W.}\ \bibnamefont {Zhou}}, \bibinfo {author} {\bibfnamefont {G.}~\bibnamefont {Cappellini}}, \bibinfo {author} {\bibfnamefont {D.}~\bibnamefont {Tusi}}, \bibinfo {author} {\bibfnamefont {L.}~\bibnamefont {Franchi}}, \bibinfo {author} {\bibfnamefont {J.}~\bibnamefont {Parravicini}}, \bibinfo {author} {\bibfnamefont {C.}~\bibnamefont {Repellin}}, \bibinfo {author} {\bibfnamefont {S.}~\bibnamefont {Greschner}}, \bibinfo {author} {\bibfnamefont {M.}~\bibnamefont {Inguscio}}, \bibinfo {author} {\bibfnamefont {T.}~\bibnamefont {Giamarchi}}, \bibinfo {author} {\bibfnamefont {M.}~\bibnamefont {Filippone}}, \bibinfo {author} {\bibfnamefont {J.}~\bibnamefont {Catani}},\ and\ \bibinfo {author} {\bibfnamefont {L.}~\bibnamefont {Fallani}},\ }\bibfield  {title} {\bibinfo {title} {{Observation of universal Hall response in strongly interacting Fermions}},\ }\href {https://doi.org/10.1126/science.add1969} {\bibfield  {journal} {\bibinfo  {journal} {Science}\ }\textbf {\bibinfo {volume} {381}},\ \bibinfo {pages} {427} (\bibinfo {year} {2023})}\BibitemShut {NoStop}%
    \bibitem [{\citenamefont {Kwan}\ \emph {et~al.}()\citenamefont {Kwan}, \citenamefont {Segura}, \citenamefont {Li}, \citenamefont {Kim}, \citenamefont {Gorshkov}, \citenamefont {Eckardt}, \citenamefont {Bakkali-Hassani},\ and\ \citenamefont {Greiner}}]{Kwan23}%
      \BibitemOpen
      \bibfield  {author} {\bibinfo {author} {\bibfnamefont {J.}~\bibnamefont {Kwan}}, \bibinfo {author} {\bibfnamefont {P.}~\bibnamefont {Segura}}, \bibinfo {author} {\bibfnamefont {Y.}~\bibnamefont {Li}}, \bibinfo {author} {\bibfnamefont {S.}~\bibnamefont {Kim}}, \bibinfo {author} {\bibfnamefont {A.~V.}\ \bibnamefont {Gorshkov}}, \bibinfo {author} {\bibfnamefont {A.}~\bibnamefont {Eckardt}}, \bibinfo {author} {\bibfnamefont {B.}~\bibnamefont {Bakkali-Hassani}},\ and\ \bibinfo {author} {\bibfnamefont {M.}~\bibnamefont {Greiner}},\ }\bibfield  {title} {\bibinfo {title} {{Realization of 1D anyons with arbitrary statistical phase}},\ }\href {https://arxiv.org/abs/2306.01737} {\bibinfo  {journal} {arxiv:2306.01737}\ }\BibitemShut {NoStop}%
    \bibitem [{\citenamefont {Viebahn}\ \emph {et~al.}(2024)\citenamefont {Viebahn}, \citenamefont {Walter}, \citenamefont {Bertok}, \citenamefont {Zhu}, \citenamefont {G\"achter}, \citenamefont {Aligia}, \citenamefont {Heidrich-Meisner},\ and\ \citenamefont {Esslinger}}]{Viebahn23}%
      \BibitemOpen
    \bibfield  {journal} {  }\bibfield  {author} {\bibinfo {author} {\bibfnamefont {K.}~\bibnamefont {Viebahn}}, \bibinfo {author} {\bibfnamefont {A.-S.}\ \bibnamefont {Walter}}, \bibinfo {author} {\bibfnamefont {E.}~\bibnamefont {Bertok}}, \bibinfo {author} {\bibfnamefont {Z.}~\bibnamefont {Zhu}}, \bibinfo {author} {\bibfnamefont {M.}~\bibnamefont {G\"achter}}, \bibinfo {author} {\bibfnamefont {A.~A.}\ \bibnamefont {Aligia}}, \bibinfo {author} {\bibfnamefont {F.}~\bibnamefont {Heidrich-Meisner}},\ and\ \bibinfo {author} {\bibfnamefont {T.}~\bibnamefont {Esslinger}},\ }\bibfield  {title} {\bibinfo {title} {Interactions enable thouless pumping in a nonsliding lattice},\ }\href {https://doi.org/10.1103/PhysRevX.14.021049} {\bibfield  {journal} {\bibinfo  {journal} {Phys. Rev. X}\ }\textbf {\bibinfo {volume} {14}},\ \bibinfo {pages} {021049} (\bibinfo {year} {2024})}\BibitemShut {NoStop}%
    \bibitem [{\citenamefont {Katz}\ \emph {et~al.}()\citenamefont {Katz}, \citenamefont {Feng}, \citenamefont {Porras},\ and\ \citenamefont {Monroe}}]{Katz24}%
      \BibitemOpen
      \bibfield  {author} {\bibinfo {author} {\bibfnamefont {O.}~\bibnamefont {Katz}}, \bibinfo {author} {\bibfnamefont {L.}~\bibnamefont {Feng}}, \bibinfo {author} {\bibfnamefont {D.}~\bibnamefont {Porras}},\ and\ \bibinfo {author} {\bibfnamefont {C.}~\bibnamefont {Monroe}},\ }\bibfield  {title} {\bibinfo {title} {Observing topological insulator phases with a programmable quantum simulator},\ }\href {https://arxiv.org/abs/2401.10362} {\bibinfo  {journal} {arxiv:2401.10362}\ }\BibitemShut {NoStop}%
    \bibitem [{\citenamefont {Schuckert}\ \emph {et~al.}(2025)\citenamefont {Schuckert}, \citenamefont {Katz}, \citenamefont {Feng}, \citenamefont {Crane}, \citenamefont {De}, \citenamefont {Hafezi}, \citenamefont {Gorshkov},\ and\ \citenamefont {Monroe}}]{schuckert2023observation-old}%
      \BibitemOpen
    \bibfield  {journal} {  }\bibfield  {author} {\bibinfo {author} {\bibfnamefont {A.}~\bibnamefont {Schuckert}}, \bibinfo {author} {\bibfnamefont {O.}~\bibnamefont {Katz}}, \bibinfo {author} {\bibfnamefont {L.}~\bibnamefont {Feng}}, \bibinfo {author} {\bibfnamefont {E.}~\bibnamefont {Crane}}, \bibinfo {author} {\bibfnamefont {A.}~\bibnamefont {De}}, \bibinfo {author} {\bibfnamefont {M.}~\bibnamefont {Hafezi}}, \bibinfo {author} {\bibfnamefont {A.~V.}\ \bibnamefont {Gorshkov}},\ and\ \bibinfo {author} {\bibfnamefont {C.}~\bibnamefont {Monroe}},\ }\bibfield  {title} {\bibinfo {title} {Observation of a finite-energy phase transition in a one-dimensional quantum simulator},\ }\href {https://doi.org/10.1038/s41567-024-02751-2} {\bibfield  {journal} {\bibinfo  {journal} {Nat. Phys.}\ } (\bibinfo {year} {2025})}\BibitemShut {NoStop}%
    \bibitem [{\citenamefont {Su}\ \emph {et~al.}(1979)\citenamefont {Su}, \citenamefont {Schrieffer},\ and\ \citenamefont {Heeger}}]{PhysRevLett.42.1698}%
      \BibitemOpen
      \bibfield  {author} {\bibinfo {author} {\bibfnamefont {W.~P.}\ \bibnamefont {Su}}, \bibinfo {author} {\bibfnamefont {J.~R.}\ \bibnamefont {Schrieffer}},\ and\ \bibinfo {author} {\bibfnamefont {A.~J.}\ \bibnamefont {Heeger}},\ }\bibfield  {title} {\bibinfo {title} {{Solitons in polyacetylene}},\ }\href {https://doi.org/10.1103/PhysRevLett.42.1698} {\bibfield  {journal} {\bibinfo  {journal} {Phys. Rev. Lett.}\ }\textbf {\bibinfo {volume} {42}},\ \bibinfo {pages} {1698} (\bibinfo {year} {1979})}\BibitemShut {NoStop}%
    \bibitem [{\citenamefont {Grusdt}\ \emph {et~al.}(2013)\citenamefont {Grusdt}, \citenamefont {H{\"o}ning},\ and\ \citenamefont {Fleischhauer}}]{grusdt2013topological}%
      \BibitemOpen
      \bibfield  {author} {\bibinfo {author} {\bibfnamefont {F.}~\bibnamefont {Grusdt}}, \bibinfo {author} {\bibfnamefont {M.}~\bibnamefont {H{\"o}ning}},\ and\ \bibinfo {author} {\bibfnamefont {M.}~\bibnamefont {Fleischhauer}},\ }\bibfield  {title} {\bibinfo {title} {{Topological edge states in the one-dimensional superlattice Bose-Hubbard model}},\ }\href {https://doi.org/10.1103/PhysRevLett.110.260405} {\bibfield  {journal} {\bibinfo  {journal} {Phys. Rev. Lett.}\ }\textbf {\bibinfo {volume} {110}},\ \bibinfo {pages} {260405} (\bibinfo {year} {2013})}\BibitemShut {NoStop}%
    \bibitem [{\citenamefont {Zhu}\ \emph {et~al.}(2013)\citenamefont {Zhu}, \citenamefont {Wang}, \citenamefont {Chan},\ and\ \citenamefont {Duan}}]{Zhu2013Topological}%
      \BibitemOpen
      \bibfield  {author} {\bibinfo {author} {\bibfnamefont {S.-L.}\ \bibnamefont {Zhu}}, \bibinfo {author} {\bibfnamefont {Z.-D.}\ \bibnamefont {Wang}}, \bibinfo {author} {\bibfnamefont {Y.-H.}\ \bibnamefont {Chan}},\ and\ \bibinfo {author} {\bibfnamefont {L.-M.}\ \bibnamefont {Duan}},\ }\bibfield  {title} {\bibinfo {title} {{Topological Bose-Mott insulators in a one-dimensional optical superlattice}},\ }\href {https://doi.org/10.1103/PhysRevLett.110.075303} {\bibfield  {journal} {\bibinfo  {journal} {Phys. Rev. Lett.}\ }\textbf {\bibinfo {volume} {110}},\ \bibinfo {pages} {075303} (\bibinfo {year} {2013})}\BibitemShut {NoStop}%
    \bibitem [{\citenamefont {Yang}\ \emph {et~al.}(2020)\citenamefont {Yang}, \citenamefont {Sun}, \citenamefont {Huang}, \citenamefont {Wang}, \citenamefont {Deng}, \citenamefont {Dai}, \citenamefont {Yuan},\ and\ \citenamefont {Pan}}]{Yang2019CoolingAE}%
      \BibitemOpen
      \bibfield  {author} {\bibinfo {author} {\bibfnamefont {B.}~\bibnamefont {Yang}}, \bibinfo {author} {\bibfnamefont {H.}~\bibnamefont {Sun}}, \bibinfo {author} {\bibfnamefont {C.-J.}\ \bibnamefont {Huang}}, \bibinfo {author} {\bibfnamefont {H.-Y.}\ \bibnamefont {Wang}}, \bibinfo {author} {\bibfnamefont {Y.}~\bibnamefont {Deng}}, \bibinfo {author} {\bibfnamefont {H.-N.}\ \bibnamefont {Dai}}, \bibinfo {author} {\bibfnamefont {Z.-S.}\ \bibnamefont {Yuan}},\ and\ \bibinfo {author} {\bibfnamefont {J.-W.}\ \bibnamefont {Pan}},\ }\bibfield  {title} {\bibinfo {title} {{Cooling and entangling ultracold atoms in optical lattices}},\ }\href {https://doi.org/10.1126/science.aaz6801} {\bibfield  {journal} {\bibinfo  {journal} {Science}\ }\textbf {\bibinfo {volume} {369}},\ \bibinfo {pages} {550 } (\bibinfo {year} {2020})}\BibitemShut {NoStop}%
    \bibitem [{\citenamefont {Wang}\ \emph {et~al.}(2023)\citenamefont {Wang}, \citenamefont {Zhang}, \citenamefont {Yao}, \citenamefont {Liu}, \citenamefont {Zhu}, \citenamefont {Zheng}, \citenamefont {Wang}, \citenamefont {Zhai}, \citenamefont {Yuan},\ and\ \citenamefont {Pan}}]{Wang2022InterrelatedTA}%
      \BibitemOpen
      \bibfield  {author} {\bibinfo {author} {\bibfnamefont {H.-Y.}\ \bibnamefont {Wang}}, \bibinfo {author} {\bibfnamefont {W.-Y.}\ \bibnamefont {Zhang}}, \bibinfo {author} {\bibfnamefont {Z.}~\bibnamefont {Yao}}, \bibinfo {author} {\bibfnamefont {Y.}~\bibnamefont {Liu}}, \bibinfo {author} {\bibfnamefont {Z.-H.}\ \bibnamefont {Zhu}}, \bibinfo {author} {\bibfnamefont {Y.-G.}\ \bibnamefont {Zheng}}, \bibinfo {author} {\bibfnamefont {X.-K.}\ \bibnamefont {Wang}}, \bibinfo {author} {\bibfnamefont {H.}~\bibnamefont {Zhai}}, \bibinfo {author} {\bibfnamefont {Z.-S.}\ \bibnamefont {Yuan}},\ and\ \bibinfo {author} {\bibfnamefont {J.-W.}\ \bibnamefont {Pan}},\ }\bibfield  {title} {\bibinfo {title} {{Interrelated thermalization and quantum criticality in a lattice gauge simulator}},\ }\href {https://doi.org/10.1103/PhysRevLett.131.050401} {\bibfield  {journal} {\bibinfo  {journal} {Phys. Rev. Lett.}\ }\textbf {\bibinfo {volume} {131}},\ \bibinfo {pages} {050401} (\bibinfo {year} {2023})}\BibitemShut {NoStop}%
    \bibitem [{\citenamefont {Caio}\ \emph {et~al.}(2015)\citenamefont {Caio}, \citenamefont {Cooper},\ and\ \citenamefont {Bhaseen}}]{Caio_2015}%
      \BibitemOpen
      \bibfield  {author} {\bibinfo {author} {\bibfnamefont {M.~D.}\ \bibnamefont {Caio}}, \bibinfo {author} {\bibfnamefont {N.~R.}\ \bibnamefont {Cooper}},\ and\ \bibinfo {author} {\bibfnamefont {M.~J.}\ \bibnamefont {Bhaseen}},\ }\bibfield  {title} {\bibinfo {title} {{Quantum quenches in Chern insulators}},\ }\href {https://doi.org/10.1103/PhysRevLett.115.236403} {\bibfield  {journal} {\bibinfo  {journal} {Phys. Rev. Lett.}\ }\textbf {\bibinfo {volume} {115}},\ \bibinfo {pages} {236403} (\bibinfo {year} {2015})}\BibitemShut {NoStop}%
    \bibitem [{\citenamefont {D’Alessio}\ and\ \citenamefont {Rigol}(2015)}]{D_Alessio_2015}%
      \BibitemOpen
      \bibfield  {author} {\bibinfo {author} {\bibfnamefont {L.}~\bibnamefont {D’Alessio}}\ and\ \bibinfo {author} {\bibfnamefont {M.}~\bibnamefont {Rigol}},\ }\bibfield  {title} {\bibinfo {title} {{Dynamical preparation of Floquet Chern insulators}},\ }\href {https://doi.org//10.1038/ncomms9336} {\bibfield  {journal} {\bibinfo  {journal} {Nat. Commun.}\ }\textbf {\bibinfo {volume} {6}},\ \bibinfo {pages} {8336} (\bibinfo {year} {2015})}\BibitemShut {NoStop}%
    \bibitem [{\citenamefont {Hu}\ \emph {et~al.}(2016)\citenamefont {Hu}, \citenamefont {Zoller},\ and\ \citenamefont {Budich}}]{PhysRevLett.117.126803}%
      \BibitemOpen
      \bibfield  {author} {\bibinfo {author} {\bibfnamefont {Y.}~\bibnamefont {Hu}}, \bibinfo {author} {\bibfnamefont {P.}~\bibnamefont {Zoller}},\ and\ \bibinfo {author} {\bibfnamefont {J.~C.}\ \bibnamefont {Budich}},\ }\bibfield  {title} {\bibinfo {title} {{Dynamical buildup of a quantized Hall response from nontopological states}},\ }\href {https://doi.org/10.1103/PhysRevLett.117.126803} {\bibfield  {journal} {\bibinfo  {journal} {Phys. Rev. Lett.}\ }\textbf {\bibinfo {volume} {117}},\ \bibinfo {pages} {126803} (\bibinfo {year} {2016})}\BibitemShut {NoStop}%
    \bibitem [{\citenamefont {Wilson}\ \emph {et~al.}(2016)\citenamefont {Wilson}, \citenamefont {Song},\ and\ \citenamefont {Refael}}]{PhysRevLett.117.235302}%
      \BibitemOpen
      \bibfield  {author} {\bibinfo {author} {\bibfnamefont {J.~H.}\ \bibnamefont {Wilson}}, \bibinfo {author} {\bibfnamefont {J.~C.~W.}\ \bibnamefont {Song}},\ and\ \bibinfo {author} {\bibfnamefont {G.}~\bibnamefont {Refael}},\ }\bibfield  {title} {\bibinfo {title} {{Remnant geometric Hall response in a quantum quench}},\ }\href {https://doi.org/10.1103/PhysRevLett.117.235302} {\bibfield  {journal} {\bibinfo  {journal} {Phys. Rev. Lett.}\ }\textbf {\bibinfo {volume} {117}},\ \bibinfo {pages} {235302} (\bibinfo {year} {2016})}\BibitemShut {NoStop}%
    \bibitem [{\citenamefont {Wang}\ \emph {et~al.}(2017)\citenamefont {Wang}, \citenamefont {Zhang}, \citenamefont {Chen}, \citenamefont {Yu},\ and\ \citenamefont {Zhai}}]{PhysRevLett.118.185701}%
      \BibitemOpen
      \bibfield  {author} {\bibinfo {author} {\bibfnamefont {C.}~\bibnamefont {Wang}}, \bibinfo {author} {\bibfnamefont {P.}~\bibnamefont {Zhang}}, \bibinfo {author} {\bibfnamefont {X.}~\bibnamefont {Chen}}, \bibinfo {author} {\bibfnamefont {J.}~\bibnamefont {Yu}},\ and\ \bibinfo {author} {\bibfnamefont {H.}~\bibnamefont {Zhai}},\ }\bibfield  {title} {\bibinfo {title} {{Scheme to measure the topological number of a Chern insulator from quench dynamics}},\ }\href {https://doi.org/10.1103/PhysRevLett.118.185701} {\bibfield  {journal} {\bibinfo  {journal} {Phys. Rev. Lett.}\ }\textbf {\bibinfo {volume} {118}},\ \bibinfo {pages} {185701} (\bibinfo {year} {2017})}\BibitemShut {NoStop}%
    \bibitem [{\citenamefont {Gong}\ and\ \citenamefont {Ueda}(2018)}]{PhysRevLett.121.250601}%
      \BibitemOpen
      \bibfield  {author} {\bibinfo {author} {\bibfnamefont {Z.}~\bibnamefont {Gong}}\ and\ \bibinfo {author} {\bibfnamefont {M.}~\bibnamefont {Ueda}},\ }\bibfield  {title} {\bibinfo {title} {Topological entanglement-spectrum crossing in quench dynamics},\ }\href {https://doi.org/10.1103/PhysRevLett.121.250601} {\bibfield  {journal} {\bibinfo  {journal} {Phys. Rev. Lett.}\ }\textbf {\bibinfo {volume} {121}},\ \bibinfo {pages} {250601} (\bibinfo {year} {2018})}\BibitemShut {NoStop}%
    \bibitem [{\citenamefont {Yang}\ \emph {et~al.}(2018)\citenamefont {Yang}, \citenamefont {Li},\ and\ \citenamefont {Chen}}]{PhysRevB.97.060304}%
      \BibitemOpen
      \bibfield  {author} {\bibinfo {author} {\bibfnamefont {C.}~\bibnamefont {Yang}}, \bibinfo {author} {\bibfnamefont {L.}~\bibnamefont {Li}},\ and\ \bibinfo {author} {\bibfnamefont {S.}~\bibnamefont {Chen}},\ }\bibfield  {title} {\bibinfo {title} {Dynamical topological invariant after a quantum quench},\ }\href {https://doi.org/10.1103/PhysRevB.97.060304} {\bibfield  {journal} {\bibinfo  {journal} {Phys. Rev. B}\ }\textbf {\bibinfo {volume} {97}},\ \bibinfo {pages} {060304} (\bibinfo {year} {2018})}\BibitemShut {NoStop}%
    \bibitem [{\citenamefont {Zhang}\ \emph {et~al.}(2018{\natexlab{b}})\citenamefont {Zhang}, \citenamefont {Zhang}, \citenamefont {Niu},\ and\ \citenamefont {Liu}}]{Zhang_2018}%
      \BibitemOpen
      \bibfield  {author} {\bibinfo {author} {\bibfnamefont {L.}~\bibnamefont {Zhang}}, \bibinfo {author} {\bibfnamefont {L.}~\bibnamefont {Zhang}}, \bibinfo {author} {\bibfnamefont {S.}~\bibnamefont {Niu}},\ and\ \bibinfo {author} {\bibfnamefont {X.-J.}\ \bibnamefont {Liu}},\ }\bibfield  {title} {\bibinfo {title} {Dynamical classification of topological quantum phases},\ }\href {https://doi.org/10.1016/j.scib.2018.09.018} {\bibfield  {journal} {\bibinfo  {journal} {Science Bulletin}\ }\textbf {\bibinfo {volume} {63}},\ \bibinfo {pages} {1385–1391} (\bibinfo {year} {2018}{\natexlab{b}})}\BibitemShut {NoStop}%
    \bibitem [{\citenamefont {Fläschner}\ \emph {et~al.}(2018)\citenamefont {Fläschner}, \citenamefont {Vogel}, \citenamefont {Tarnowski}, \citenamefont {Rem}, \citenamefont {Lühmann}, \citenamefont {Heyl}, \citenamefont {Budich}, \citenamefont {Mathey}, \citenamefont {Sengstock},\ and\ \citenamefont {Weitenberg}}]{flaschner2018observation}%
      \BibitemOpen
      \bibfield  {author} {\bibinfo {author} {\bibfnamefont {N.}~\bibnamefont {Fläschner}}, \bibinfo {author} {\bibfnamefont {D.}~\bibnamefont {Vogel}}, \bibinfo {author} {\bibfnamefont {M.}~\bibnamefont {Tarnowski}}, \bibinfo {author} {\bibfnamefont {B.~S.}\ \bibnamefont {Rem}}, \bibinfo {author} {\bibfnamefont {D.-S.}\ \bibnamefont {Lühmann}}, \bibinfo {author} {\bibfnamefont {M.}~\bibnamefont {Heyl}}, \bibinfo {author} {\bibfnamefont {J.~C.}\ \bibnamefont {Budich}}, \bibinfo {author} {\bibfnamefont {L.}~\bibnamefont {Mathey}}, \bibinfo {author} {\bibfnamefont {K.}~\bibnamefont {Sengstock}},\ and\ \bibinfo {author} {\bibfnamefont {C.}~\bibnamefont {Weitenberg}},\ }\bibfield  {title} {\bibinfo {title} {Observation of dynamical vortices after quenches in a system with topology},\ }\href {https://doi.org/10.1038/s41567-017-0013-8} {\bibfield  {journal} {\bibinfo  {journal} {Nat. Phys.}\ }\textbf {\bibinfo {volume} {14}},\ \bibinfo {pages} {265} (\bibinfo {year} {2018})}\BibitemShut {NoStop}%
    \bibitem [{\citenamefont {Tarnowski}\ \emph {et~al.}(2019)\citenamefont {Tarnowski}, \citenamefont {Ünal}, \citenamefont {Fläschner}, \citenamefont {Rem}, \citenamefont {Eckardt}, \citenamefont {Sengstock},\ and\ \citenamefont {Weitenberg}}]{tarnowski2019measuring}%
      \BibitemOpen
      \bibfield  {author} {\bibinfo {author} {\bibfnamefont {M.}~\bibnamefont {Tarnowski}}, \bibinfo {author} {\bibfnamefont {F.~N.}\ \bibnamefont {Ünal}}, \bibinfo {author} {\bibfnamefont {N.}~\bibnamefont {Fläschner}}, \bibinfo {author} {\bibfnamefont {B.~S.}\ \bibnamefont {Rem}}, \bibinfo {author} {\bibfnamefont {A.}~\bibnamefont {Eckardt}}, \bibinfo {author} {\bibfnamefont {K.}~\bibnamefont {Sengstock}},\ and\ \bibinfo {author} {\bibfnamefont {C.}~\bibnamefont {Weitenberg}},\ }\bibfield  {title} {\bibinfo {title} {Measuring topology from dynamics by obtaining the {{Chern}} number from a linking number},\ }\href {https://doi.org/10.1038/s41467-019-09668-y} {\bibfield  {journal} {\bibinfo  {journal} {Nat. Commun.}\ }\textbf {\bibinfo {volume} {10}},\ \bibinfo {pages} {1728} (\bibinfo {year} {2019})}\BibitemShut {NoStop}%
    \bibitem [{\citenamefont {McGinley}\ and\ \citenamefont {Cooper}(2018)}]{McGinley_2018}%
      \BibitemOpen
      \bibfield  {author} {\bibinfo {author} {\bibfnamefont {M.}~\bibnamefont {McGinley}}\ and\ \bibinfo {author} {\bibfnamefont {N.~R.}\ \bibnamefont {Cooper}},\ }\bibfield  {title} {\bibinfo {title} {{Topology of one-dimensional quantum systems out of equilibrium}},\ }\href {https://doi.org/10.1103/PhysRevLett.121.090401} {\bibfield  {journal} {\bibinfo  {journal} {Phys. Rev. Lett.}\ }\textbf {\bibinfo {volume} {121}},\ \bibinfo {pages} {090401} (\bibinfo {year} {2018})}\BibitemShut {NoStop}%
    \bibitem [{\citenamefont {McGinley}\ and\ \citenamefont {Cooper}(2019)}]{McGinley_2019}%
      \BibitemOpen
      \bibfield  {author} {\bibinfo {author} {\bibfnamefont {M.}~\bibnamefont {McGinley}}\ and\ \bibinfo {author} {\bibfnamefont {N.~R.}\ \bibnamefont {Cooper}},\ }\bibfield  {title} {\bibinfo {title} {{Interacting symmetry-protected topological phases out of equilibrium}},\ }\href {https://doi.org/10.1103/PhysRevResearch.1.033204} {\bibfield  {journal} {\bibinfo  {journal} {Phys. Rev. Res.}\ }\textbf {\bibinfo {volume} {1}},\ \bibinfo {pages} {033204} (\bibinfo {year} {2019})}\BibitemShut {NoStop}%
    \bibitem [{\citenamefont {Zhang}\ \emph {et~al.}(2021)\citenamefont {Zhang}, \citenamefont {Zhang}, \citenamefont {Hu}, \citenamefont {Niu},\ and\ \citenamefont {Liu}}]{PhysRevB.103.224308}%
      \BibitemOpen
      \bibfield  {author} {\bibinfo {author} {\bibfnamefont {L.}~\bibnamefont {Zhang}}, \bibinfo {author} {\bibfnamefont {L.}~\bibnamefont {Zhang}}, \bibinfo {author} {\bibfnamefont {Y.}~\bibnamefont {Hu}}, \bibinfo {author} {\bibfnamefont {S.}~\bibnamefont {Niu}},\ and\ \bibinfo {author} {\bibfnamefont {X.-J.}\ \bibnamefont {Liu}},\ }\bibfield  {title} {\bibinfo {title} {Nonequilibrium characterization of equilibrium correlated quantum phases},\ }\href {https://doi.org/10.1103/PhysRevB.103.224308} {\bibfield  {journal} {\bibinfo  {journal} {Phys. Rev. B}\ }\textbf {\bibinfo {volume} {103}},\ \bibinfo {pages} {224308} (\bibinfo {year} {2021})}\BibitemShut {NoStop}%
    \bibitem [{\citenamefont {Pastori}\ \emph {et~al.}(2020)\citenamefont {Pastori}, \citenamefont {Barbarino},\ and\ \citenamefont {Budich}}]{PhysRevResearch.2.033259}%
      \BibitemOpen
      \bibfield  {author} {\bibinfo {author} {\bibfnamefont {L.}~\bibnamefont {Pastori}}, \bibinfo {author} {\bibfnamefont {S.}~\bibnamefont {Barbarino}},\ and\ \bibinfo {author} {\bibfnamefont {J.~C.}\ \bibnamefont {Budich}},\ }\bibfield  {title} {\bibinfo {title} {Signatures of topology in quantum quench dynamics and their interrelation},\ }\href {https://doi.org/10.1103/PhysRevResearch.2.033259} {\bibfield  {journal} {\bibinfo  {journal} {Phys. Rev. Res.}\ }\textbf {\bibinfo {volume} {2}},\ \bibinfo {pages} {033259} (\bibinfo {year} {2020})}\BibitemShut {NoStop}%
    \bibitem [{\citenamefont {Haldane}(1983)}]{Haldane1983NonlinearFT}%
      \BibitemOpen
      \bibfield  {author} {\bibinfo {author} {\bibfnamefont {F.~D.~M.}\ \bibnamefont {Haldane}},\ }\bibfield  {title} {\bibinfo {title} {{Nonlinear field theory of large-spin Heisenberg antiferromagnets: Semiclassically quantized solitons of the one-dimensional easy-axis N{\'e}el state}},\ }\href {https://doi.org/10.1103/PhysRevLett.50.1153} {\bibfield  {journal} {\bibinfo  {journal} {Phys. Rev. Lett.}\ }\textbf {\bibinfo {volume} {50}},\ \bibinfo {pages} {1153} (\bibinfo {year} {1983})}\BibitemShut {NoStop}%
    \bibitem [{\citenamefont {Chen}\ \emph {et~al.}(2012)\citenamefont {Chen}, \citenamefont {Gu}, \citenamefont {Liu},\ and\ \citenamefont {Wen}}]{ChenX2012}%
      \BibitemOpen
      \bibfield  {author} {\bibinfo {author} {\bibfnamefont {X.}~\bibnamefont {Chen}}, \bibinfo {author} {\bibfnamefont {Z.-C.}\ \bibnamefont {Gu}}, \bibinfo {author} {\bibfnamefont {Z.-X.}\ \bibnamefont {Liu}},\ and\ \bibinfo {author} {\bibfnamefont {X.-G.}\ \bibnamefont {Wen}},\ }\bibfield  {title} {\bibinfo {title} {Symmetry-protected topological orders in interacting bosonic systems},\ }\href {https://doi.org/10.1126/science.1227224} {\bibfield  {journal} {\bibinfo  {journal} {Science}\ }\textbf {\bibinfo {volume} {338}},\ \bibinfo {pages} {1604} (\bibinfo {year} {2012})}\BibitemShut {NoStop}%
    \bibitem [{\citenamefont {Dalla~Torre}\ \emph {et~al.}(2006)\citenamefont {Dalla~Torre}, \citenamefont {Berg},\ and\ \citenamefont {Altman}}]{Torre06}%
      \BibitemOpen
      \bibfield  {author} {\bibinfo {author} {\bibfnamefont {E.~G.}\ \bibnamefont {Dalla~Torre}}, \bibinfo {author} {\bibfnamefont {E.}~\bibnamefont {Berg}},\ and\ \bibinfo {author} {\bibfnamefont {E.}~\bibnamefont {Altman}},\ }\bibfield  {title} {\bibinfo {title} {{Hidden order in 1D Bose insulators}},\ }\href {https://doi.org/10.1103/PhysRevLett.97.260401} {\bibfield  {journal} {\bibinfo  {journal} {Phys. Rev. Lett.}\ }\textbf {\bibinfo {volume} {97}},\ \bibinfo {pages} {260401} (\bibinfo {year} {2006})}\BibitemShut {NoStop}%
    \bibitem [{\citenamefont {Berg}\ \emph {et~al.}(2008)\citenamefont {Berg}, \citenamefont {Dalla~Torre}, \citenamefont {Giamarchi},\ and\ \citenamefont {Altman}}]{Erez08}%
      \BibitemOpen
      \bibfield  {author} {\bibinfo {author} {\bibfnamefont {E.}~\bibnamefont {Berg}}, \bibinfo {author} {\bibfnamefont {E.~G.}\ \bibnamefont {Dalla~Torre}}, \bibinfo {author} {\bibfnamefont {T.}~\bibnamefont {Giamarchi}},\ and\ \bibinfo {author} {\bibfnamefont {E.}~\bibnamefont {Altman}},\ }\bibfield  {title} {\bibinfo {title} {{Rise and fall of hidden string order of lattice bosons}},\ }\href {https://doi.org/10.1103/PhysRevB.77.245119} {\bibfield  {journal} {\bibinfo  {journal} {Phys. Rev. B}\ }\textbf {\bibinfo {volume} {77}},\ \bibinfo {pages} {245119} (\bibinfo {year} {2008})}\BibitemShut {NoStop}%
    \bibitem [{\citenamefont {Mazza}\ \emph {et~al.}(2014)\citenamefont {Mazza}, \citenamefont {Rossini}, \citenamefont {Endres},\ and\ \citenamefont {Fazio}}]{Mazza14}%
      \BibitemOpen
      \bibfield  {author} {\bibinfo {author} {\bibfnamefont {L.}~\bibnamefont {Mazza}}, \bibinfo {author} {\bibfnamefont {D.}~\bibnamefont {Rossini}}, \bibinfo {author} {\bibfnamefont {M.}~\bibnamefont {Endres}},\ and\ \bibinfo {author} {\bibfnamefont {R.}~\bibnamefont {Fazio}},\ }\bibfield  {title} {\bibinfo {title} {{Out-of-equilibrium dynamics and thermalization of string order}},\ }\href {https://doi.org/10.1103/PhysRevB.90.020301} {\bibfield  {journal} {\bibinfo  {journal} {Phys. Rev. B}\ }\textbf {\bibinfo {volume} {90}},\ \bibinfo {pages} {020301} (\bibinfo {year} {2014})}\BibitemShut {NoStop}%
    \bibitem [{\citenamefont {den Nijs}\ and\ \citenamefont {Rommelse}(1989)}]{Nijs89}%
      \BibitemOpen
      \bibfield  {author} {\bibinfo {author} {\bibfnamefont {M.}~\bibnamefont {den Nijs}}\ and\ \bibinfo {author} {\bibfnamefont {K.}~\bibnamefont {Rommelse}},\ }\bibfield  {title} {\bibinfo {title} {{Preroughening transitions in crystal surfaces and valence-bond phases in quantum spin chains}},\ }\href {https://doi.org/10.1103/PhysRevB.40.4709} {\bibfield  {journal} {\bibinfo  {journal} {Phys. Rev. B}\ }\textbf {\bibinfo {volume} {40}},\ \bibinfo {pages} {4709} (\bibinfo {year} {1989})}\BibitemShut {NoStop}%
    \bibitem [{\citenamefont {Kennedy}\ and\ \citenamefont {Tasaki}(1992)}]{Kennedy92}%
      \BibitemOpen
      \bibfield  {author} {\bibinfo {author} {\bibfnamefont {T.}~\bibnamefont {Kennedy}}\ and\ \bibinfo {author} {\bibfnamefont {H.}~\bibnamefont {Tasaki}},\ }\bibfield  {title} {\bibinfo {title} {{Hidden ${\mathrm{Z}}_{2}$\ifmmode\times\else\texttimes\fi{}${\mathrm{Z}}_{2}$ symmetry breaking in Haldane-gap antiferromagnets}},\ }\href {https://doi.org/10.1103/PhysRevB.45.304} {\bibfield  {journal} {\bibinfo  {journal} {Phys. Rev. B}\ }\textbf {\bibinfo {volume} {45}},\ \bibinfo {pages} {304} (\bibinfo {year} {1992})}\BibitemShut {NoStop}%
    \bibitem [{SM()}]{SM}%
      \BibitemOpen
      \href@noop {} {}\bibinfo {note} {See Supplemental Material for ground-state phase diagram and topological properties, more discussions about relaxation dynamics of string order, entanglement entropy, initial-state dependence, and entanglement spectrum, and many-body dynamics based on analytical and effective models. Supplemental Material includes Refs.~\cite{PhysRevLett.69.2863,PhysRevB.82.012405,PhysRevLett.108.116401,PhysRevB.75.144420,PhysRevB.87.054402,Bahovadinov_2019,PhysRevB.86.125441, PhysRevLett.101.010504,PhysRevB.81.064439,PhysRevLett.113.020401,PhysRevX.6.041033, Islam_2015,PhysRevLett.109.020505,Essler_Frahm_Göhmann_Klümper_Korepin_2005,auerbach2012interacting,PhysRevLett.106.140405}}\BibitemShut {NoStop}%
    \bibitem [{\citenamefont {White}(1992)}]{PhysRevLett.69.2863}%
      \BibitemOpen
      \bibfield  {author} {\bibinfo {author} {\bibfnamefont {S.~R.}\ \bibnamefont {White}},\ }\bibfield  {title} {\bibinfo {title} {Density matrix formulation for quantum renormalization groups},\ }\href {https://doi.org/10.1103/PhysRevLett.69.2863} {\bibfield  {journal} {\bibinfo  {journal} {Phys. Rev. Lett.}\ }\textbf {\bibinfo {volume} {69}},\ \bibinfo {pages} {2863} (\bibinfo {year} {1992})}\BibitemShut {NoStop}%
    \bibitem [{\citenamefont {Song}\ \emph {et~al.}(2010)\citenamefont {Song}, \citenamefont {Rachel},\ and\ \citenamefont {Le~Hur}}]{PhysRevB.82.012405}%
      \BibitemOpen
      \bibfield  {author} {\bibinfo {author} {\bibfnamefont {H.~F.}\ \bibnamefont {Song}}, \bibinfo {author} {\bibfnamefont {S.}~\bibnamefont {Rachel}},\ and\ \bibinfo {author} {\bibfnamefont {K.}~\bibnamefont {Le~Hur}},\ }\bibfield  {title} {\bibinfo {title} {General relation between entanglement and fluctuations in one dimension},\ }\href {https://doi.org/10.1103/PhysRevB.82.012405} {\bibfield  {journal} {\bibinfo  {journal} {Phys. Rev. B}\ }\textbf {\bibinfo {volume} {82}},\ \bibinfo {pages} {012405} (\bibinfo {year} {2010})}\BibitemShut {NoStop}%
    \bibitem [{\citenamefont {Rachel}\ \emph {et~al.}(2012)\citenamefont {Rachel}, \citenamefont {Laflorencie}, \citenamefont {Song},\ and\ \citenamefont {Le~Hur}}]{PhysRevLett.108.116401}%
      \BibitemOpen
      \bibfield  {author} {\bibinfo {author} {\bibfnamefont {S.}~\bibnamefont {Rachel}}, \bibinfo {author} {\bibfnamefont {N.}~\bibnamefont {Laflorencie}}, \bibinfo {author} {\bibfnamefont {H.~F.}\ \bibnamefont {Song}},\ and\ \bibinfo {author} {\bibfnamefont {K.}~\bibnamefont {Le~Hur}},\ }\bibfield  {title} {\bibinfo {title} {Detecting quantum critical points using bipartite fluctuations},\ }\href {https://doi.org/10.1103/PhysRevLett.108.116401} {\bibfield  {journal} {\bibinfo  {journal} {Phys. Rev. Lett.}\ }\textbf {\bibinfo {volume} {108}},\ \bibinfo {pages} {116401} (\bibinfo {year} {2012})}\BibitemShut {NoStop}%
    \bibitem [{\citenamefont {Anfuso}\ and\ \citenamefont {Rosch}(2007)}]{PhysRevB.75.144420}%
      \BibitemOpen
      \bibfield  {author} {\bibinfo {author} {\bibfnamefont {F.}~\bibnamefont {Anfuso}}\ and\ \bibinfo {author} {\bibfnamefont {A.}~\bibnamefont {Rosch}},\ }\bibfield  {title} {\bibinfo {title} {{String order and adiabatic continuity of Haldane chains and band insulators}},\ }\href {https://doi.org/10.1103/PhysRevB.75.144420} {\bibfield  {journal} {\bibinfo  {journal} {Phys. Rev. B}\ }\textbf {\bibinfo {volume} {75}},\ \bibinfo {pages} {144420} (\bibinfo {year} {2007})}\BibitemShut {NoStop}%
    \bibitem [{\citenamefont {Wang}\ \emph {et~al.}(2013)\citenamefont {Wang}, \citenamefont {Li},\ and\ \citenamefont {Cho}}]{PhysRevB.87.054402}%
      \BibitemOpen
      \bibfield  {author} {\bibinfo {author} {\bibfnamefont {H.~T.}\ \bibnamefont {Wang}}, \bibinfo {author} {\bibfnamefont {B.}~\bibnamefont {Li}},\ and\ \bibinfo {author} {\bibfnamefont {S.~Y.}\ \bibnamefont {Cho}},\ }\bibfield  {title} {\bibinfo {title} {{Topological quantum phase transition in bond-alternating spin-$\frac{1}{2}$ Heisenberg chains}},\ }\href {https://doi.org/10.1103/PhysRevB.87.054402} {\bibfield  {journal} {\bibinfo  {journal} {Phys. Rev. B}\ }\textbf {\bibinfo {volume} {87}},\ \bibinfo {pages} {054402} (\bibinfo {year} {2013})}\BibitemShut {NoStop}%
    \bibitem [{\citenamefont {Bahovadinov}\ \emph {et~al.}(2019)\citenamefont {Bahovadinov}, \citenamefont {Gülseren},\ and\ \citenamefont {Schnack}}]{Bahovadinov_2019}%
      \BibitemOpen
      \bibfield  {author} {\bibinfo {author} {\bibfnamefont {M.~S.}\ \bibnamefont {Bahovadinov}}, \bibinfo {author} {\bibfnamefont {O.}~\bibnamefont {Gülseren}},\ and\ \bibinfo {author} {\bibfnamefont {J.}~\bibnamefont {Schnack}},\ }\bibfield  {title} {\bibinfo {title} {Local entanglement and string order parameter in dimerized models},\ }\href {https://doi.org/10.1088/1361-648X/ab41b5} {\bibfield  {journal} {\bibinfo  {journal} {Journal of Physics: Condensed Matter}\ }\textbf {\bibinfo {volume} {31}},\ \bibinfo {pages} {505602} (\bibinfo {year} {2019})}\BibitemShut {NoStop}%
    \bibitem [{\citenamefont {Pollmann}\ and\ \citenamefont {Turner}(2012)}]{PhysRevB.86.125441}%
      \BibitemOpen
      \bibfield  {author} {\bibinfo {author} {\bibfnamefont {F.}~\bibnamefont {Pollmann}}\ and\ \bibinfo {author} {\bibfnamefont {A.~M.}\ \bibnamefont {Turner}},\ }\bibfield  {title} {\bibinfo {title} {Detection of symmetry-protected topological phases in one dimension},\ }\href {https://doi.org/10.1103/PhysRevB.86.125441} {\bibfield  {journal} {\bibinfo  {journal} {Phys. Rev. B}\ }\textbf {\bibinfo {volume} {86}},\ \bibinfo {pages} {125441} (\bibinfo {year} {2012})}\BibitemShut {NoStop}%
    \bibitem [{\citenamefont {Li}\ and\ \citenamefont {Haldane}(2008)}]{PhysRevLett.101.010504}%
      \BibitemOpen
      \bibfield  {author} {\bibinfo {author} {\bibfnamefont {H.}~\bibnamefont {Li}}\ and\ \bibinfo {author} {\bibfnamefont {F.~D.~M.}\ \bibnamefont {Haldane}},\ }\bibfield  {title} {\bibinfo {title} {{Entanglement spectrum as a generalization of entanglement entropy: Identification of topological order in non-abelian fractional quantum Hall effect states}},\ }\href {https://doi.org/10.1103/PhysRevLett.101.010504} {\bibfield  {journal} {\bibinfo  {journal} {Phys. Rev. Lett.}\ }\textbf {\bibinfo {volume} {101}},\ \bibinfo {pages} {010504} (\bibinfo {year} {2008})}\BibitemShut {NoStop}%
    \bibitem [{\citenamefont {Pollmann}\ \emph {et~al.}(2010)\citenamefont {Pollmann}, \citenamefont {Turner}, \citenamefont {Berg},\ and\ \citenamefont {Oshikawa}}]{PhysRevB.81.064439}%
      \BibitemOpen
      \bibfield  {author} {\bibinfo {author} {\bibfnamefont {F.}~\bibnamefont {Pollmann}}, \bibinfo {author} {\bibfnamefont {A.~M.}\ \bibnamefont {Turner}}, \bibinfo {author} {\bibfnamefont {E.}~\bibnamefont {Berg}},\ and\ \bibinfo {author} {\bibfnamefont {M.}~\bibnamefont {Oshikawa}},\ }\bibfield  {title} {\bibinfo {title} {Entanglement spectrum of a topological phase in one dimension},\ }\href {https://doi.org/10.1103/PhysRevB.81.064439} {\bibfield  {journal} {\bibinfo  {journal} {Phys. Rev. B}\ }\textbf {\bibinfo {volume} {81}},\ \bibinfo {pages} {064439} (\bibinfo {year} {2010})}\BibitemShut {NoStop}%
    \bibitem [{\citenamefont {Ejima}\ \emph {et~al.}(2014)\citenamefont {Ejima}, \citenamefont {Lange},\ and\ \citenamefont {Fehske}}]{PhysRevLett.113.020401}%
      \BibitemOpen
      \bibfield  {author} {\bibinfo {author} {\bibfnamefont {S.}~\bibnamefont {Ejima}}, \bibinfo {author} {\bibfnamefont {F.}~\bibnamefont {Lange}},\ and\ \bibinfo {author} {\bibfnamefont {H.}~\bibnamefont {Fehske}},\ }\bibfield  {title} {\bibinfo {title} {{Spectral and entanglement properties of the Bosonic Haldane insulator}},\ }\href {https://doi.org/10.1103/PhysRevLett.113.020401} {\bibfield  {journal} {\bibinfo  {journal} {Phys. Rev. Lett.}\ }\textbf {\bibinfo {volume} {113}},\ \bibinfo {pages} {020401} (\bibinfo {year} {2014})}\BibitemShut {NoStop}%
    \bibitem [{\citenamefont {Pichler}\ \emph {et~al.}(2016)\citenamefont {Pichler}, \citenamefont {Zhu}, \citenamefont {Seif}, \citenamefont {Zoller},\ and\ \citenamefont {Hafezi}}]{PhysRevX.6.041033}%
      \BibitemOpen
      \bibfield  {author} {\bibinfo {author} {\bibfnamefont {H.}~\bibnamefont {Pichler}}, \bibinfo {author} {\bibfnamefont {G.}~\bibnamefont {Zhu}}, \bibinfo {author} {\bibfnamefont {A.}~\bibnamefont {Seif}}, \bibinfo {author} {\bibfnamefont {P.}~\bibnamefont {Zoller}},\ and\ \bibinfo {author} {\bibfnamefont {M.}~\bibnamefont {Hafezi}},\ }\bibfield  {title} {\bibinfo {title} {{Measurement protocol for the entanglement spectrum of cold atoms}},\ }\href {https://doi.org/10.1103/PhysRevX.6.041033} {\bibfield  {journal} {\bibinfo  {journal} {Phys. Rev. X}\ }\textbf {\bibinfo {volume} {6}},\ \bibinfo {pages} {041033} (\bibinfo {year} {2016})}\BibitemShut {NoStop}%
    \bibitem [{\citenamefont {Islam}\ \emph {et~al.}(2015)\citenamefont {Islam}, \citenamefont {Ma}, \citenamefont {Preiss}, \citenamefont {Eric~Tai}, \citenamefont {Lukin}, \citenamefont {Rispoli},\ and\ \citenamefont {Greiner}}]{Islam_2015}%
      \BibitemOpen
      \bibfield  {author} {\bibinfo {author} {\bibfnamefont {R.}~\bibnamefont {Islam}}, \bibinfo {author} {\bibfnamefont {R.}~\bibnamefont {Ma}}, \bibinfo {author} {\bibfnamefont {P.~M.}\ \bibnamefont {Preiss}}, \bibinfo {author} {\bibfnamefont {M.}~\bibnamefont {Eric~Tai}}, \bibinfo {author} {\bibfnamefont {A.}~\bibnamefont {Lukin}}, \bibinfo {author} {\bibfnamefont {M.}~\bibnamefont {Rispoli}},\ and\ \bibinfo {author} {\bibfnamefont {M.}~\bibnamefont {Greiner}},\ }\bibfield  {title} {\bibinfo {title} {Measuring entanglement entropy in a quantum many-body system},\ }\href {https://doi.org/10.1038/nature15750} {\bibfield  {journal} {\bibinfo  {journal} {Nature}\ }\textbf {\bibinfo {volume} {528}},\ \bibinfo {pages} {77–83} (\bibinfo {year} {2015})}\BibitemShut {NoStop}%
    \bibitem [{\citenamefont {Daley}\ \emph {et~al.}(2012)\citenamefont {Daley}, \citenamefont {Pichler}, \citenamefont {Schachenmayer},\ and\ \citenamefont {Zoller}}]{PhysRevLett.109.020505}%
      \BibitemOpen
      \bibfield  {author} {\bibinfo {author} {\bibfnamefont {A.~J.}\ \bibnamefont {Daley}}, \bibinfo {author} {\bibfnamefont {H.}~\bibnamefont {Pichler}}, \bibinfo {author} {\bibfnamefont {J.}~\bibnamefont {Schachenmayer}},\ and\ \bibinfo {author} {\bibfnamefont {P.}~\bibnamefont {Zoller}},\ }\bibfield  {title} {\bibinfo {title} {{Measuring entanglement growth in quench dynamics of Bosons in an optical lattice}},\ }\href {https://doi.org/10.1103/PhysRevLett.109.020505} {\bibfield  {journal} {\bibinfo  {journal} {Phys. Rev. Lett.}\ }\textbf {\bibinfo {volume} {109}},\ \bibinfo {pages} {020505} (\bibinfo {year} {2012})}\BibitemShut {NoStop}%
    \bibitem [{\citenamefont {Essler}\ \emph {et~al.}(2005)\citenamefont {Essler}, \citenamefont {Frahm}, \citenamefont {Göhmann}, \citenamefont {Klümper},\ and\ \citenamefont {Korepin}}]{Essler_Frahm_Göhmann_Klümper_Korepin_2005}%
      \BibitemOpen
      \bibfield  {author} {\bibinfo {author} {\bibfnamefont {F.~H.~L.}\ \bibnamefont {Essler}}, \bibinfo {author} {\bibfnamefont {H.}~\bibnamefont {Frahm}}, \bibinfo {author} {\bibfnamefont {F.}~\bibnamefont {Göhmann}}, \bibinfo {author} {\bibfnamefont {A.}~\bibnamefont {Klümper}},\ and\ \bibinfo {author} {\bibfnamefont {V.~E.}\ \bibnamefont {Korepin}},\ }\href {https://doi.org/10.1017/CBO9780511534843} { {\bibinfo {title} {{The One-Dimensional Hubbard Model}}}}\ (\bibinfo  {publisher} {Cambridge University Press},\ \bibinfo {year} {2005})\BibitemShut {NoStop}%
    \bibitem [{\citenamefont {Auerbach}(2012)}]{auerbach2012interacting}%
      \BibitemOpen
      \bibfield  {author} {\bibinfo {author} {\bibfnamefont {A.}~\bibnamefont {Auerbach}},\ }\href {https://doi.org/10.1007/978-1-4612-0869-3} { {\bibinfo {title} {{Interacting Electrons and Quantum Magnetism}}}}\ (\bibinfo  {publisher} {Springer New York},\ \bibinfo {year} {2012})\BibitemShut {NoStop}%
    \bibitem [{\citenamefont {Cassidy}\ \emph {et~al.}(2011)\citenamefont {Cassidy}, \citenamefont {Clark},\ and\ \citenamefont {Rigol}}]{PhysRevLett.106.140405}%
      \BibitemOpen
      \bibfield  {author} {\bibinfo {author} {\bibfnamefont {A.~C.}\ \bibnamefont {Cassidy}}, \bibinfo {author} {\bibfnamefont {C.~W.}\ \bibnamefont {Clark}},\ and\ \bibinfo {author} {\bibfnamefont {M.}~\bibnamefont {Rigol}},\ }\bibfield  {title} {\bibinfo {title} {{Generalized thermalization in an integrable lattice system}},\ }\href {https://doi.org/10.1103/PhysRevLett.106.140405} {\bibfield  {journal} {\bibinfo  {journal} {Phys. Rev. Lett.}\ }\textbf {\bibinfo {volume} {106}},\ \bibinfo {pages} {140405} (\bibinfo {year} {2011})}\BibitemShut {NoStop}%
    \bibitem [{\citenamefont {Weinberg}\ and\ \citenamefont {Bukov}(2017)}]{Phillip2017QuSpin}%
      \BibitemOpen
      \bibfield  {author} {\bibinfo {author} {\bibfnamefont {P.}~\bibnamefont {Weinberg}}\ and\ \bibinfo {author} {\bibfnamefont {M.}~\bibnamefont {Bukov}},\ }\bibfield  {title} {\bibinfo {title} {{QuSpin: a Python package for dynamics and exact diagonalisation of quantum many body systems part I: spin chains}},\ }\href {https://doi.org/10.21468/SciPostPhys.2.1.003} {\bibfield  {journal} {\bibinfo  {journal} {SciPost Phys.}\ }\textbf {\bibinfo {volume} {2}},\ \bibinfo {pages} {003} (\bibinfo {year} {2017})}\BibitemShut {NoStop}%
    \bibitem [{\citenamefont {Haegeman}\ \emph {et~al.}(2011)\citenamefont {Haegeman}, \citenamefont {Cirac}, \citenamefont {Osborne}, \citenamefont {Pi\ifmmode~\check{z}\else \v{z}\fi{}orn}, \citenamefont {Verschelde},\ and\ \citenamefont {Verstraete}}]{Haegeman2011Time}%
      \BibitemOpen
      \bibfield  {author} {\bibinfo {author} {\bibfnamefont {J.}~\bibnamefont {Haegeman}}, \bibinfo {author} {\bibfnamefont {J.~I.}\ \bibnamefont {Cirac}}, \bibinfo {author} {\bibfnamefont {T.~J.}\ \bibnamefont {Osborne}}, \bibinfo {author} {\bibfnamefont {I.}~\bibnamefont {Pi\ifmmode~\check{z}\else \v{z}\fi{}orn}}, \bibinfo {author} {\bibfnamefont {H.}~\bibnamefont {Verschelde}},\ and\ \bibinfo {author} {\bibfnamefont {F.}~\bibnamefont {Verstraete}},\ }\bibfield  {title} {\bibinfo {title} {{Time-dependent variational principle for quantum lattices}},\ }\href {https://doi.org/10.1103/PhysRevLett.107.070601} {\bibfield  {journal} {\bibinfo  {journal} {Phys. Rev. Lett.}\ }\textbf {\bibinfo {volume} {107}},\ \bibinfo {pages} {070601} (\bibinfo {year} {2011})}\BibitemShut {NoStop}%
    \bibitem [{\citenamefont {Goto}\ and\ \citenamefont {Danshita}(2019)}]{Goto2019Performance}%
      \BibitemOpen
      \bibfield  {author} {\bibinfo {author} {\bibfnamefont {S.}~\bibnamefont {Goto}}\ and\ \bibinfo {author} {\bibfnamefont {I.}~\bibnamefont {Danshita}},\ }\bibfield  {title} {\bibinfo {title} {{Performance of the time-dependent variational principle for matrix product states in the long-time evolution of a pure state}},\ }\href {https://doi.org/10.1103/PhysRevB.99.054307} {\bibfield  {journal} {\bibinfo  {journal} {Phys. Rev. B}\ }\textbf {\bibinfo {volume} {99}},\ \bibinfo {pages} {054307} (\bibinfo {year} {2019})}\BibitemShut {NoStop}%
    \bibitem [{\citenamefont {Paeckel}\ \emph {et~al.}(2019)\citenamefont {Paeckel}, \citenamefont {Köhler}, \citenamefont {Swoboda}, \citenamefont {Manmana}, \citenamefont {Schollwöck},\ and\ \citenamefont {Hubig}}]{Sebastian2019Time}%
      \BibitemOpen
      \bibfield  {author} {\bibinfo {author} {\bibfnamefont {S.}~\bibnamefont {Paeckel}}, \bibinfo {author} {\bibfnamefont {T.}~\bibnamefont {Köhler}}, \bibinfo {author} {\bibfnamefont {A.}~\bibnamefont {Swoboda}}, \bibinfo {author} {\bibfnamefont {S.~R.}\ \bibnamefont {Manmana}}, \bibinfo {author} {\bibfnamefont {U.}~\bibnamefont {Schollwöck}},\ and\ \bibinfo {author} {\bibfnamefont {C.}~\bibnamefont {Hubig}},\ }\bibfield  {title} {\bibinfo {title} {{Time-evolution methods for matrix-product states}},\ }\href {https://doi.org/https://doi.org/10.1016/j.aop.2019.167998} {\bibfield  {journal} {\bibinfo  {journal} {Ann. Phys.}\ }\textbf {\bibinfo {volume} {411}},\ \bibinfo {pages} {167998} (\bibinfo {year} {2019})}\BibitemShut {NoStop}%
    \bibitem [{\citenamefont {Barthel}(2016)}]{Barthel2016Matrix}%
      \BibitemOpen
      \bibfield  {author} {\bibinfo {author} {\bibfnamefont {T.}~\bibnamefont {Barthel}},\ }\bibfield  {title} {\bibinfo {title} {{Matrix product purifications for canonical ensembles and quantum number distributions}},\ }\href {https://doi.org/10.1103/PhysRevB.94.115157} {\bibfield  {journal} {\bibinfo  {journal} {Phys. Rev. B}\ }\textbf {\bibinfo {volume} {94}},\ \bibinfo {pages} {115157} (\bibinfo {year} {2016})}\BibitemShut {NoStop}%
    \bibitem [{\citenamefont {Feiguin}\ and\ \citenamefont {White}(2005)}]{Feiguin2005Finite}%
      \BibitemOpen
      \bibfield  {author} {\bibinfo {author} {\bibfnamefont {A.~E.}\ \bibnamefont {Feiguin}}\ and\ \bibinfo {author} {\bibfnamefont {S.~R.}\ \bibnamefont {White}},\ }\bibfield  {title} {\bibinfo {title} {{Finite-temperature density matrix renormalization using an enlarged Hilbert space}},\ }\href {https://doi.org/10.1103/PhysRevB.72.220401} {\bibfield  {journal} {\bibinfo  {journal} {Phys. Rev. B}\ }\textbf {\bibinfo {volume} {72}},\ \bibinfo {pages} {220401} (\bibinfo {year} {2005})}\BibitemShut {NoStop}%
    \bibitem [{\citenamefont {Fishman}\ \emph {et~al.}(2022)\citenamefont {Fishman}, \citenamefont {White},\ and\ \citenamefont {Stoudenmire}}]{Matthew2022The}%
      \BibitemOpen
      \bibfield  {author} {\bibinfo {author} {\bibfnamefont {M.}~\bibnamefont {Fishman}}, \bibinfo {author} {\bibfnamefont {S.~R.}\ \bibnamefont {White}},\ and\ \bibinfo {author} {\bibfnamefont {E.~M.}\ \bibnamefont {Stoudenmire}},\ }\bibfield  {title} {\bibinfo {title} {{The ITensor software library for Tensor Network calculations}},\ }\href {https://doi.org/10.21468/SciPostPhysCodeb.4} {\bibfield  {journal} {\bibinfo  {journal} {SciPost Phys. Codebases}\ ,\ \bibinfo {pages} {4}} (\bibinfo {year} {2022})}\BibitemShut {NoStop}%
    \bibitem [{\citenamefont {Calabrese}\ and\ \citenamefont {Cardy}(2005)}]{Calabrese_2005}%
      \BibitemOpen
      \bibfield  {author} {\bibinfo {author} {\bibfnamefont {P.}~\bibnamefont {Calabrese}}\ and\ \bibinfo {author} {\bibfnamefont {J.}~\bibnamefont {Cardy}},\ }\bibfield  {title} {\bibinfo {title} {{Evolution of entanglement entropy in one-dimensional systems}},\ }\href {https://doi.org/10.1088/1742-5468/2005/04/P04010} {\bibfield  {journal} {\bibinfo  {journal} {J. Stat. Mech.: Theory Exp.}\ }\textbf {\bibinfo {volume} {2005}},\ \bibinfo {pages} {P04010 (2005)}}\BibitemShut {NoStop}%
    \bibitem [{\citenamefont {Legeza}\ and\ \citenamefont {S\'olyom}(2006)}]{PhysRevLett.96.116401}%
      \BibitemOpen
      \bibfield  {author} {\bibinfo {author} {\bibfnamefont {O.}~\bibnamefont {Legeza}}\ and\ \bibinfo {author} {\bibfnamefont {J.}~\bibnamefont {S\'olyom}},\ }\bibfield  {title} {\bibinfo {title} {{Two-site entropy and quantum phase transitions in low-dimensional models}},\ }\href {https://doi.org/10.1103/PhysRevLett.96.116401} {\bibfield  {journal} {\bibinfo  {journal} {Phys. Rev. Lett.}\ }\textbf {\bibinfo {volume} {96}},\ \bibinfo {pages} {116401} (\bibinfo {year} {2006})}\BibitemShut {NoStop}%
    \bibitem [{\citenamefont {Anderlini}\ \emph {et~al.}(2007)\citenamefont {Anderlini}, \citenamefont {Lee}, \citenamefont {Brown}, \citenamefont {Sebby-Strabley}, \citenamefont {Phillips},\ and\ \citenamefont {Porto}}]{Anderlini_2007}%
      \BibitemOpen
      \bibfield  {author} {\bibinfo {author} {\bibfnamefont {M.}~\bibnamefont {Anderlini}}, \bibinfo {author} {\bibfnamefont {P.~J.}\ \bibnamefont {Lee}}, \bibinfo {author} {\bibfnamefont {B.~L.}\ \bibnamefont {Brown}}, \bibinfo {author} {\bibfnamefont {J.}~\bibnamefont {Sebby-Strabley}}, \bibinfo {author} {\bibfnamefont {W.~D.}\ \bibnamefont {Phillips}},\ and\ \bibinfo {author} {\bibfnamefont {J.~V.}\ \bibnamefont {Porto}},\ }\bibfield  {title} {\bibinfo {title} {Controlled exchange interaction between pairs of neutral atoms in an optical lattice},\ }\href {https://doi.org/10.1038/nature06011} {\bibfield  {journal} {\bibinfo  {journal} {Nature}\ }\textbf {\bibinfo {volume} {448}},\ \bibinfo {pages} {452} (\bibinfo {year} {2007})}\BibitemShut {NoStop}%
    \bibitem [{\citenamefont {Mazurenko}\ \emph {et~al.}(2017)\citenamefont {Mazurenko}, \citenamefont {Chiu}, \citenamefont {Ji}, \citenamefont {Parsons}, \citenamefont {Kanasz-Nagy}, \citenamefont {Schmidt}, \citenamefont {Grusdt}, \citenamefont {Demler}, \citenamefont {Greif},\ and\ \citenamefont {Greiner}}]{2017A}%
      \BibitemOpen
      \bibfield  {author} {\bibinfo {author} {\bibfnamefont {A.}~\bibnamefont {Mazurenko}}, \bibinfo {author} {\bibfnamefont {C.~S.}\ \bibnamefont {Chiu}}, \bibinfo {author} {\bibfnamefont {G.}~\bibnamefont {Ji}}, \bibinfo {author} {\bibfnamefont {M.~F.}\ \bibnamefont {Parsons}}, \bibinfo {author} {\bibfnamefont {M.}~\bibnamefont {Kanasz-Nagy}}, \bibinfo {author} {\bibfnamefont {R.}~\bibnamefont {Schmidt}}, \bibinfo {author} {\bibfnamefont {F.}~\bibnamefont {Grusdt}}, \bibinfo {author} {\bibfnamefont {E.}~\bibnamefont {Demler}}, \bibinfo {author} {\bibfnamefont {D.}~\bibnamefont {Greif}},\ and\ \bibinfo {author} {\bibfnamefont {M.}~\bibnamefont {Greiner}},\ }\bibfield  {title} {\bibinfo {title} {{A cold-atom Fermi-Hubbard antiferromagnet}},\ }\href {https://doi.org/10.1038/nature22362} {\bibfield  {journal} {\bibinfo  {journal} {Nature}\ }\textbf {\bibinfo {volume} {545}},\ \bibinfo {pages} {462} (\bibinfo {year} {2017})}\BibitemShut {NoStop}%
    \bibitem [{\citenamefont {Navon}\ \emph {et~al.}(2021)\citenamefont {Navon}, \citenamefont {Smith},\ and\ \citenamefont {Hadzibabic}}]{2021Quantum}%
      \BibitemOpen
      \bibfield  {author} {\bibinfo {author} {\bibfnamefont {N.}~\bibnamefont {Navon}}, \bibinfo {author} {\bibfnamefont {R.~P.}\ \bibnamefont {Smith}},\ and\ \bibinfo {author} {\bibfnamefont {Z.}~\bibnamefont {Hadzibabic}},\ }\bibfield  {title} {\bibinfo {title} {{Quantum gases in optical boxes}},\ }\href {https://doi.org/10.1038/s41567-021-01403-z} {\bibfield  {journal} {\bibinfo  {journal} {Nat. Phys.}\ }\textbf {\bibinfo {volume} {17}},\ \bibinfo {pages} {1334} (\bibinfo {year} {2021})}\BibitemShut {NoStop}%
    \bibitem [{\citenamefont {Gross}\ and\ \citenamefont {Bakr}(2021)}]{Gross_2021}%
      \BibitemOpen
      \bibfield  {author} {\bibinfo {author} {\bibfnamefont {C.}~\bibnamefont {Gross}}\ and\ \bibinfo {author} {\bibfnamefont {W.~S.}\ \bibnamefont {Bakr}},\ }\bibfield  {title} {\bibinfo {title} {{Quantum gas microscopy for single atom and spin detection}},\ }\href {https://doi.org/10.1038/s41567-021-01370-5} {\bibfield  {journal} {\bibinfo  {journal} {Nat. Phys.}\ }\textbf {\bibinfo {volume} {17}},\ \bibinfo {pages} {1316} (\bibinfo {year} {2021})}\BibitemShut {NoStop}%
    \bibitem [{\citenamefont {Moeckel}\ and\ \citenamefont {Kehrein}(2008)}]{PhysRevLett.100.175702}%
      \BibitemOpen
      \bibfield  {author} {\bibinfo {author} {\bibfnamefont {M.}~\bibnamefont {Moeckel}}\ and\ \bibinfo {author} {\bibfnamefont {S.}~\bibnamefont {Kehrein}},\ }\bibfield  {title} {\bibinfo {title} {{Interaction quench in the Hubbard model}},\ }\href {https://doi.org/10.1103/PhysRevLett.100.175702} {\bibfield  {journal} {\bibinfo  {journal} {Phys. Rev. Lett.}\ }\textbf {\bibinfo {volume} {100}},\ \bibinfo {pages} {175702} (\bibinfo {year} {2008})}\BibitemShut {NoStop}%
    \end{thebibliography}
\end{document}